\documentclass[compsoc, conference, a4paper, 10pt, times]{IEEEtran}

\usepackage{cite}
\usepackage{amsmath,amssymb,amsfonts}
\usepackage{algorithmic}
\usepackage{enumitem}
\usepackage{graphicx}
\usepackage{textcomp}
\usepackage{xcolor}
\usepackage{float}
\usepackage{booktabs}
\usepackage[hidelinks]{hyperref}

\begin{document}

\title{Results of the 2023 Census of Beat Saber Users: Virtual Reality Gaming Population Insights and Factors Affecting Virtual Reality E-Sports Performance}


\author{\IEEEauthorblockN{Vivek Nair}
\IEEEauthorblockA{\textit{UC Berkeley} \\
Berkeley, California \\
vcn@berkeley.edu}
\and
\IEEEauthorblockN{Viktor Radulov}
\IEEEauthorblockA{\textit{BeatLeader} \\
Kyiv, Ukraine \\
golova@golova.dev}
\and
\IEEEauthorblockN{James F. O'Brien}
\IEEEauthorblockA{\textit{UC Berkeley} \\
Berkeley, California \\
job@berkeley.edu}
}

\maketitle

\begin{abstract}
The emergence of affordable standalone virtual reality (VR) devices has allowed VR technology to reach mass-market adoption in recent years, driven primarily by the popularity of VR gaming applications such as Beat Saber. However, despite being the top-grossing VR application to date and the most popular VR e-sport, the population of over 6 million Beat Saber users has not yet been widely studied. In this report, we present a large-scale comprehensive survey of Beat Saber players (N=1,006) that sheds light on several important aspects of this population, including their background, biometrics, demographics, health information, behavioral patterns, and technical device specifications. We further provide insights into the emerging field of VR \mbox{e-sports} by analyzing correlations between responses and an authoritative measure of in-game performance.
\end{abstract}


\begin{IEEEkeywords}
virtual reality, gaming, esports, beat saber, census, survey, demographics, biometrics, health
\end{IEEEkeywords}

\section{Introduction}

``Beat Saber'' is a popular virtual reality rhythm game published by Beat Games where players slice blocks in time to a musical track with a pair of sabers they hold in each hand \cite{studio_beat_nodate}.
Beat Saber is the most popular and highest-grossing VR application, with over 6.2 million copies sold.

Beat Saber play is segmented into ``maps,'' which consist of a music track and a corresponding series of blocks that are synchronized to the song. The game includes approximately 200 official maps, but there are over 100,000 user-created ``custom'' maps for players to choose from \cite{noauthor_beatsaver_nodate}. After selecting and playing a map, the player is presented with a score based on how accurately they cut each block, with additional points for wider swings \cite{teknozfr_beat_2019}.

``BeatLeader'' is an open-source Beat Saber extension that maintains leaderboards of high scores for custom Beat Saber maps \cite{noauthor_beatleader_nodate}. Internally, Beat Leader generates a composite score known as ``performance points'' (PP) for each player, based on the score and difficulty of each map in their play history. Beat Saber players may choose to install the BeatLeader extension in order to compete with other players to achieve a higher ``rank'' on the BeatLeader leaderboards based on their PP. As of April 2023, nearly 100,000 users have installed the extension and uploaded one or more Beat Saber scores to BeatLeader. 

Despite the recent widespread attention and adoption, the exact constituency of BeatLeader's user base, and the factors influencing rank and PP, are not well understood. 
In an effort to increase understanding of this community, this report summarizes results from a census of BeatLeader users (and of Beat Saber players in general).

\section{Protocol}
We conducted an online survey consisting of about 50 questions, including questions about each participant's background, demographics, technical device specifications, behavioral patterns, health information, biometrics, and anthropometric measurements. Participation in the survey was voluntary, with all questions in the survey being optional, and no consequences for non-participation.

Beat Saber players were invited to participate in the survey via an announcement released through the official social media channels for BeatLeader. Participants were not compensated, but were given the option to add an ``achievement'' to their BeatLeader profile page, which cannot be exchanged for anything of monetary value.

The survey was conducted from April 15th, 2023 to May 1st, 2023, with 1,006 responses collected in that time. Our hope is to repeat the survey periodically to track trends and developments as the community evolves. \\

\noindent \textbf{Ethics}
We constructed our survey with significant attention to ethical considerations. Accordingly, we refrained from asking questions that could be viewed as highly sensitive, and did not solicit responses from vulnerable populations, including minors under the age of 18. Participants were required to read and agree to a thorough informed consent document prior to inclusion in the study.

An additional source of data was scoring information collected by BeatLeader. This data was already broadly, publicly available prior to this study, and was made available to us in accordance with BeatLeader's privacy policy.

This study has been reviewed and approved by UC Berkeley's Committee for Protection of Human Subjects (CPHS) as protocol \#2023-03-16120. \\

\noindent \textbf{Acknowledgments}
We appreciate the support of Beni Issler, Atticus Cull, Džiugas Ramonas, Louis Rosenberg, and Dawn Song.
This work was funded by the National Science Foundation, the National Physical Science Consortium, the Fannie and John Hertz Foundation, and the Center for Responsible Decentralized Intelligence.

\eject

\clearpage

\section{Response Distributions}

In this section, we present verbatim the questions asked of all survey participants, and provide the raw response tallies and sample distributions corresponding to all non-empty responses.

\subsection{Surveyed Population}

The first group of questions aims to characterize the population from which responses were sampled. Players were asked to list all ScoreSaber \cite{noauthor_scoresaber_nodate} and BeatLeader account(s) they had used to upload scores. While targeted broadly at Beat Saber users, the responses primarily originate from users of the BeatLeader extension, due to the promotion of the survey via BeatLeader's social media.

We expect that in comparison to the general population of Beat Saber players, the BeatLeader population, and thus survey sample, will have a bias towards greater skill level; players usually install the BeatLeader extension only if they are interested in competitively sharing their scores.

\bigskip

\noindent \textbf{Mods.} Have you ever played BeatSaber with the ScoreSaber and/or BeatLeader mods installed?
\\ \noindent \includegraphics[width=\linewidth]{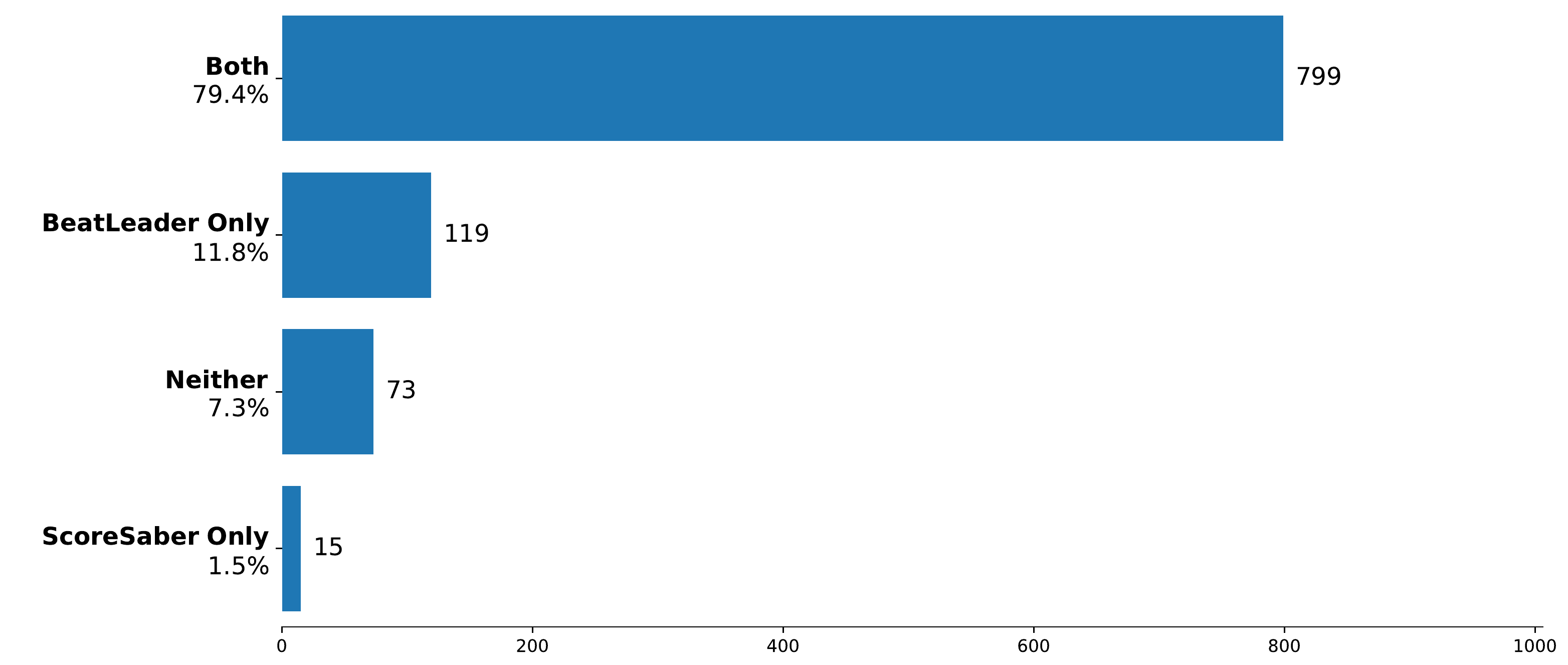}

\noindent \textbf{Secondary Accounts.} Have you ever submitted a score using a BeatLeader or ScoreSaber account not listed above?
\\ \noindent \includegraphics[width=\linewidth]{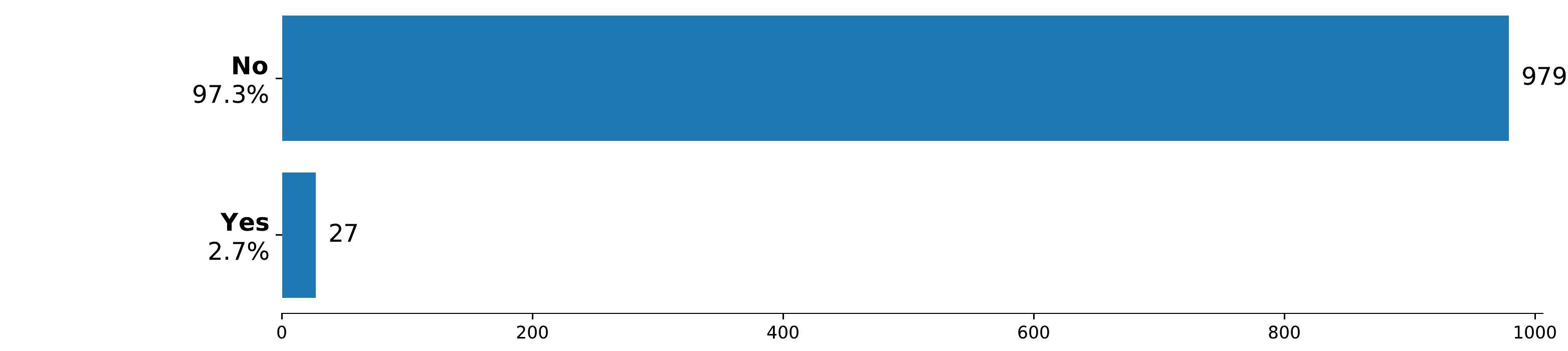}

\noindent \textbf{Multiple Users.} Has any person other than yourself ever submitted a score to any of the BeatLeader or ScoreSaber accounts listed above?
\\ \noindent \includegraphics[width=\linewidth]{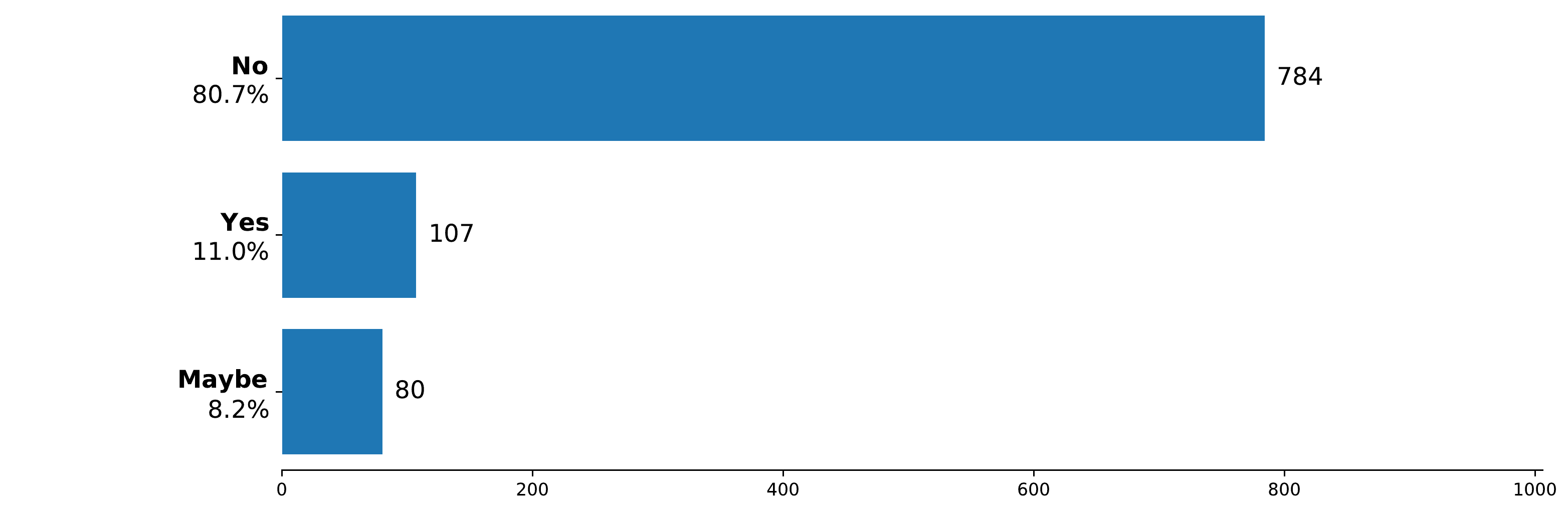}

\noindent \textbf{Play Time.} To the nearest hour, how many total hours have you spent playing Beat Saber?
\\ \noindent \includegraphics[width=\linewidth]{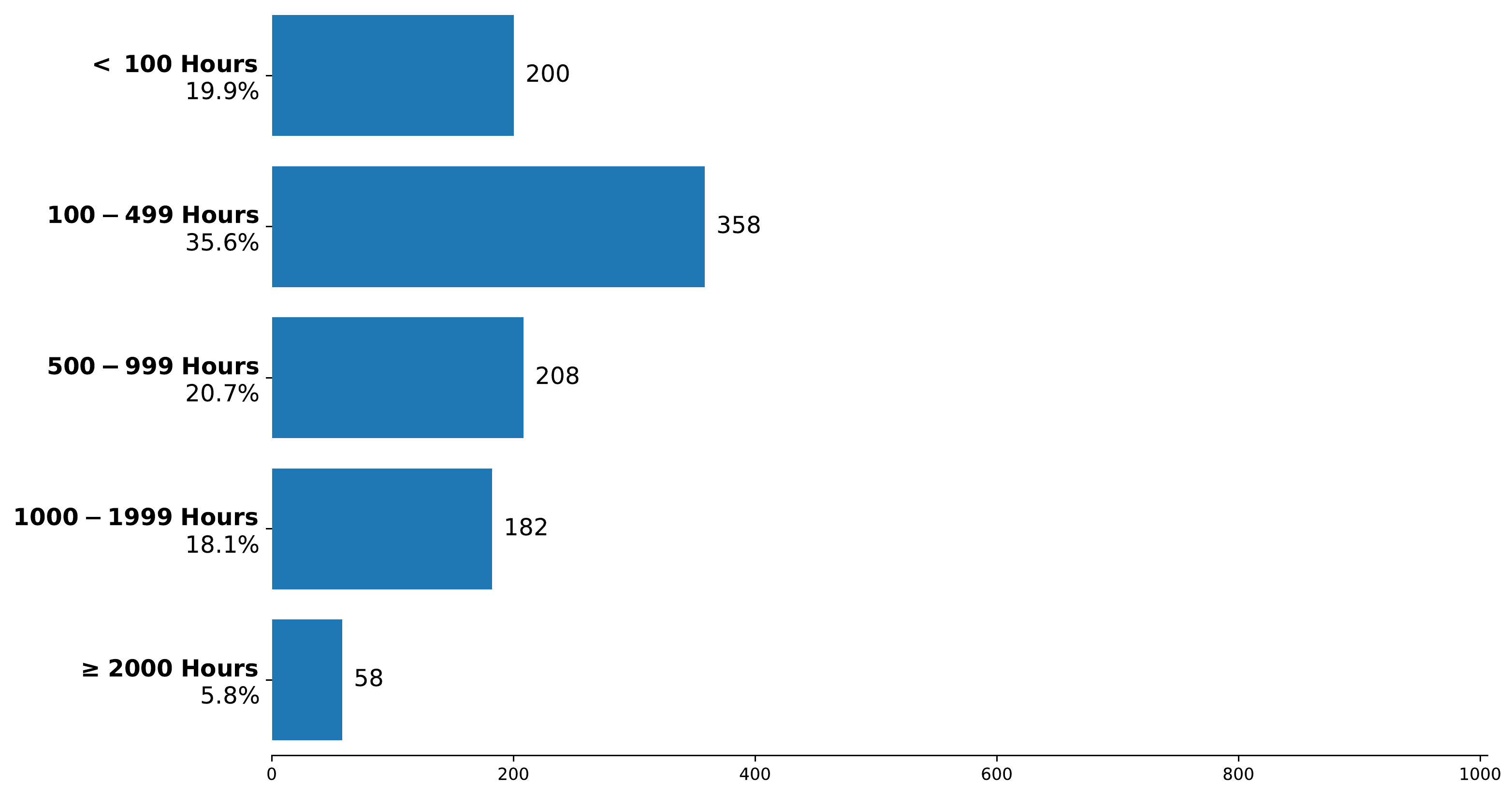}

\eject

\subsection{Demographics}

The second set of questions surveyed the demographics of the players. The questions in this section were adapted from the official questionnaire of the 2020 United States Census. We note that the results may be affected by the fact that the survey was only offered in English. 

\bigskip

\noindent \textbf{Sex.} What is your sex?
\\ \noindent \includegraphics[width=\linewidth]{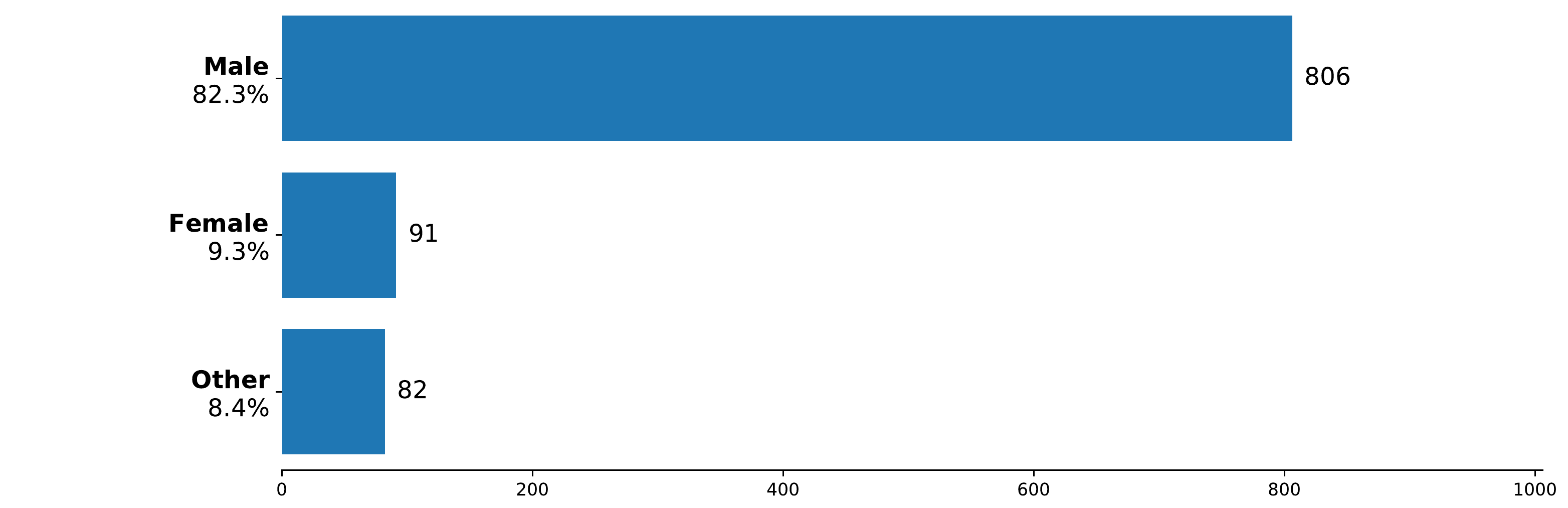}

\noindent \textbf{Age.} What is your age in years?
\\ \noindent \includegraphics[width=\linewidth]{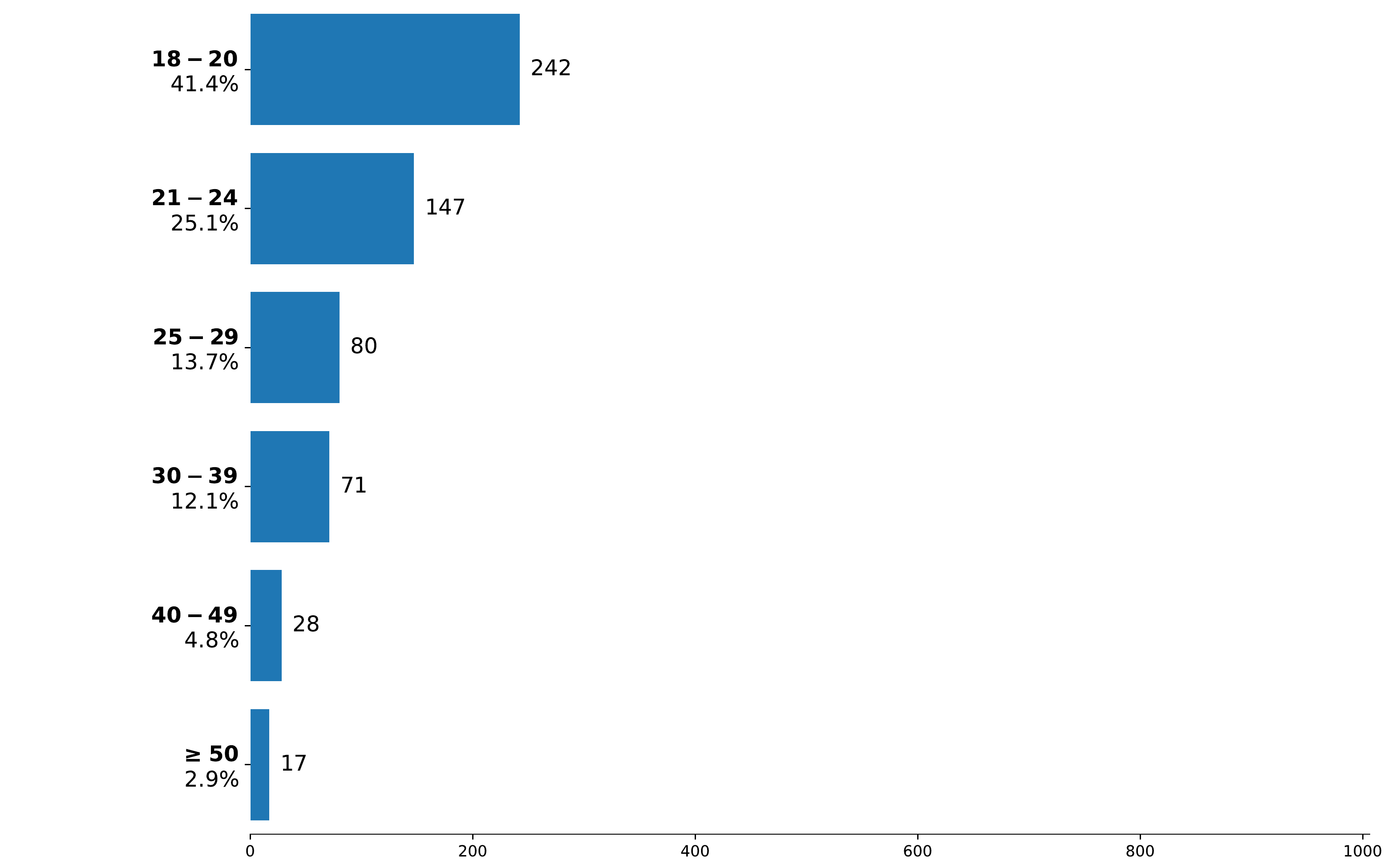}

\noindent \textbf{Employment Status.} Which of the following options best represents your current employment status?
\\ \noindent \includegraphics[width=\linewidth]{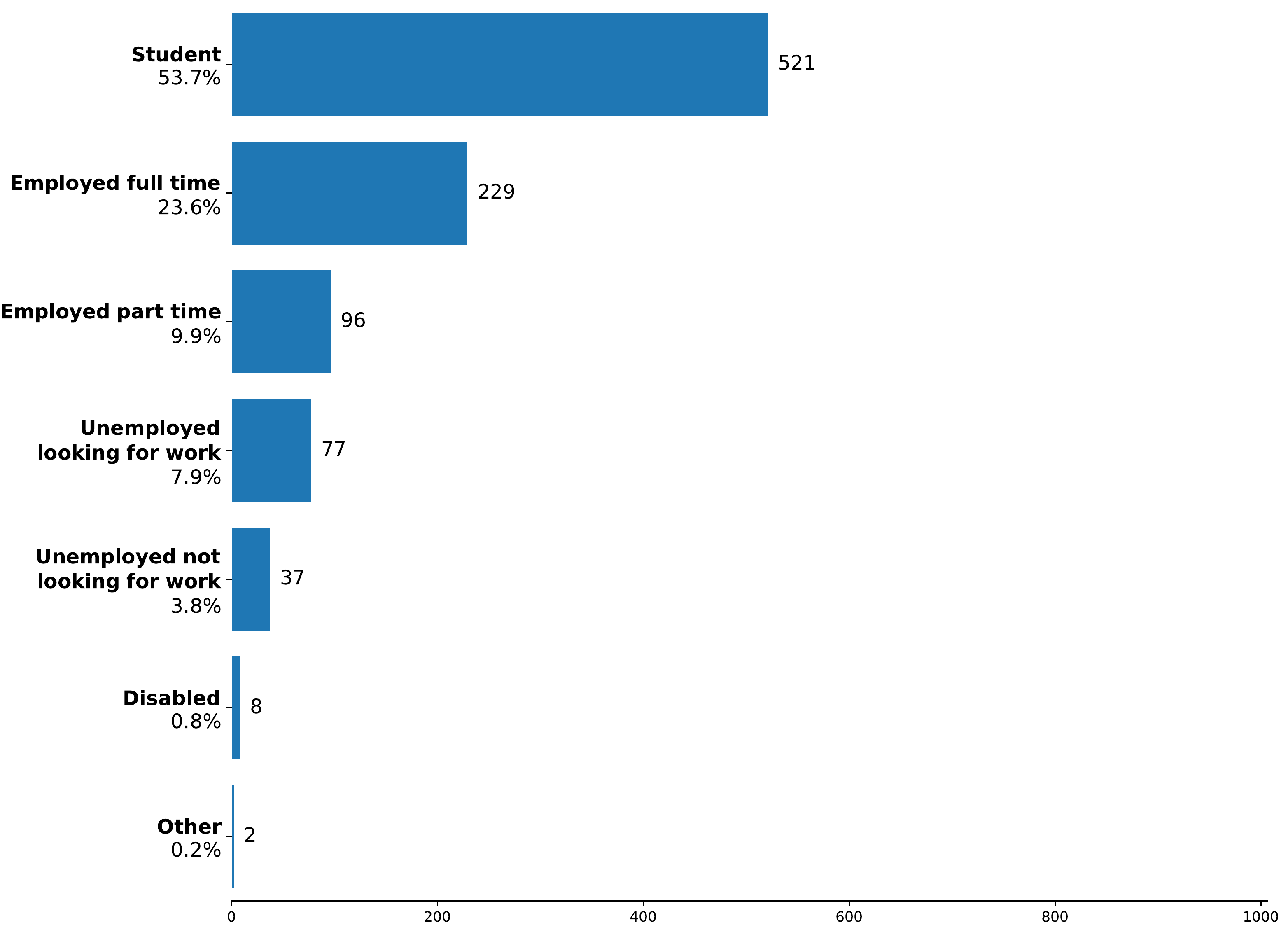}

\noindent \textbf{Marital Status.} Which of the following options best represents your current marital status?
\\ \noindent \includegraphics[width=\linewidth]{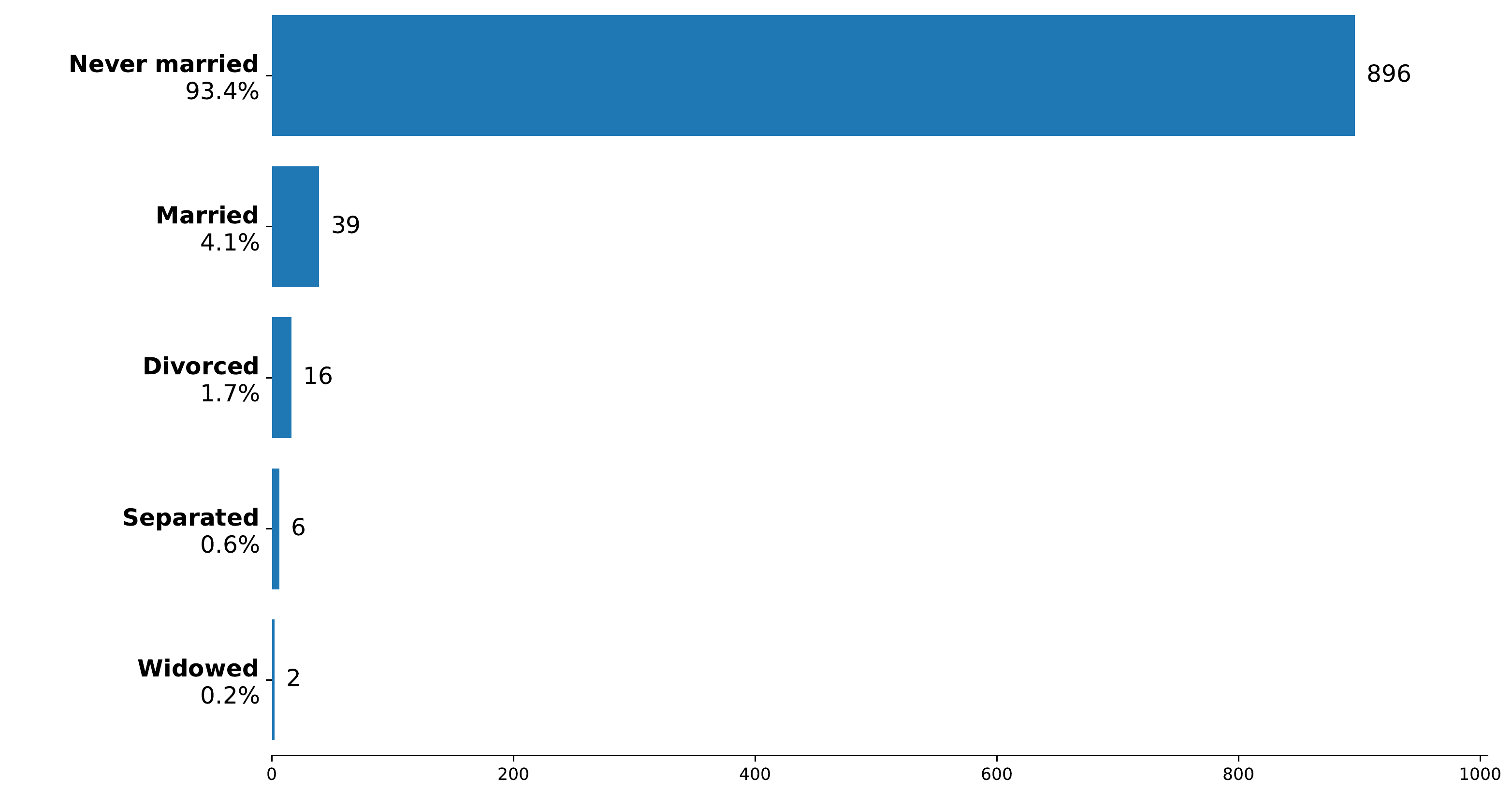}

\eject

\noindent \textbf{Languages.} Which languages do you speak fluently? If multiple, list all languages spoken in order of proficiency.
\\ \noindent \includegraphics[width=\linewidth]{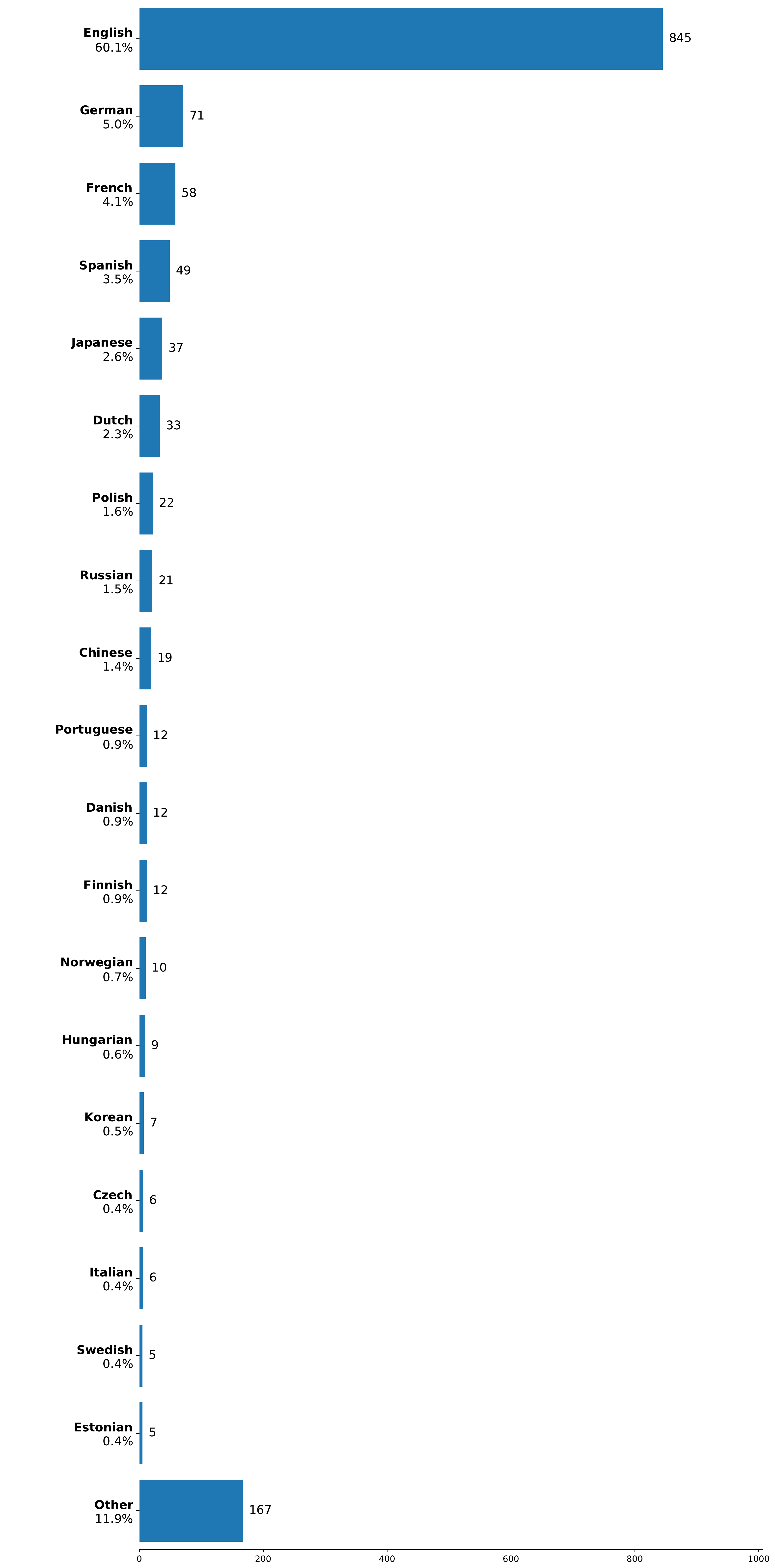}

\noindent \textbf{Educational Status.} What is the highest degree or level of school you have completed?
\\ \noindent \includegraphics[width=\linewidth]{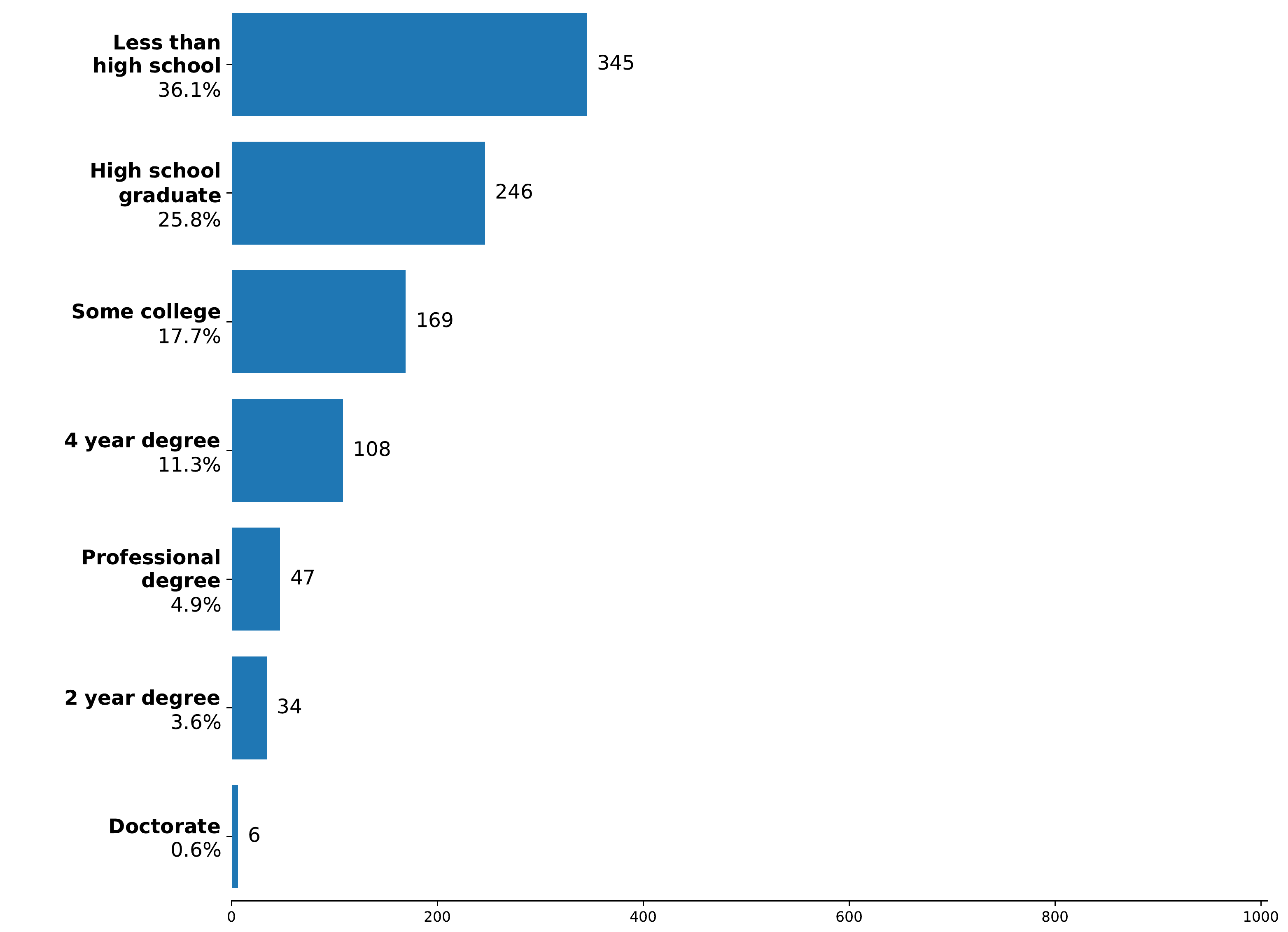}

\eject

\noindent \textbf{Income.} Which of the following options best represents your total gross income in 2022? Convert your answer to United States Dollars (USD).
\\ \noindent \includegraphics[width=\linewidth]{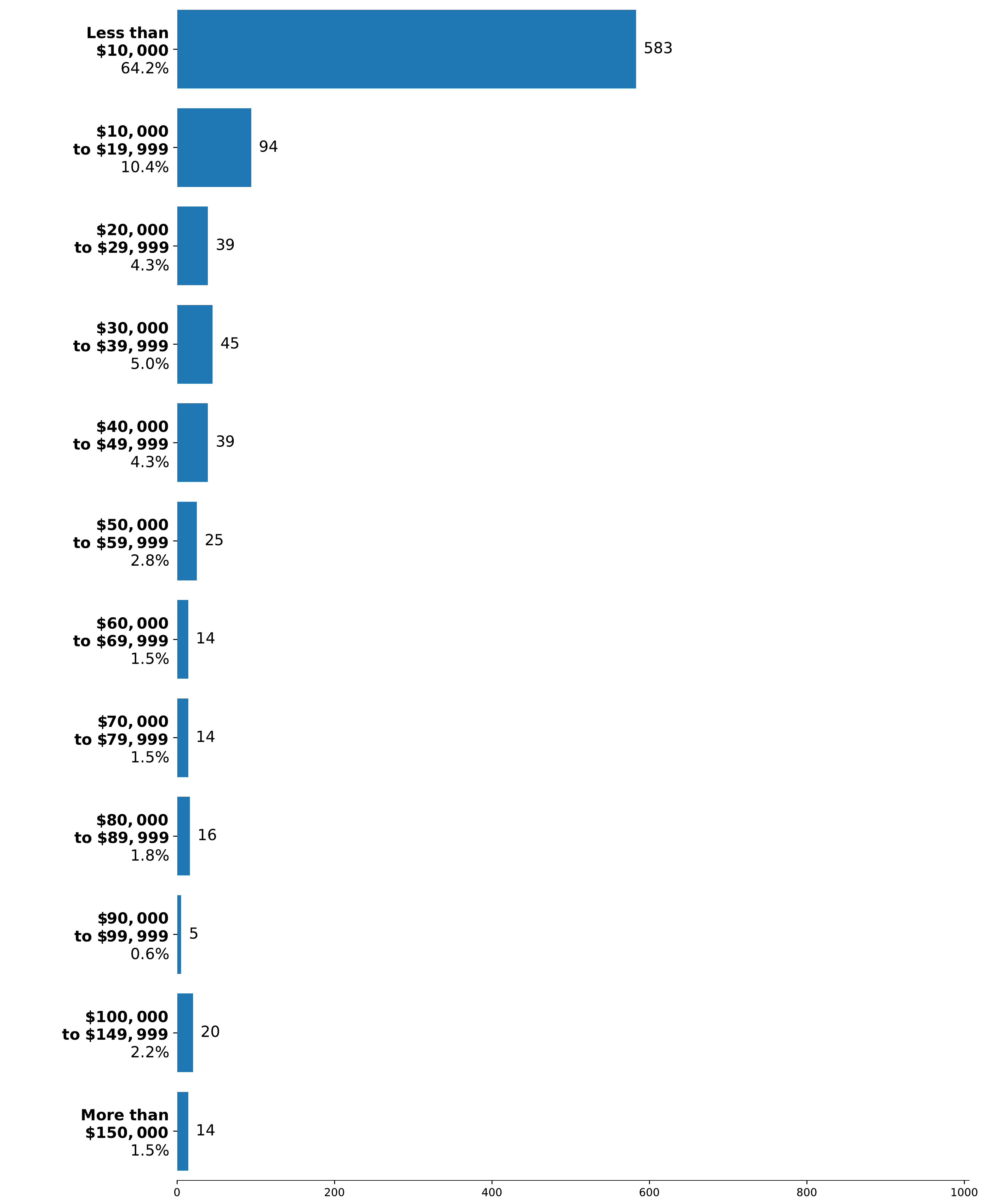}

\noindent \textbf{Ethnicity.} What is your ethnicity?
\\ \noindent \includegraphics[width=\linewidth]{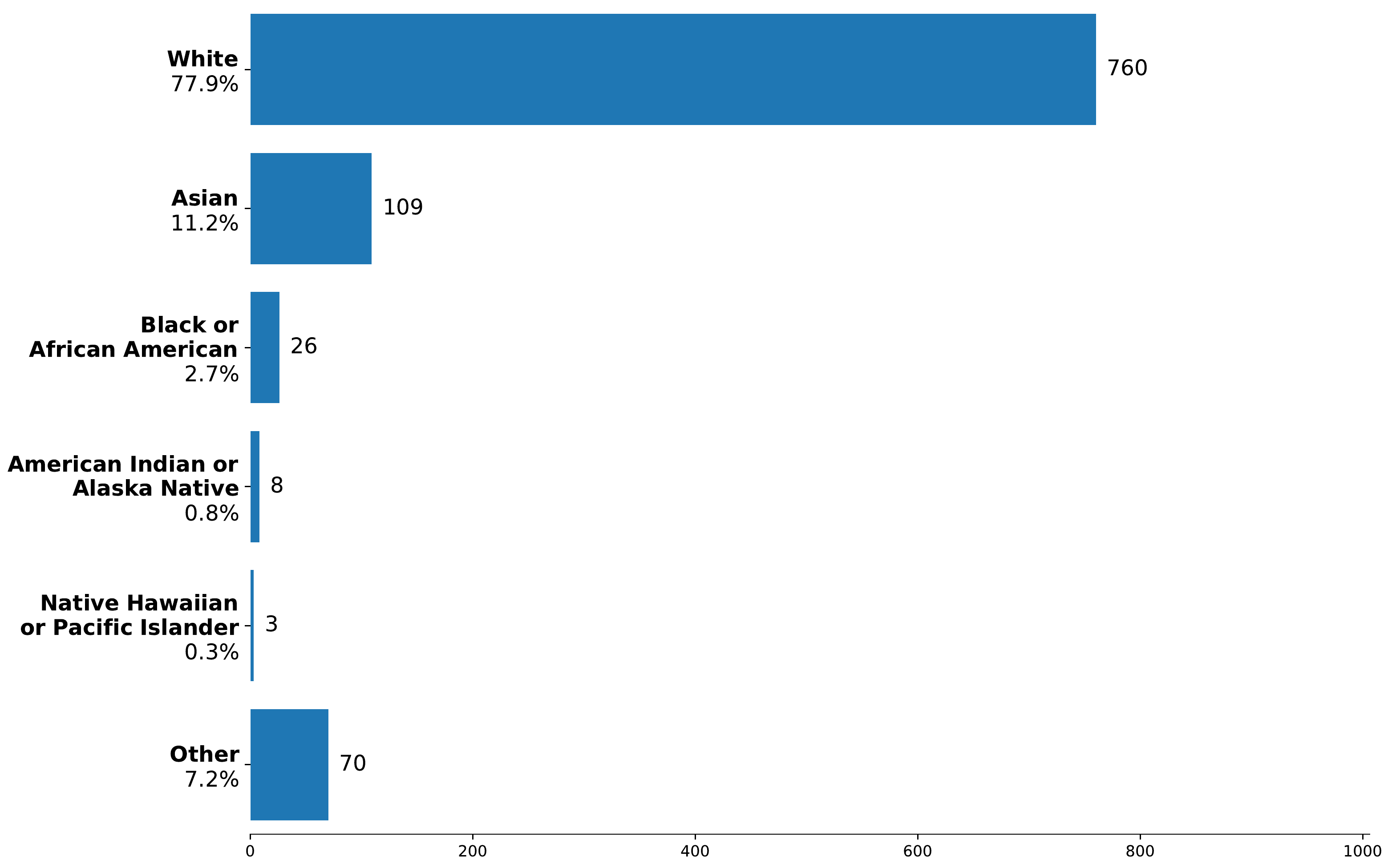}

\noindent \textbf{Political Orientation.} Which of the following generally best represents your political views?
\\ \noindent \includegraphics[width=\linewidth]{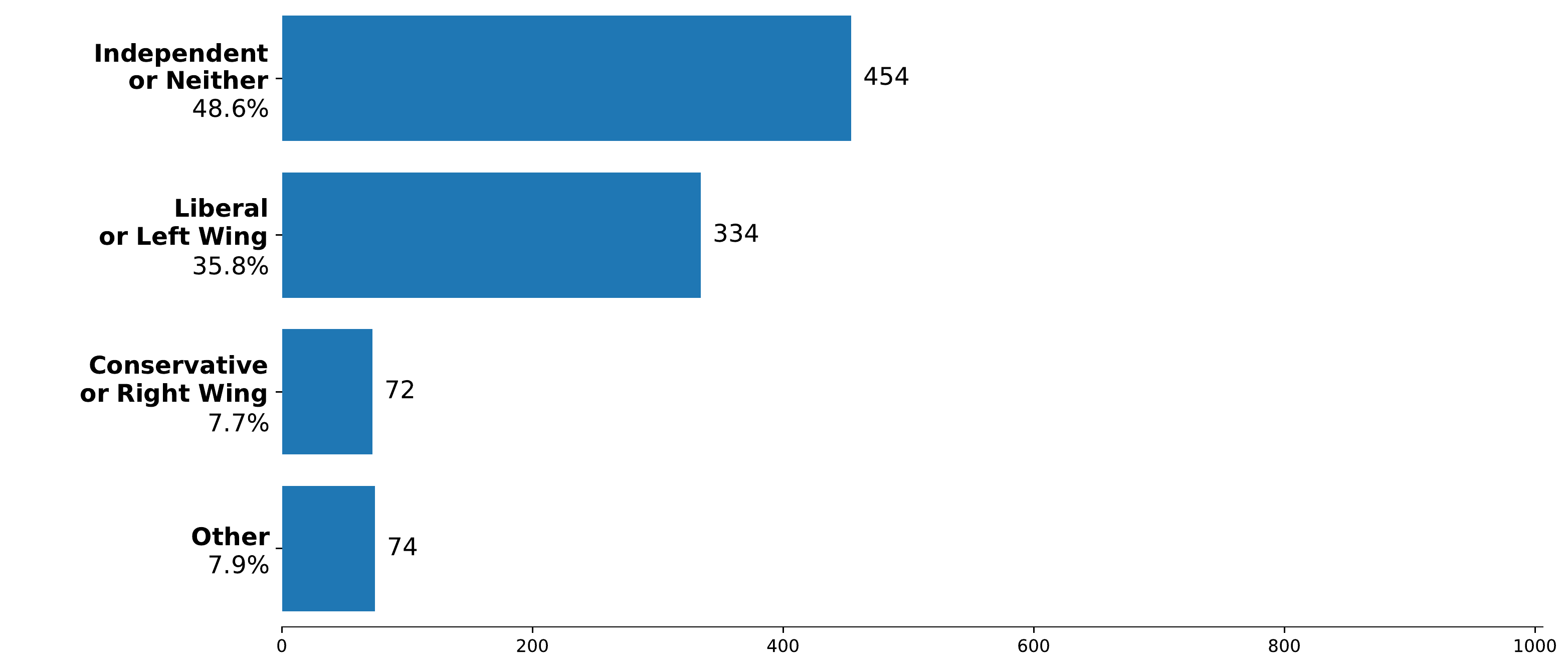}

\eject

\subsection{Technical Specifications}


Users of tethered VR devices (such as Valve Index) were asked to upload a copy of their Steam system report, which details the hardware and software configuration of their machine. Users of standalone VR devices, such as Oculus Quest, are not included in this section, unless used in combination with a PC (e.g., via Quest Link).

\bigskip

\noindent \textbf{CPU Brand.} According to the Steam system report, what is the vendor of the CPU in the user's PC?
\\ \noindent \includegraphics[width=\linewidth]{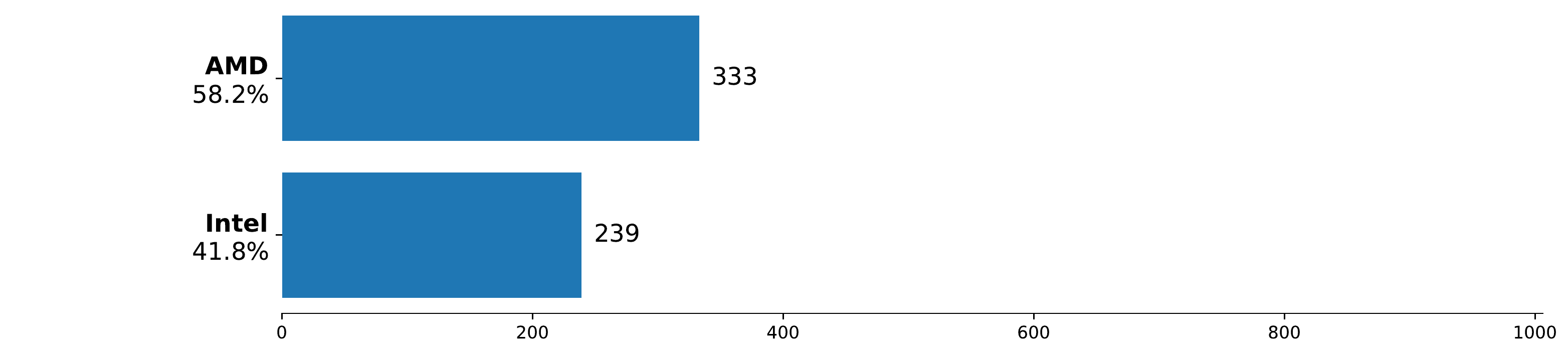}

\noindent \textbf{Logical Cores.} According to the Steam system report, how many logical CPU cores are in the user's PC?
\\ \noindent \includegraphics[width=\linewidth]{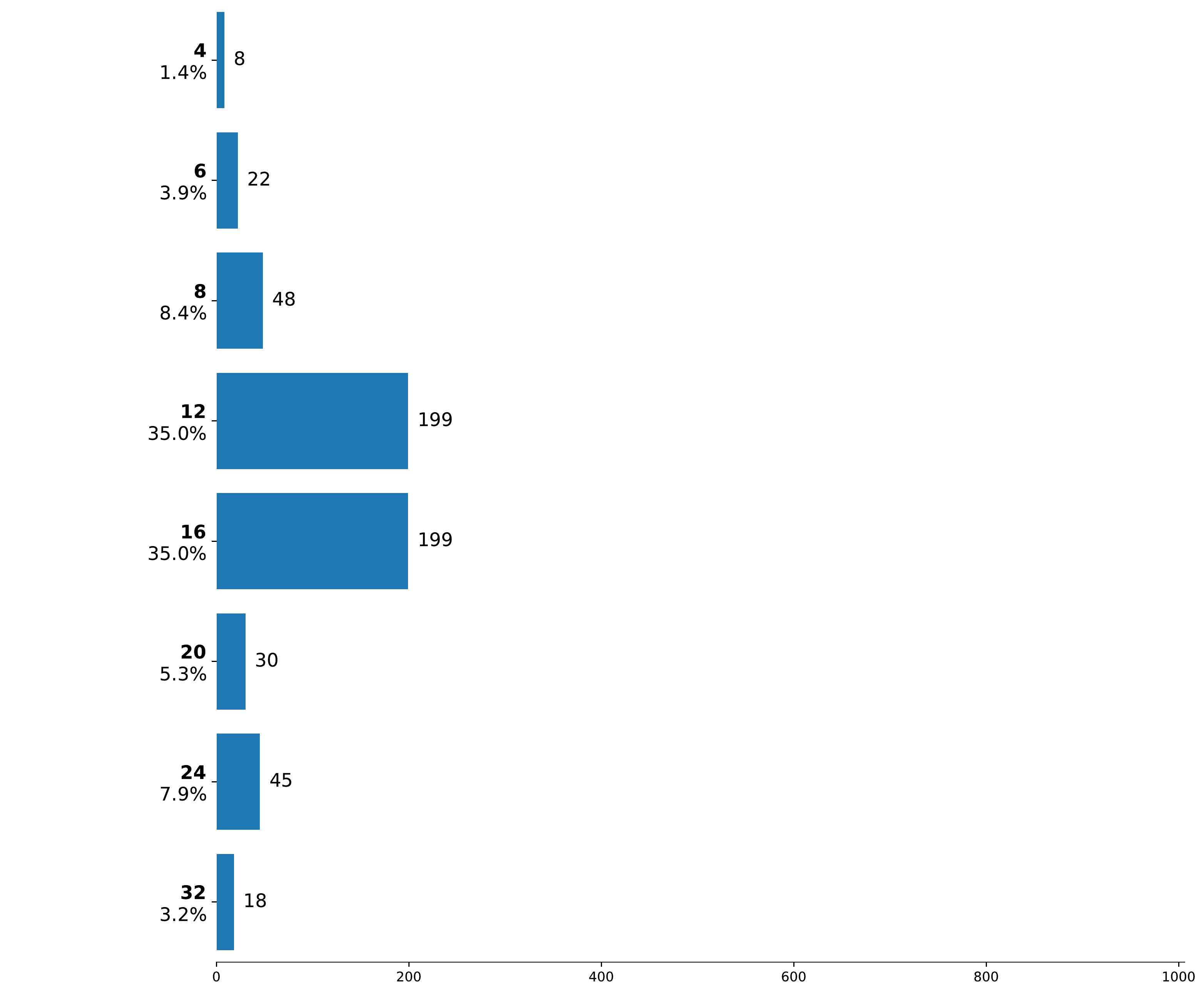}

\noindent \textbf{CPU Speed.} According to the Steam system report, what is the base CPU clock speed in the user's PC?
\\ \noindent \includegraphics[width=\linewidth]{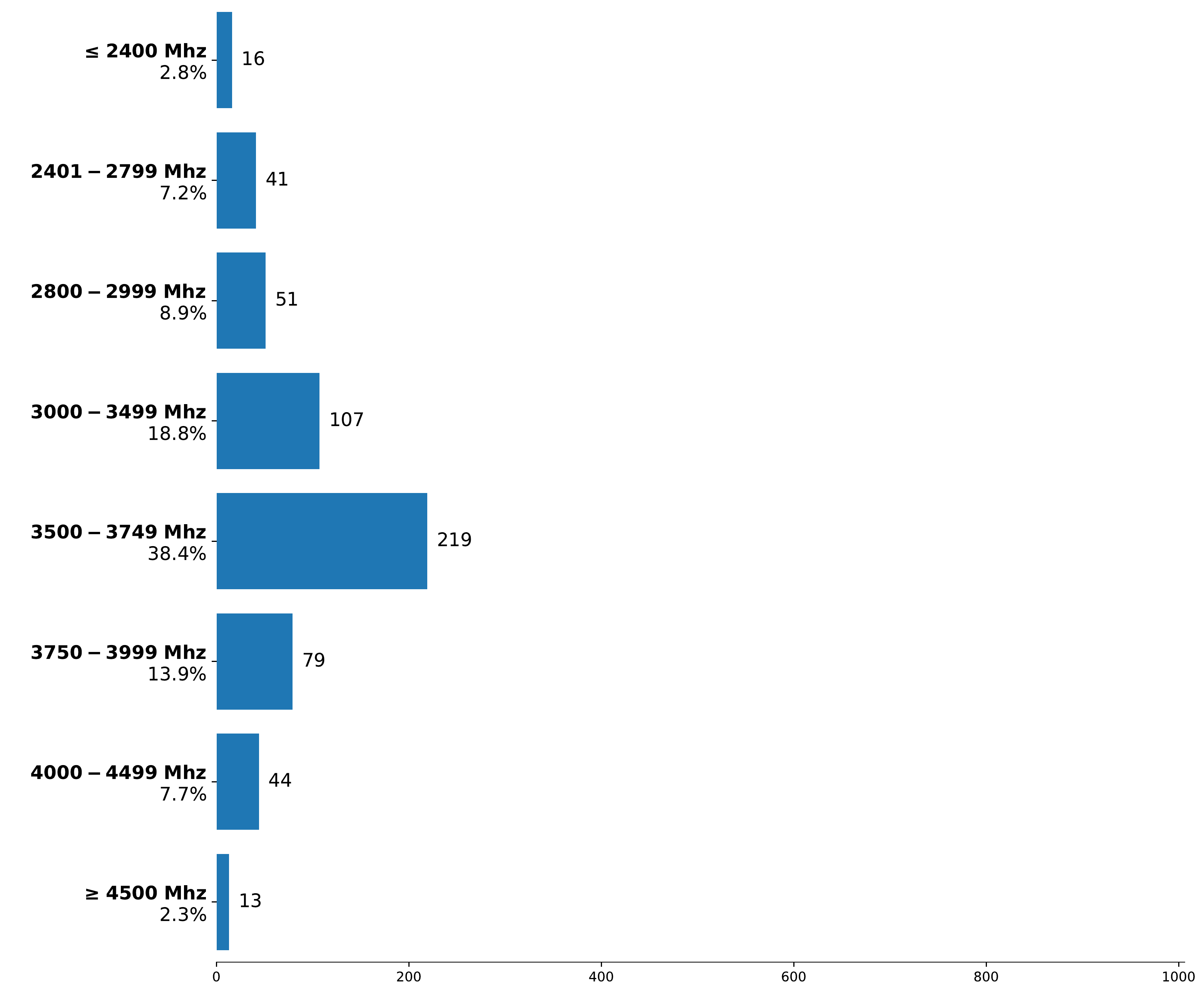}

\noindent \textbf{Form Factor.} According to the Steam system report, is the user's PC a laptop or desktop?
\\ \noindent \includegraphics[width=\linewidth]{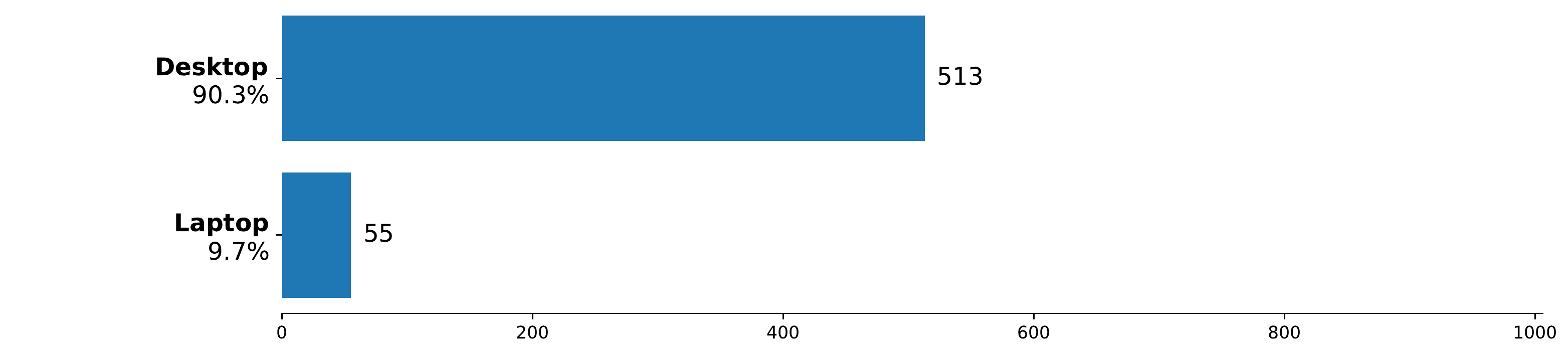}

\eject

\noindent \textbf{Operating System.} According to the Steam system report, what is the operating system of the user's PC?
\\ \noindent \includegraphics[width=\linewidth]{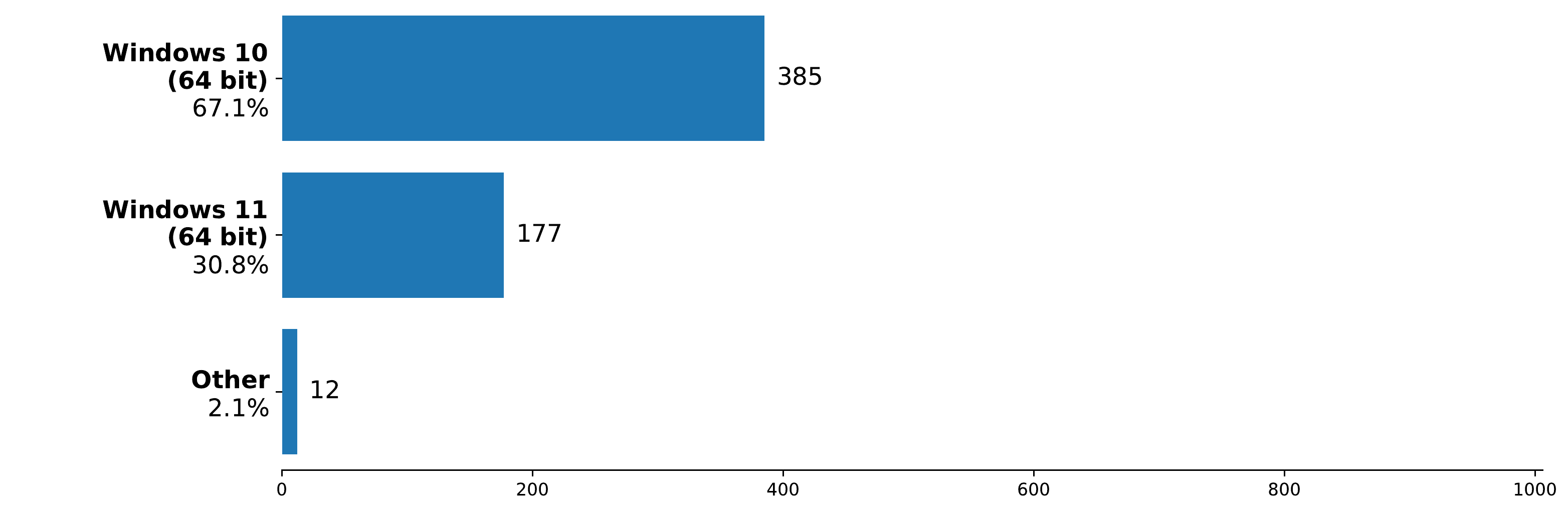}

\noindent \textbf{System Memory.} According to the Steam system report, how much RAM is in the user's PC?
\\ \noindent \includegraphics[width=\linewidth]{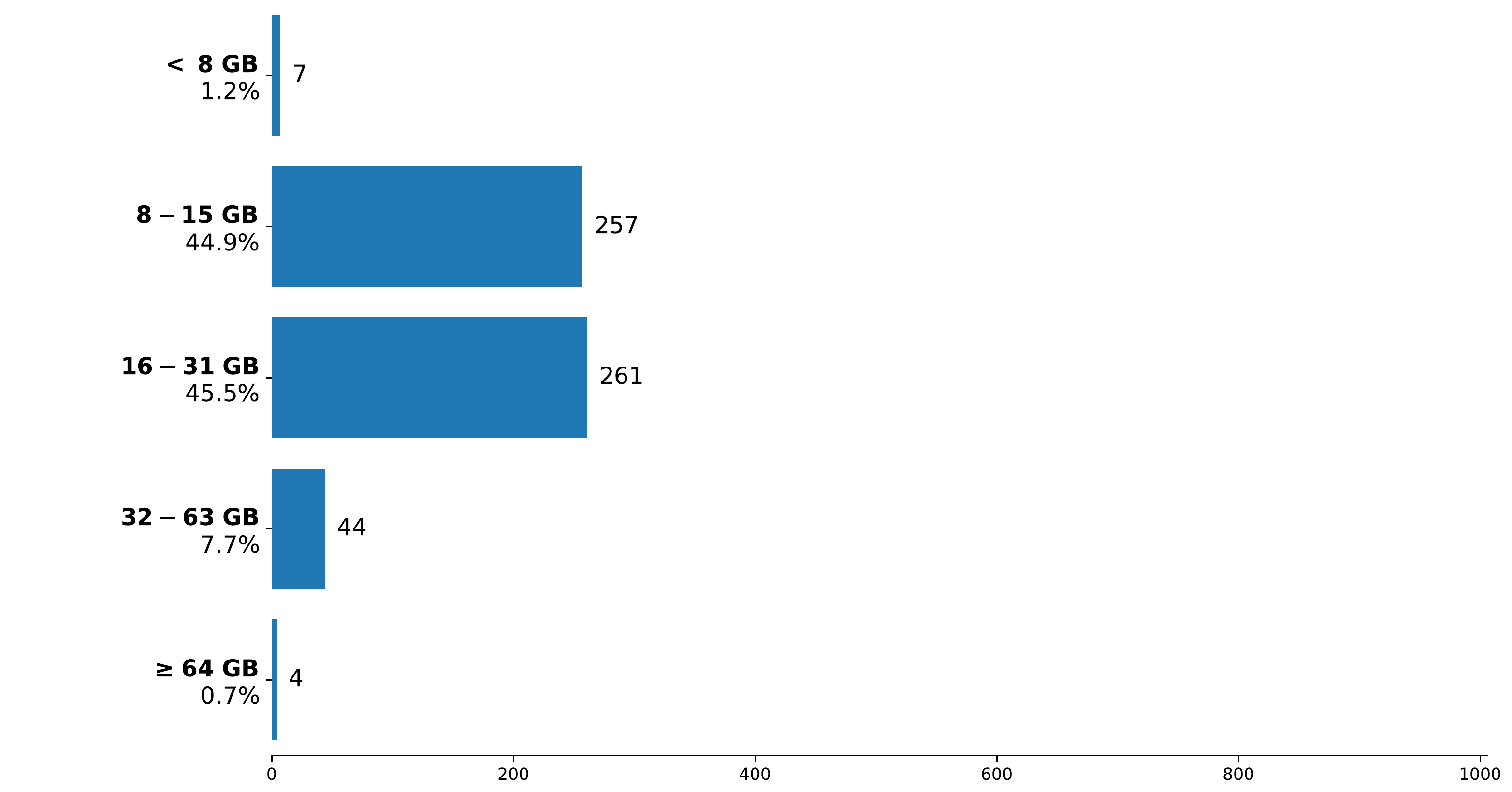}

\noindent \textbf{Drive Space.} According to the Steam system report, how much empty disk space is in the user's PC?
\\ \noindent \includegraphics[width=\linewidth]{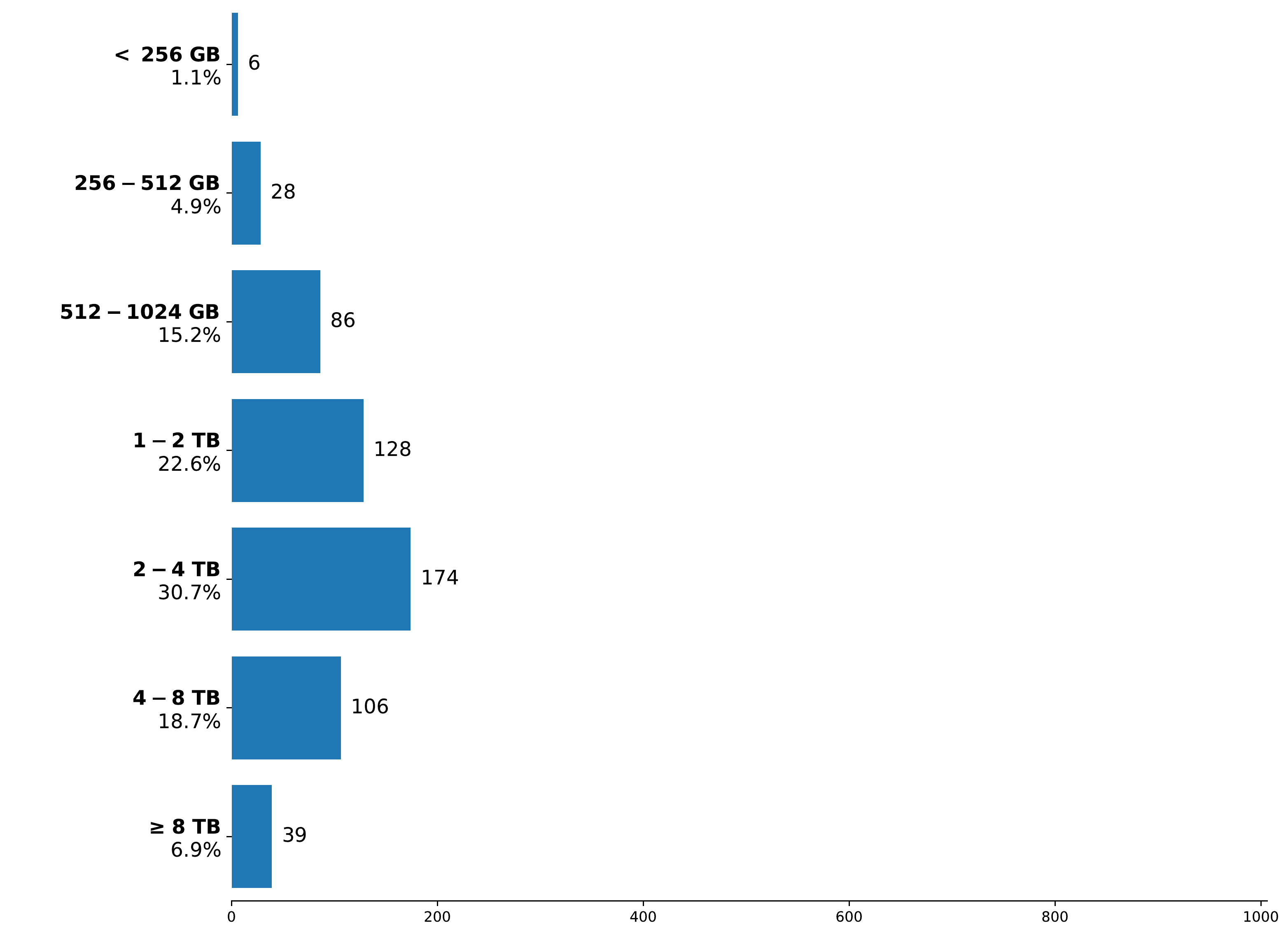}

\noindent \textbf{Base Stations.} According to the Steam system report, how many lighthouses or base stations does the user have?
\\ \noindent \includegraphics[width=\linewidth]{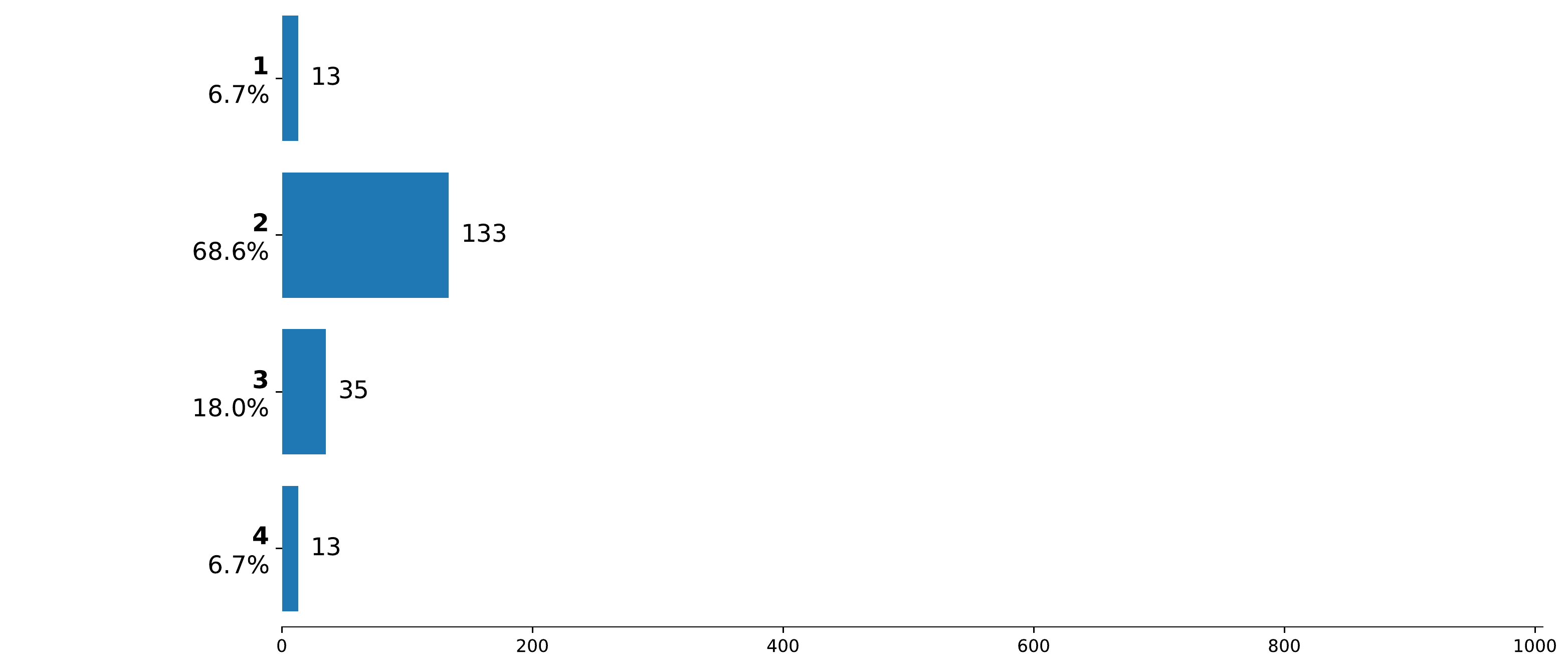}

\eject

\noindent \textbf{Graphics Card.} According to the Steam system report, what is the primary GPU of the user's PC?
\\ \noindent \includegraphics[width=\linewidth]{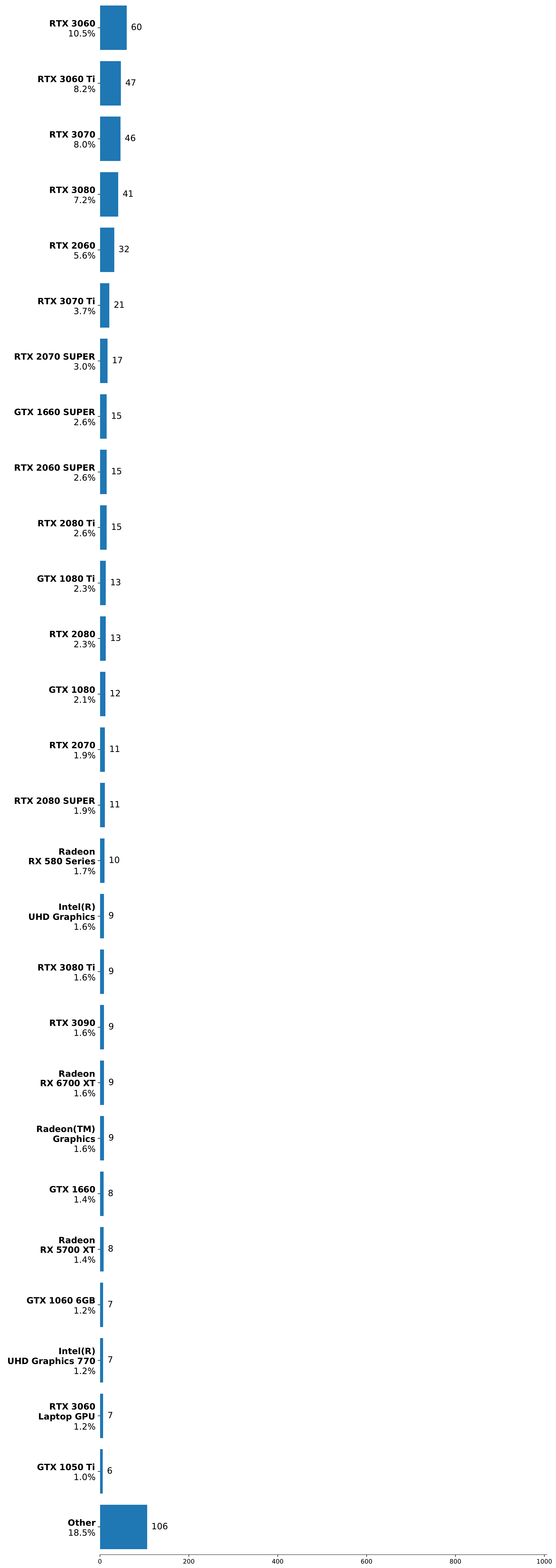}

\eject

\subsection{Background}

We asked users a variety of questions about their musical and athletic background that could plausibly have an effect on Beat Saber performance.

\bigskip

\noindent \textbf{Music.} Have you ever skillfully played a musical instrument?
\\ \noindent \includegraphics[width=\linewidth]{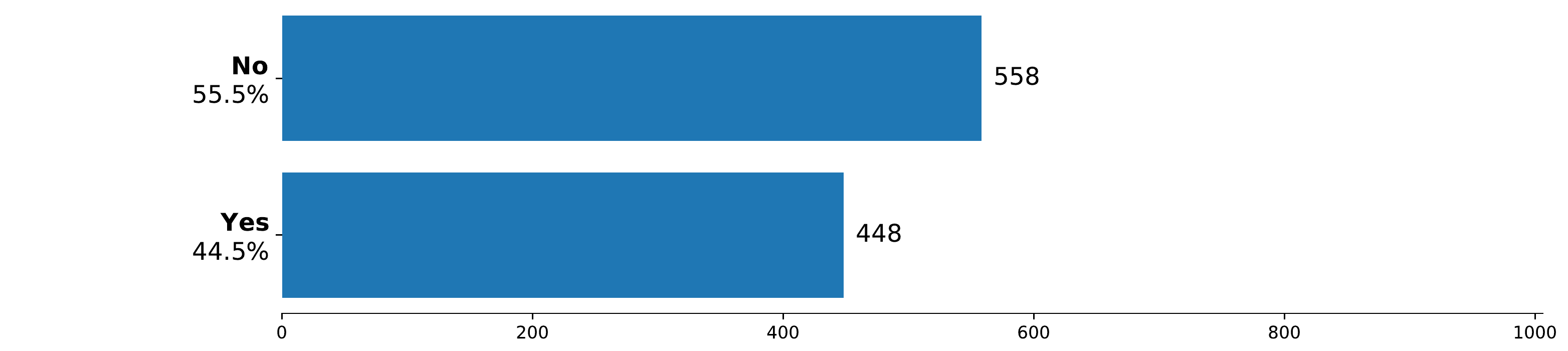}

\noindent \textbf{Music.} If you have ever skillfully played a musical instrument, list the instrument(s).
\\ \noindent \includegraphics[width=\linewidth]{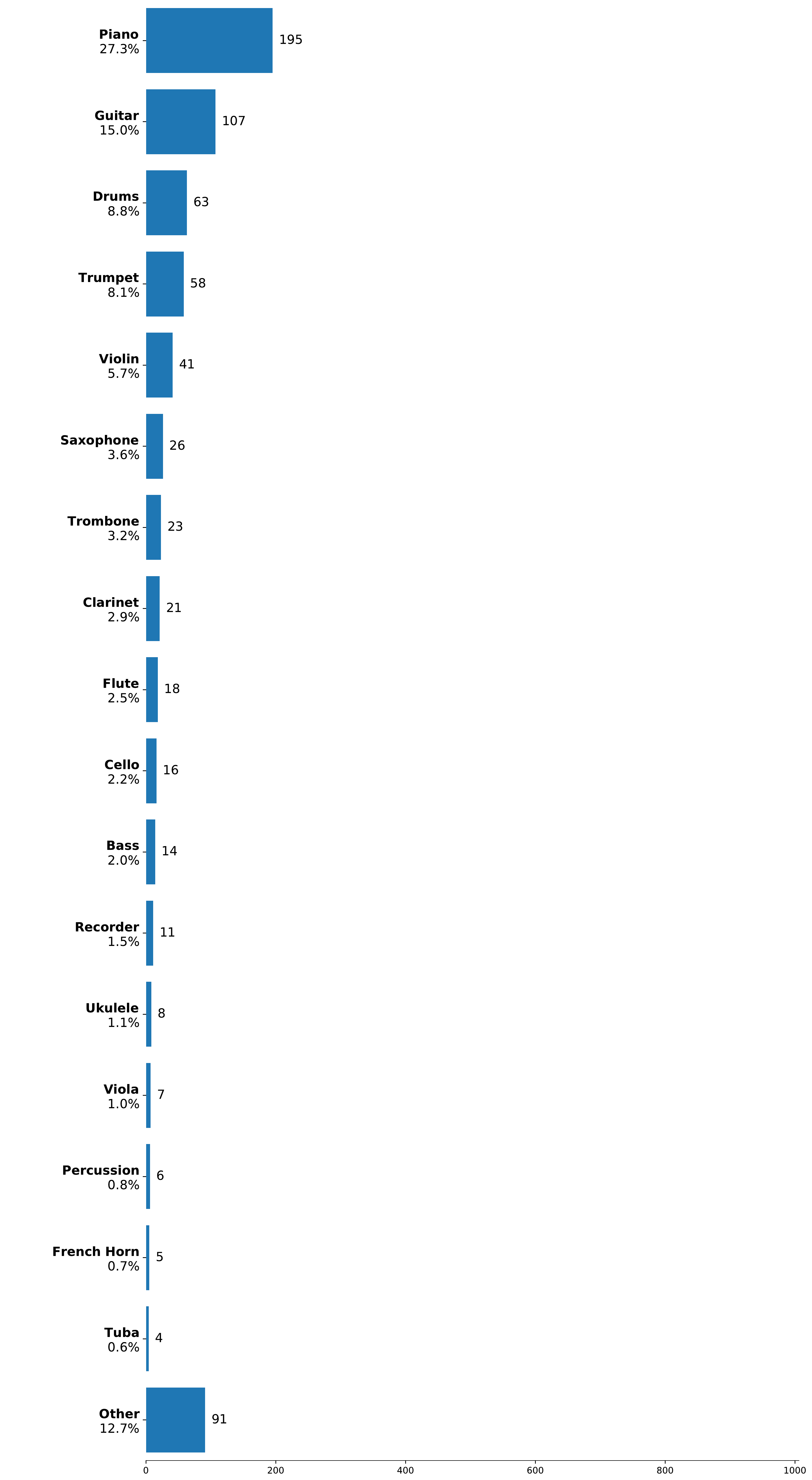}

\noindent \textbf{Dance.} Have you ever skillfully practiced or exhibited a recognized form of dance?
\\ \noindent \includegraphics[width=\linewidth]{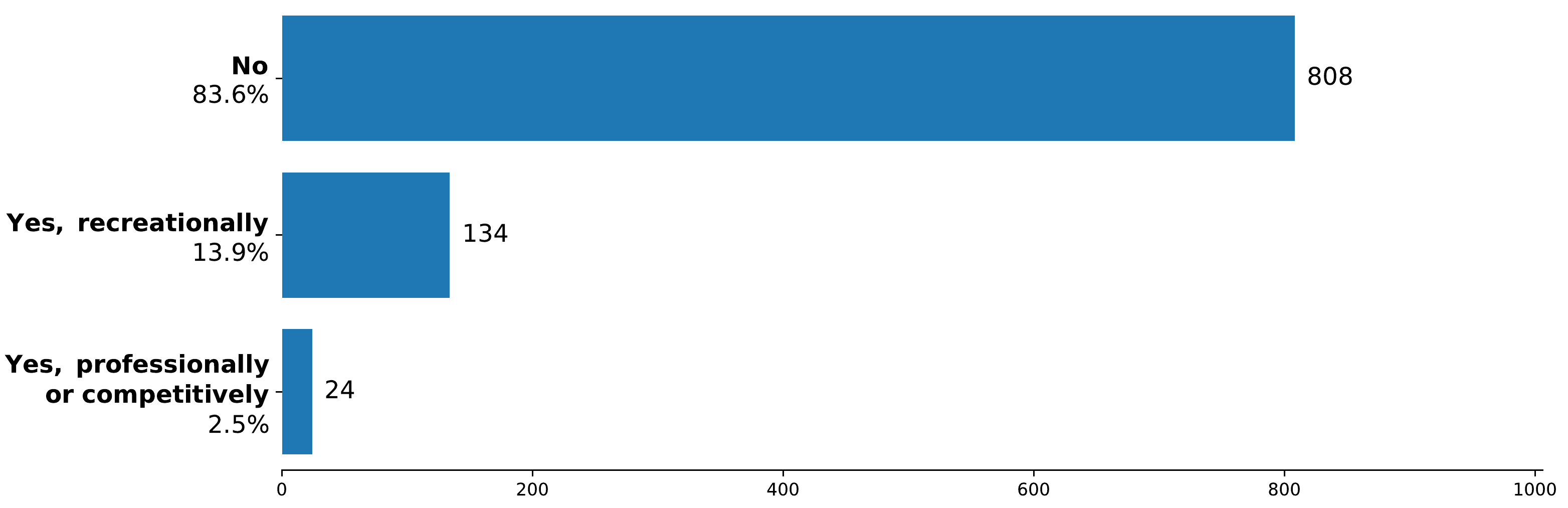}

\eject

\noindent \textbf{Rhythm Games.} Have you ever played a rhythm game other than Beat Saber?
\\ \noindent \includegraphics[width=\linewidth]{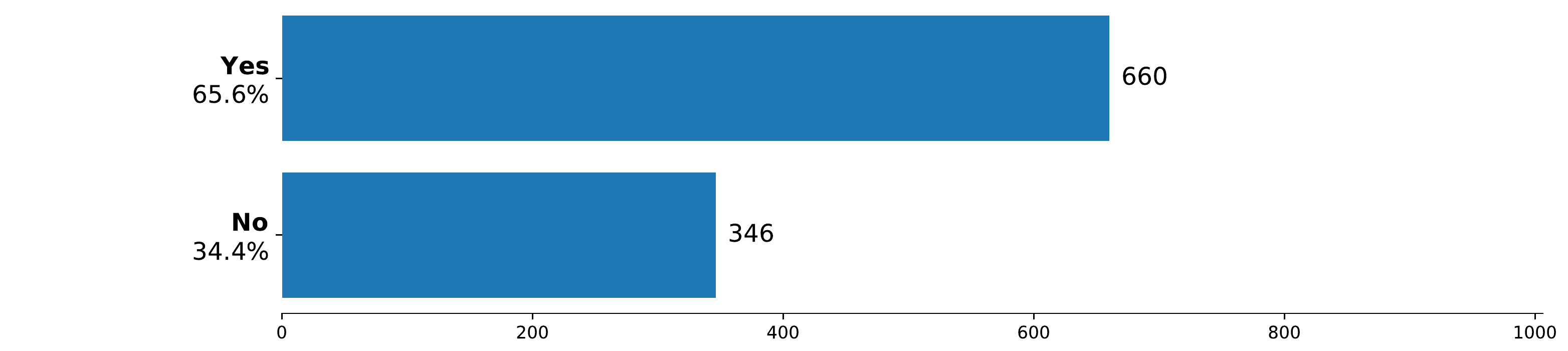}

\noindent \textbf{Rhythm Games.} If you have ever played a rhythm game other than Beat Saber, list the game(s).
\\ \noindent \includegraphics[width=\linewidth]{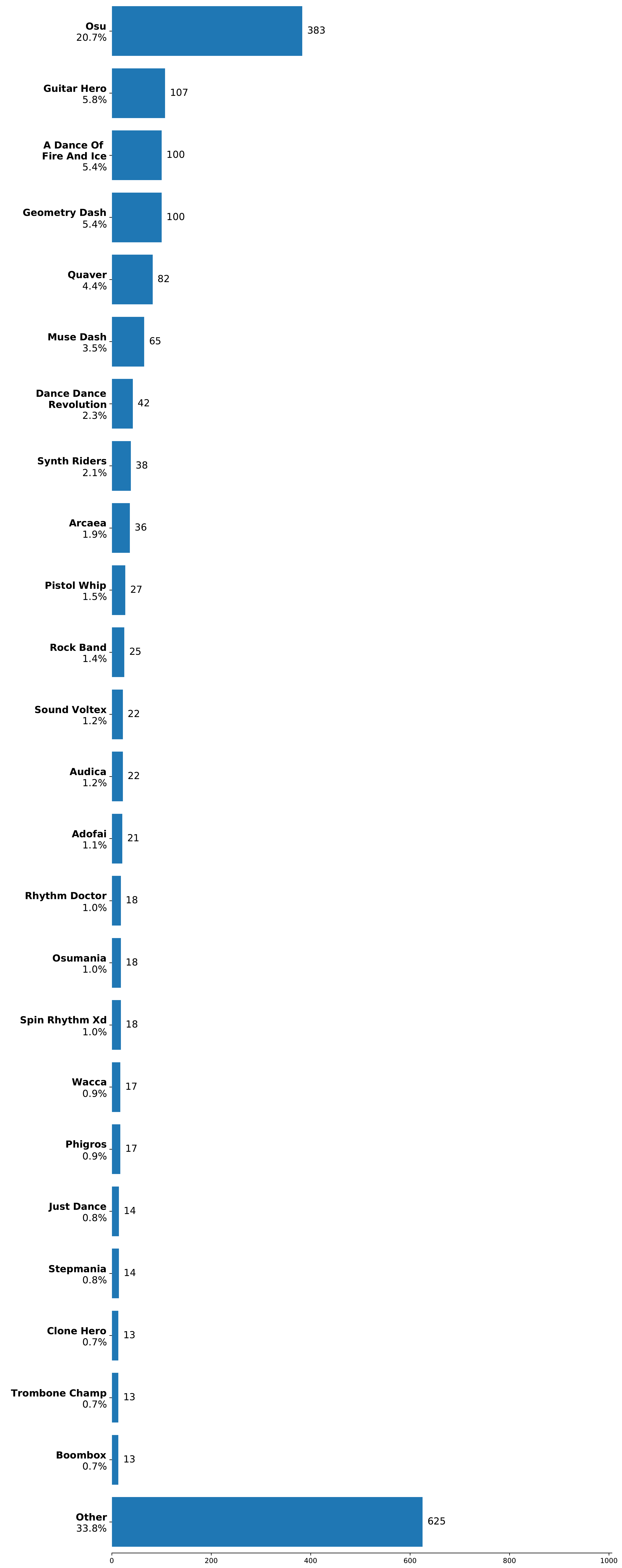}

\eject

\noindent \textbf{Athletics.} Have you ever competitively participated in an individual or team-based athletic sport?
\\ \noindent \includegraphics[width=\linewidth]{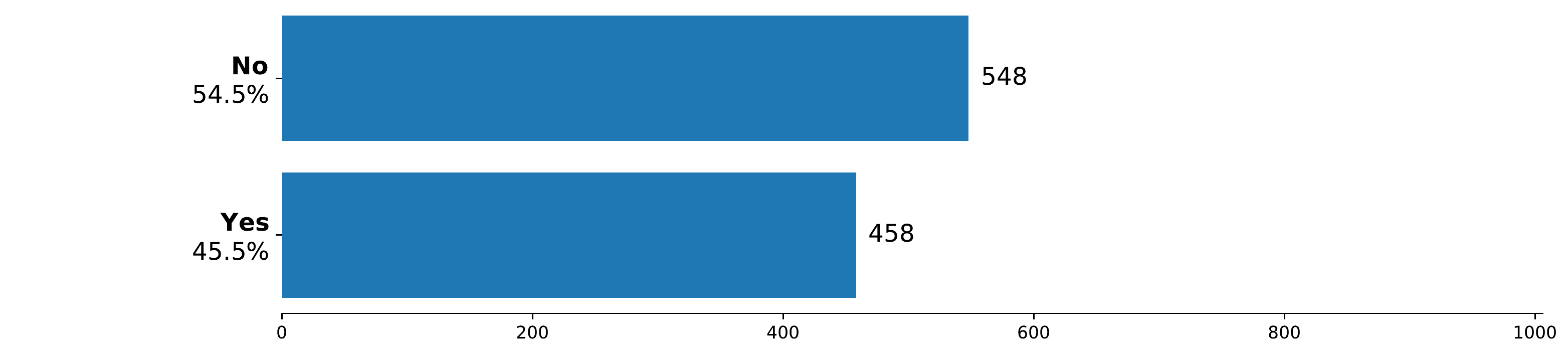}

\noindent \textbf{Athletics.} If you have ever competitively participated in an individual or team-based athletic sport, list the sport(s).
\\ \noindent \includegraphics[width=\linewidth]{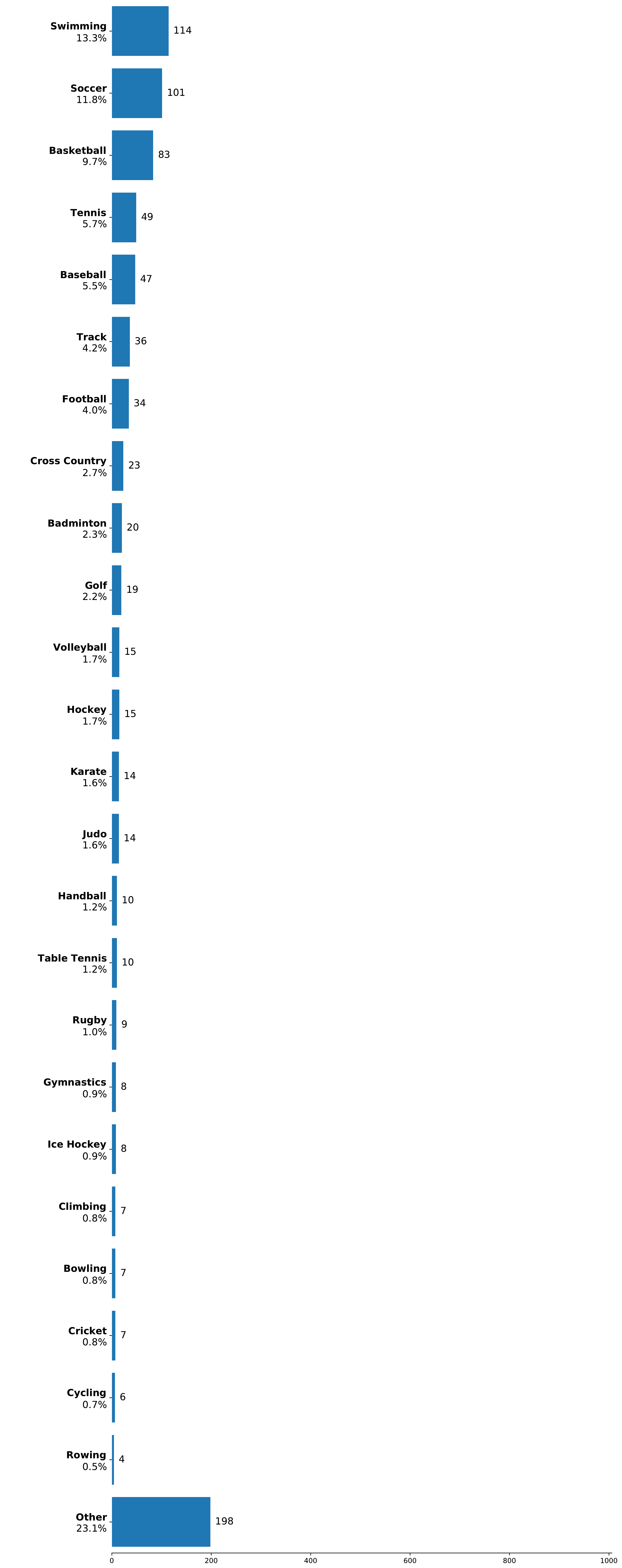}

\eject

\subsection{Health}

In this section, we asked participants a number of questions related to their personal health, eyesight, and disability status. It is our hope that the responses to this section, and their correlations to performance, can motivate improvements to the accessibility of VR applications.

\bigskip

\noindent \textbf{Eyesight.} Do you regularly wear prescription glasses or contact lenses?
\\ \noindent \includegraphics[width=\linewidth]{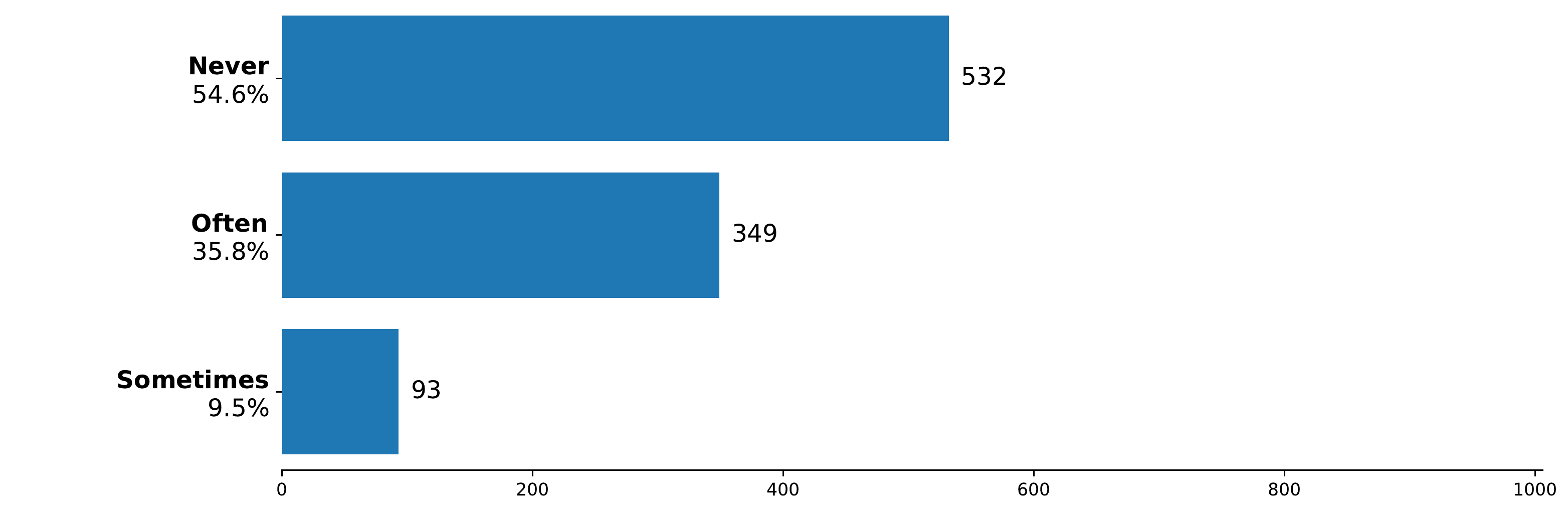}

\noindent \textbf{Lenses.} Do you usually wear prescription  glasses or contact lenses while playing Beat Saber?
\\ \noindent \includegraphics[width=\linewidth]{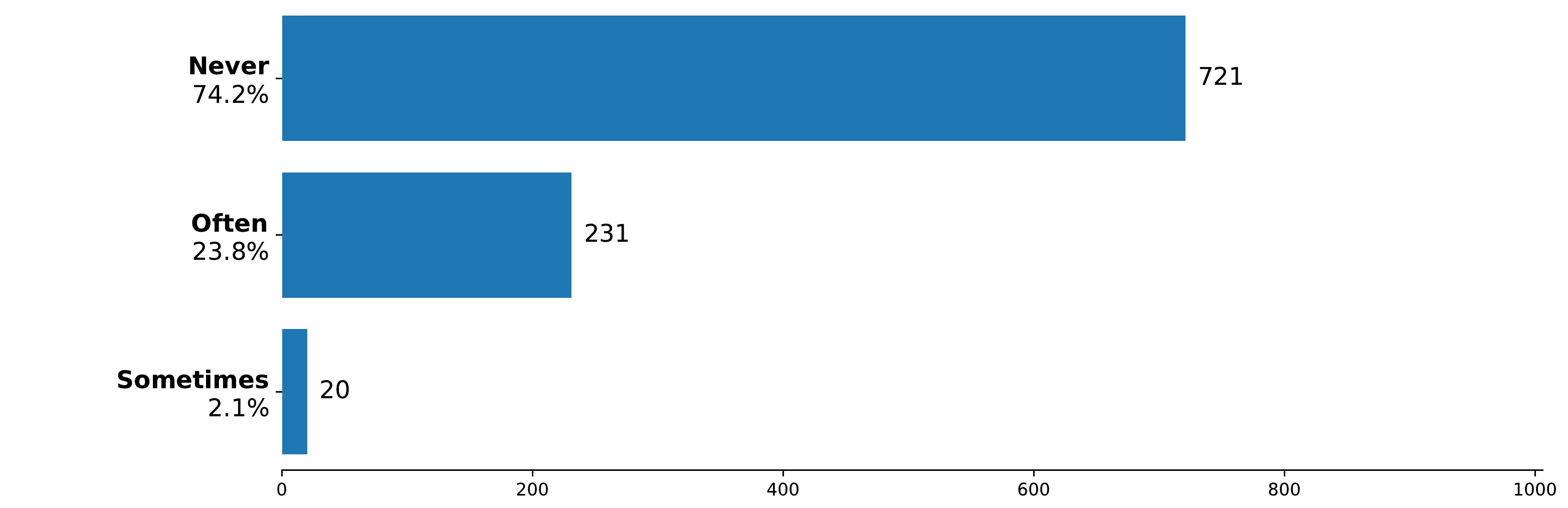}

\noindent \textbf{Color Blindness.} Do you have any form of color blindness or color vision deficiency?
\\ \noindent \includegraphics[width=\linewidth]{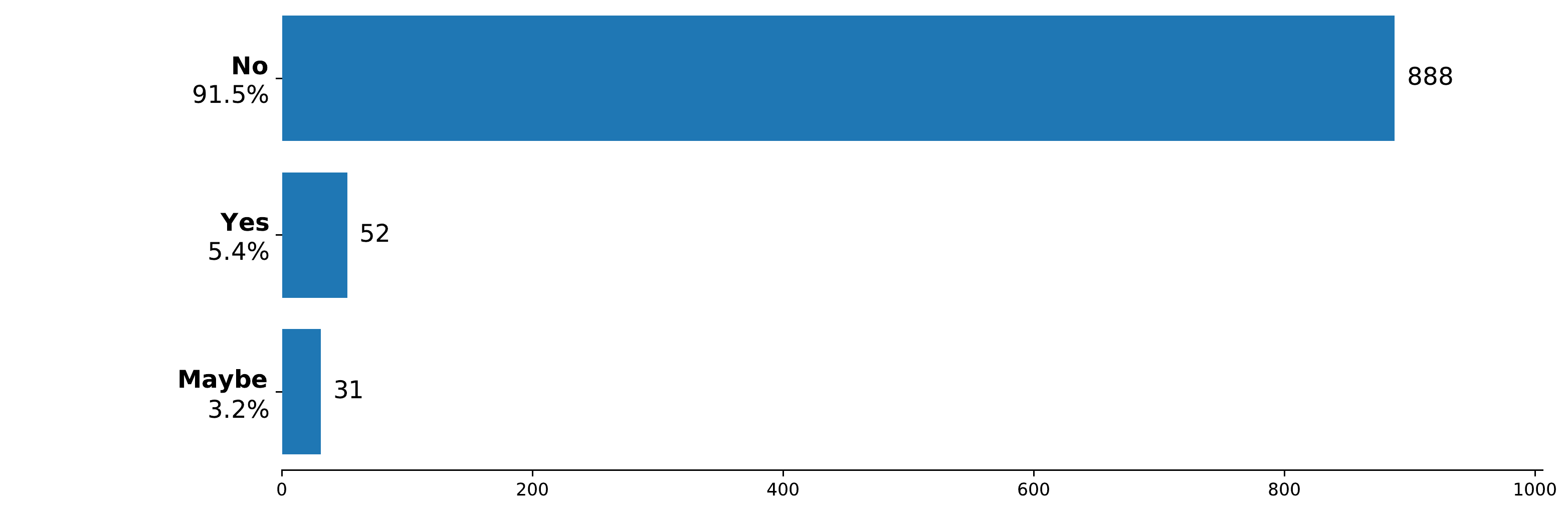}

\noindent \textbf{Physical Disabilities.} Have you ever been diagnosed with a physical disability or other physical health condition?
\\ \noindent \includegraphics[width=\linewidth]{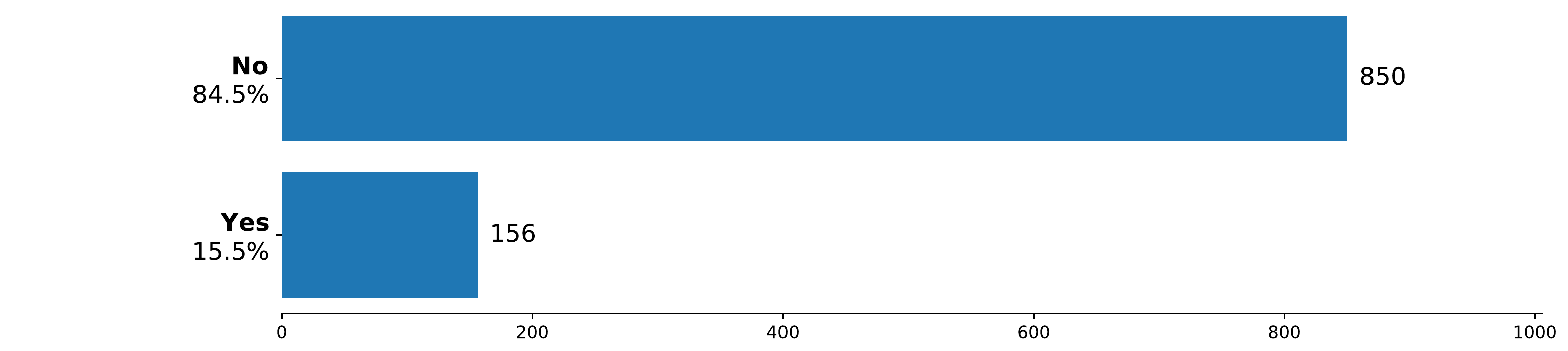}

\noindent \textbf{Mental Disabilities.} Have you ever been diagnosed with a mental disability or other mental health condition?
\\ \noindent \includegraphics[width=\linewidth]{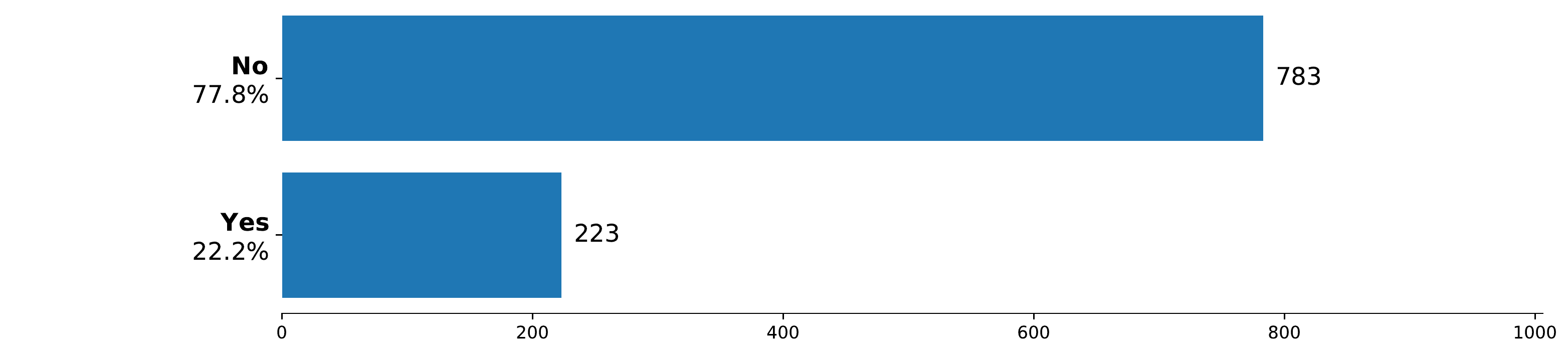}

\noindent \textbf{Illness.} In the past year, have you experienced COVID-19?
\\ \noindent \includegraphics[width=\linewidth]{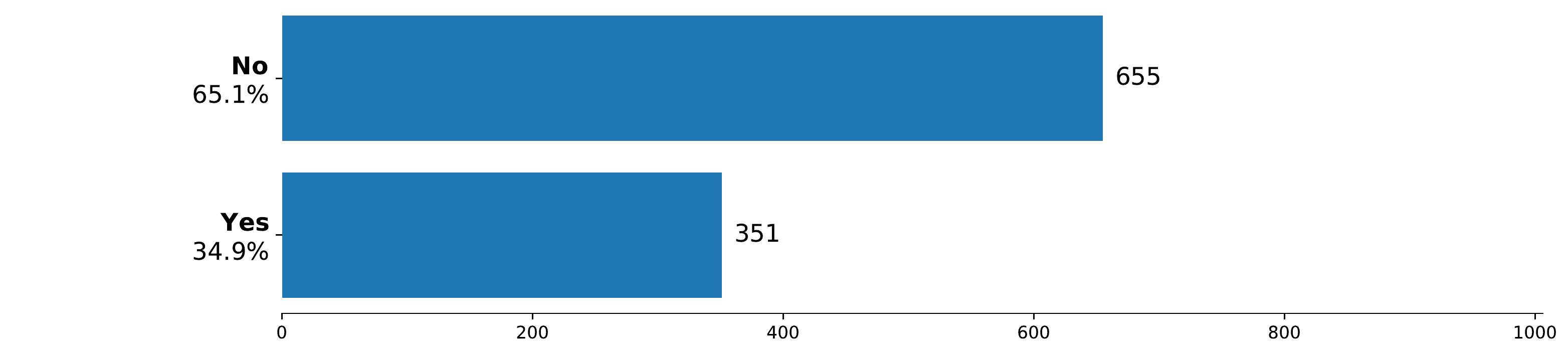}

\eject

\subsection{Behavior Patterns}

Next, we asked participants a series of questions regarding their general habits and learned behaviors before and during a typical Beat Saber play session.

\bigskip

\noindent \includegraphics[width=\linewidth]{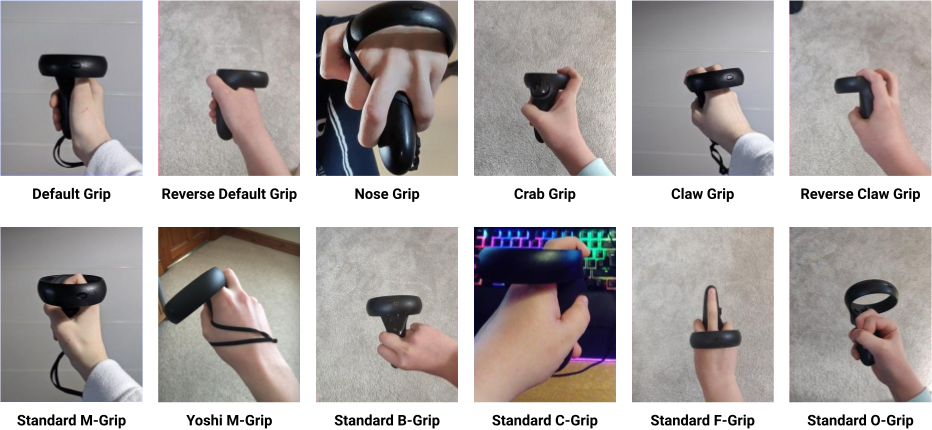}

\bigskip

\noindent \textbf{Grip.} Which of the following grips do you prefer to use on standalone VR devices (e.g., Oculus Quest 2, PICO Neo3, etc.)?
\\ \noindent \includegraphics[width=\linewidth]{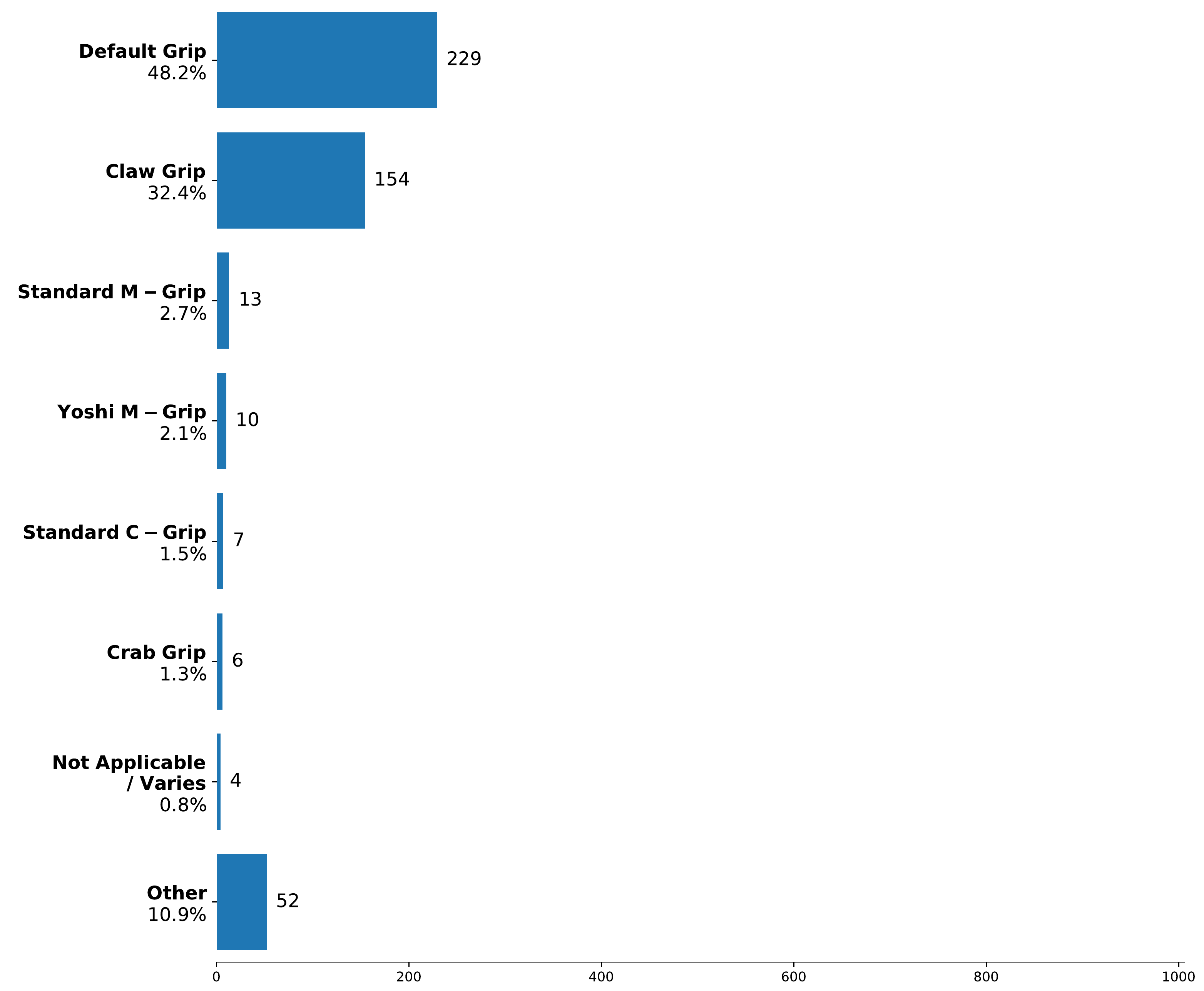}

\bigskip

\noindent \textbf{Preparation.} Which of the following activities, if any, do you perform immediately before playing Beat Saber? Select all that apply.
\\ \noindent \includegraphics[width=\linewidth]{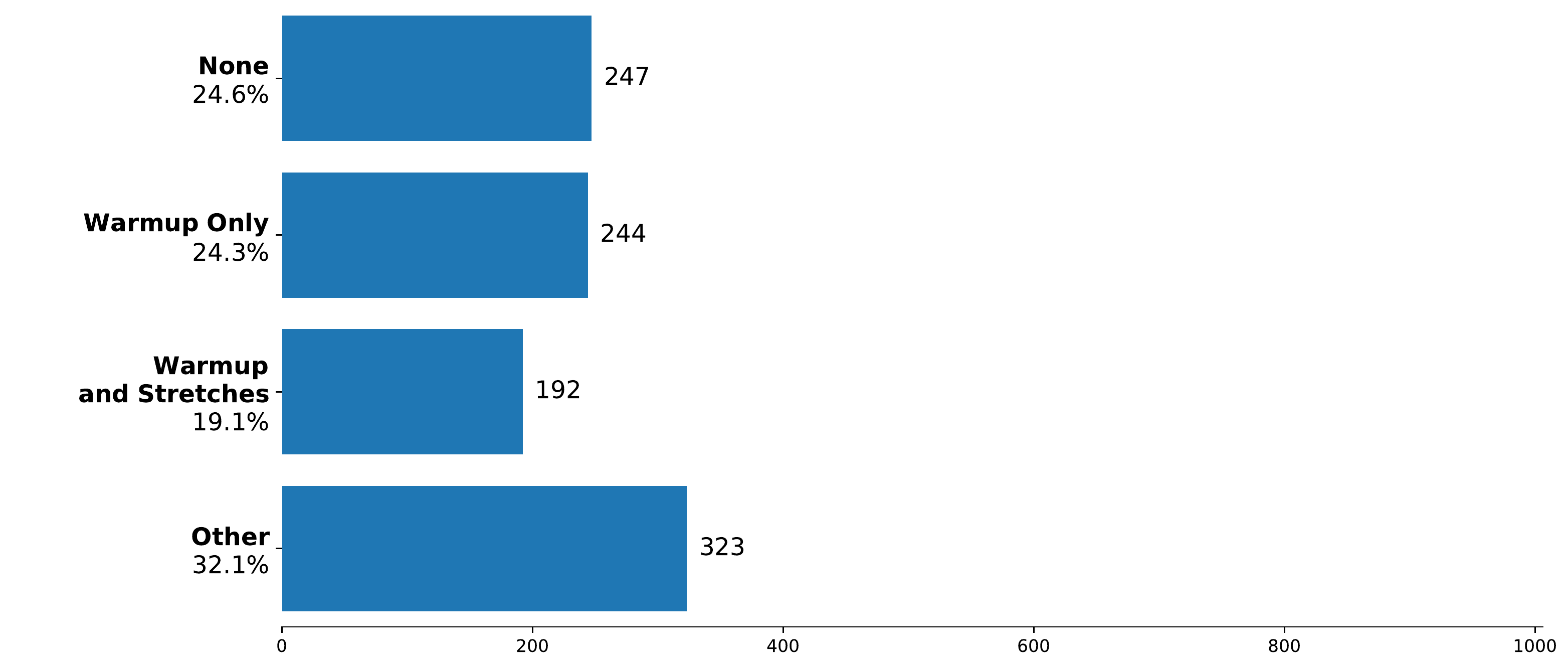}

\eject

\noindent \textbf{Physical Fitness.} How would you rate your current level of overall physical fitness?
\\ \noindent \includegraphics[width=\linewidth]{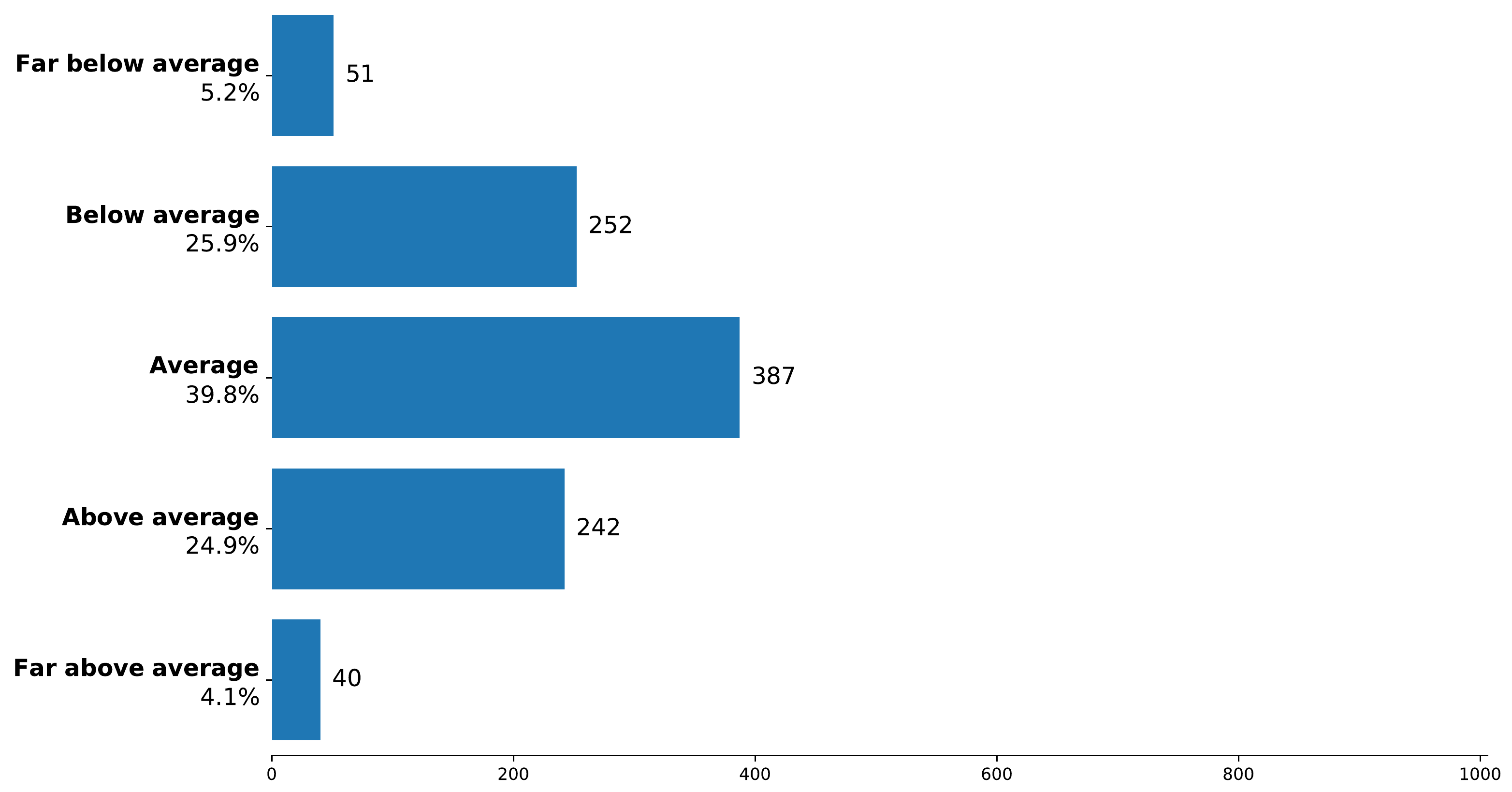}

\noindent \textbf{Caffinated Items.} Approximately how many caffeinated foods or beverages (e.g., Coffee, Black Tea, Energy Drinks, etc.) do you consume on a regular basis?
\\ \noindent \includegraphics[width=\linewidth]{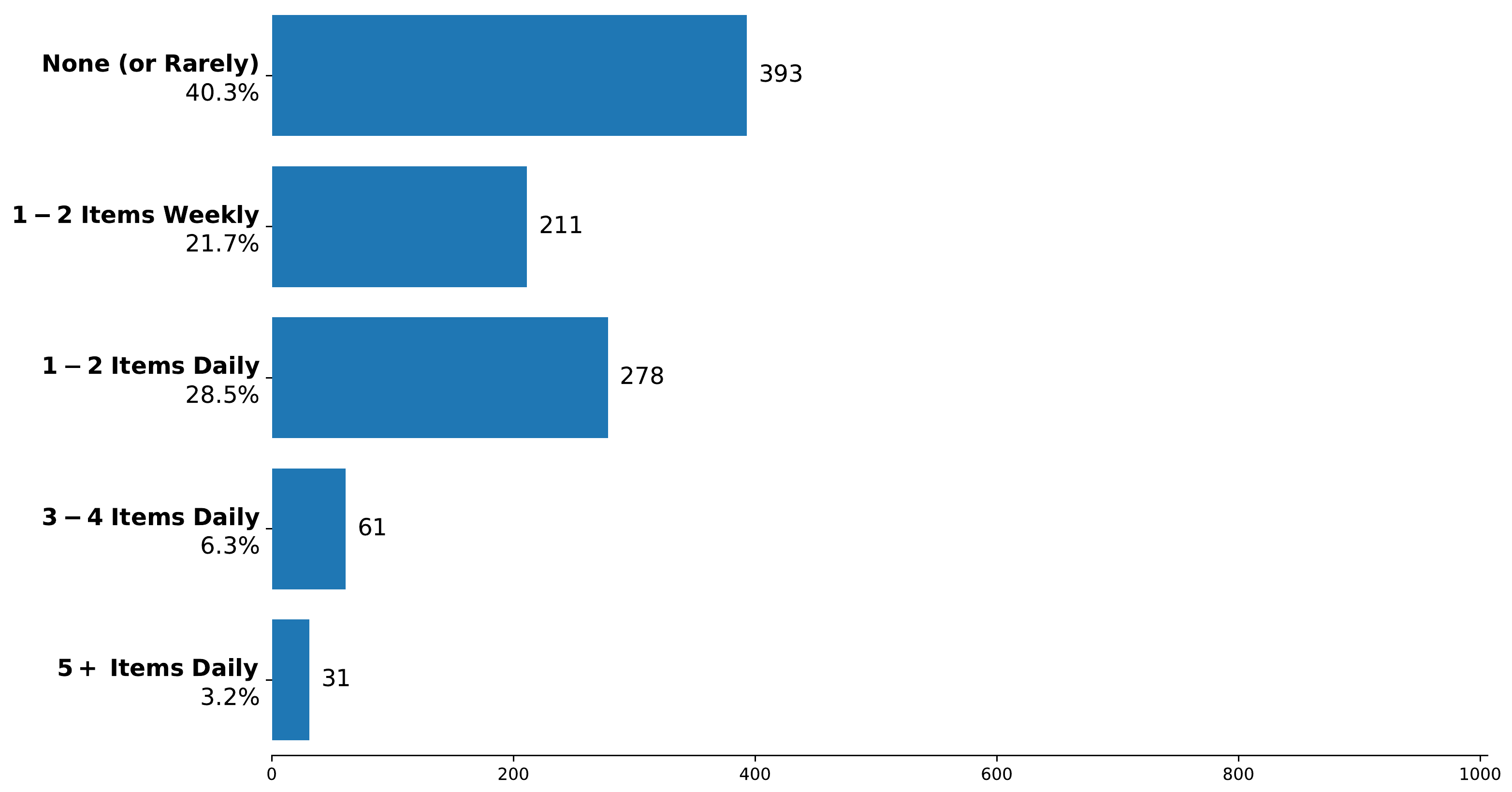}

\noindent \textbf{Caffeine Consumption.} Do you usually consume caffeine in the 3 hours before starting to play Beat Saber?
\\ \noindent \includegraphics[width=\linewidth]{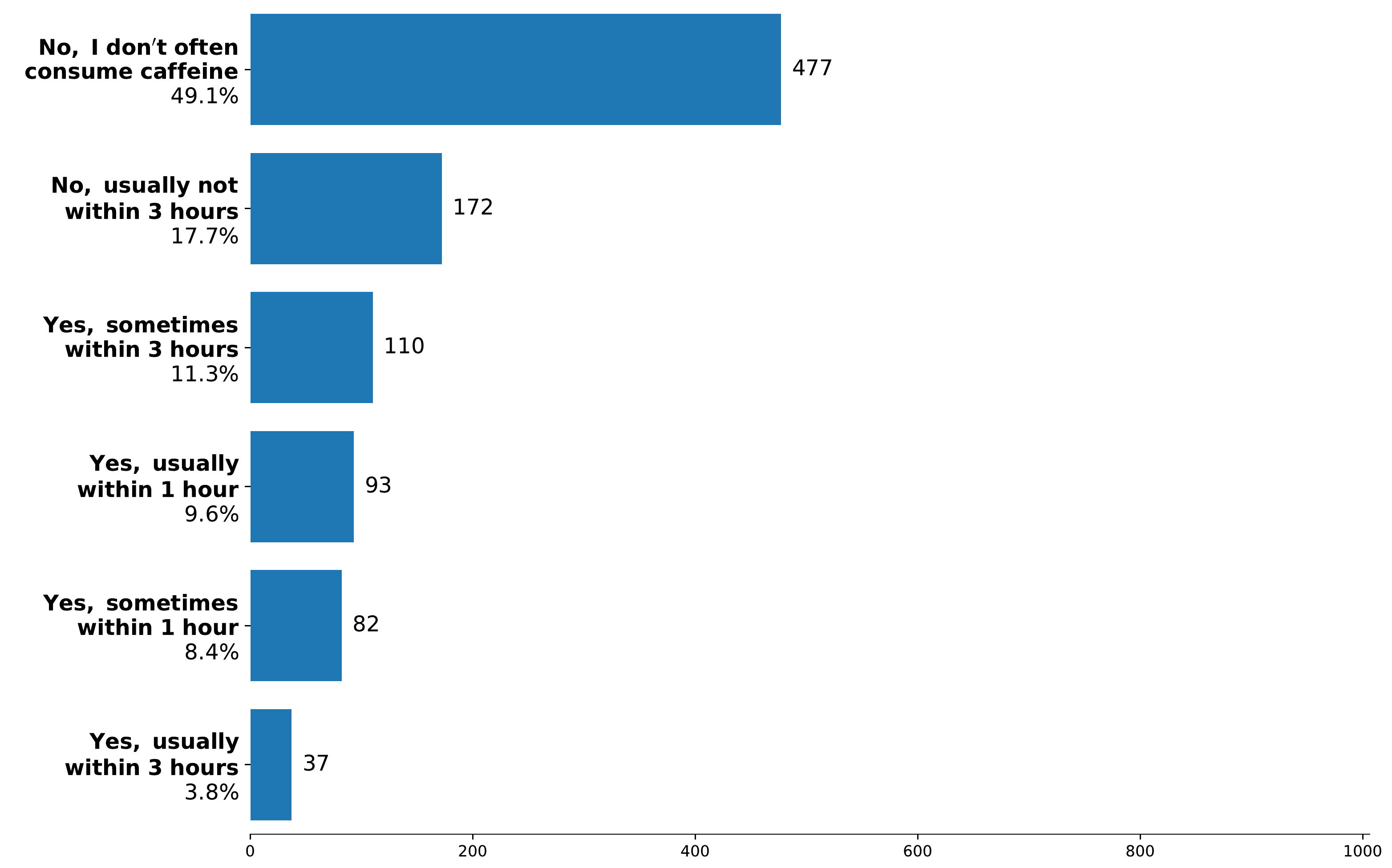}

\noindent \textbf{Substance Use.} How often do you play Beat Saber while under the influence of an intoxicating substance (including alcohol)?
\\ \noindent \includegraphics[width=\linewidth]{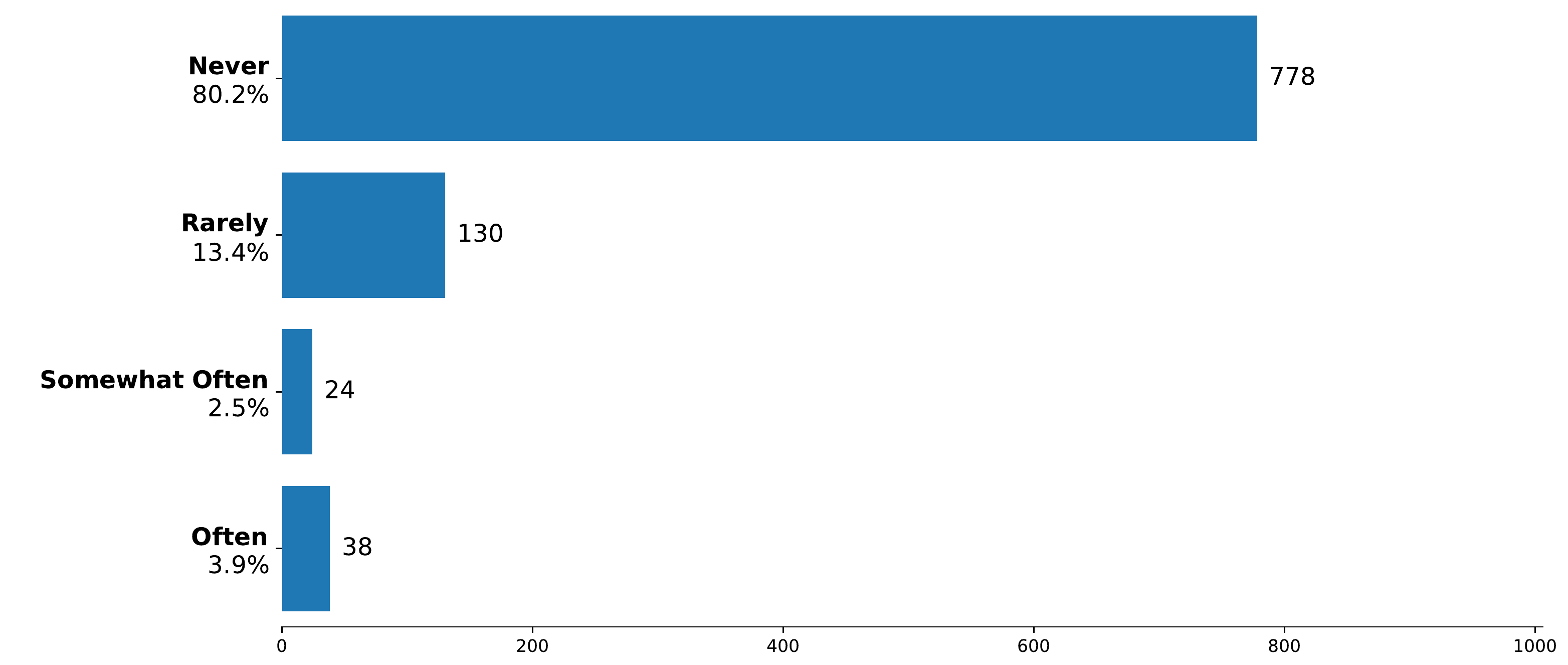}

\eject

\subsection{Environment}

In this section, we asked questions about the location in which users typically play Beat Saber.

\bigskip

\noindent \textbf{Venue.} In which location do you most often play Beat Saber?
\\ \noindent \includegraphics[width=\linewidth]{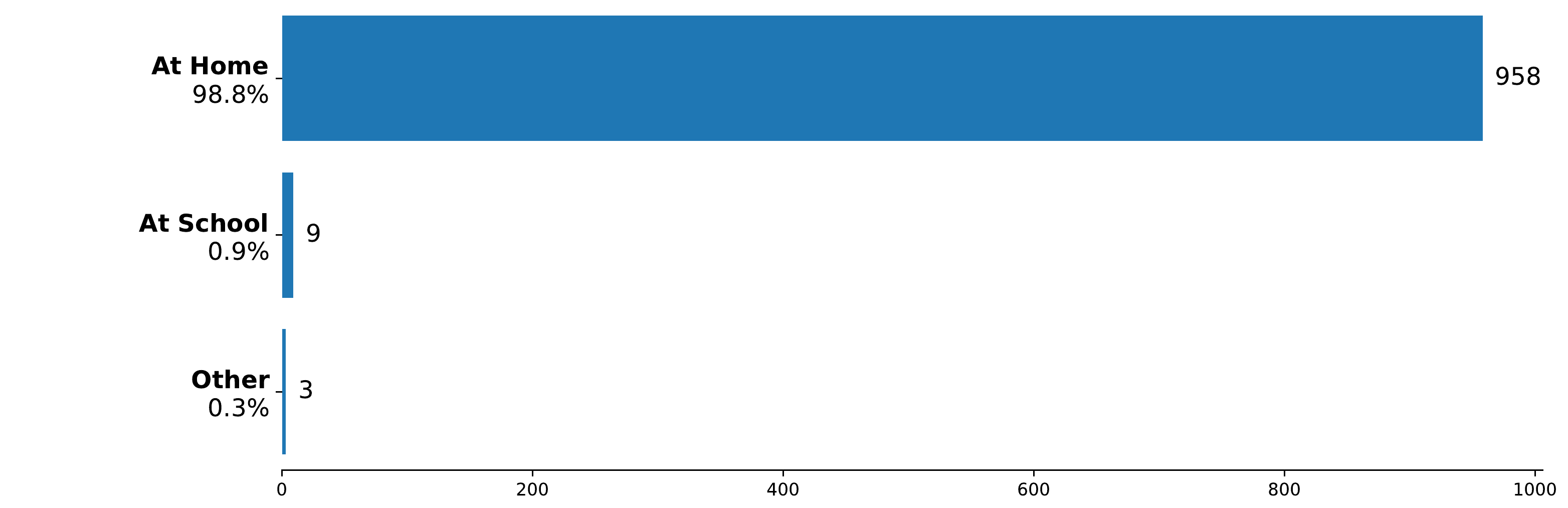}

\noindent \textbf{Room Size.} What are the dimensions of the play area in which you most often play Beat Saber?
\\ \noindent \includegraphics[width=\linewidth]{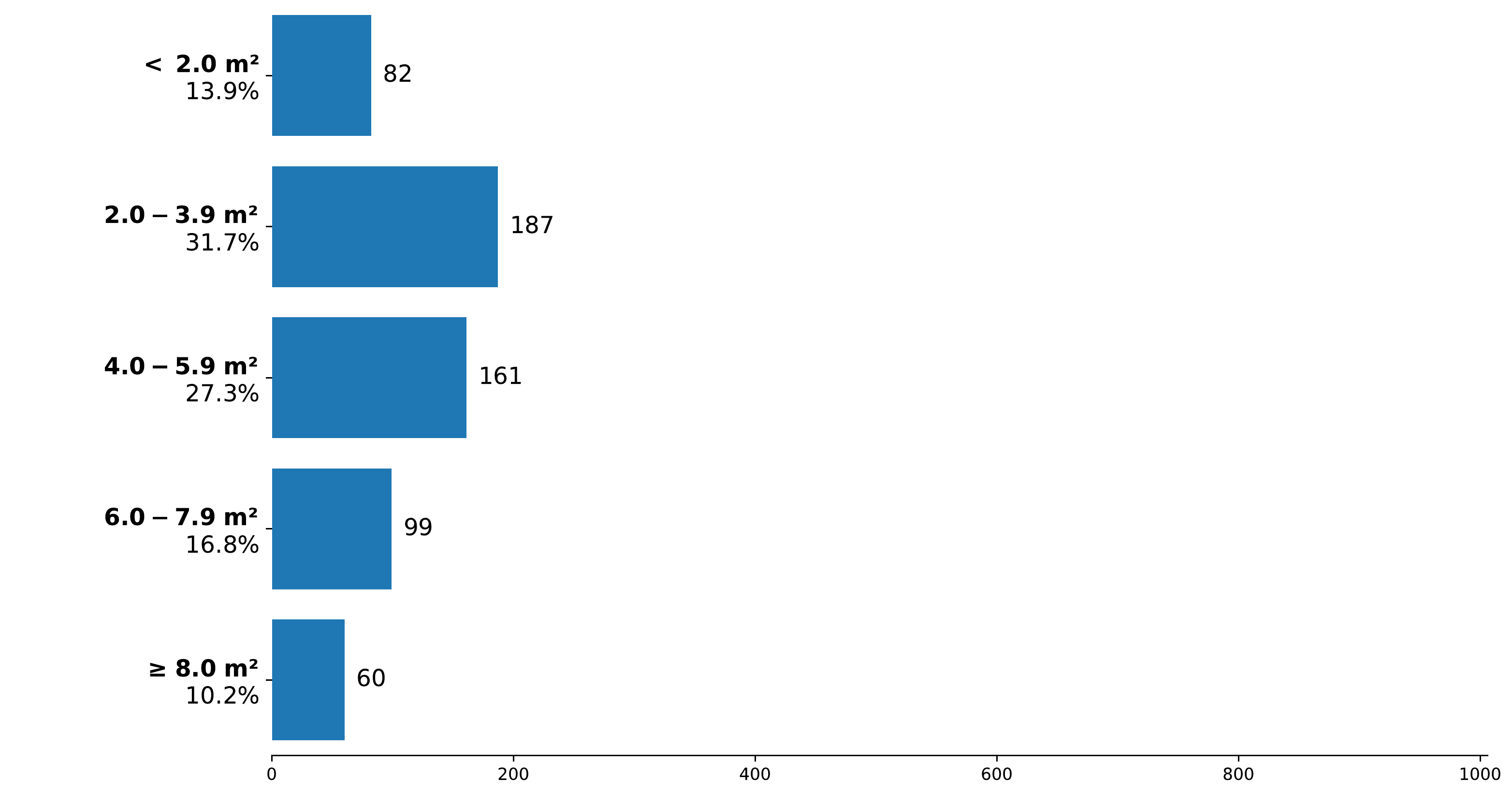}

\noindent \textbf{Location.} What is the name of the country in which you most often play Beat Saber?
\\ \noindent \includegraphics[width=\linewidth]{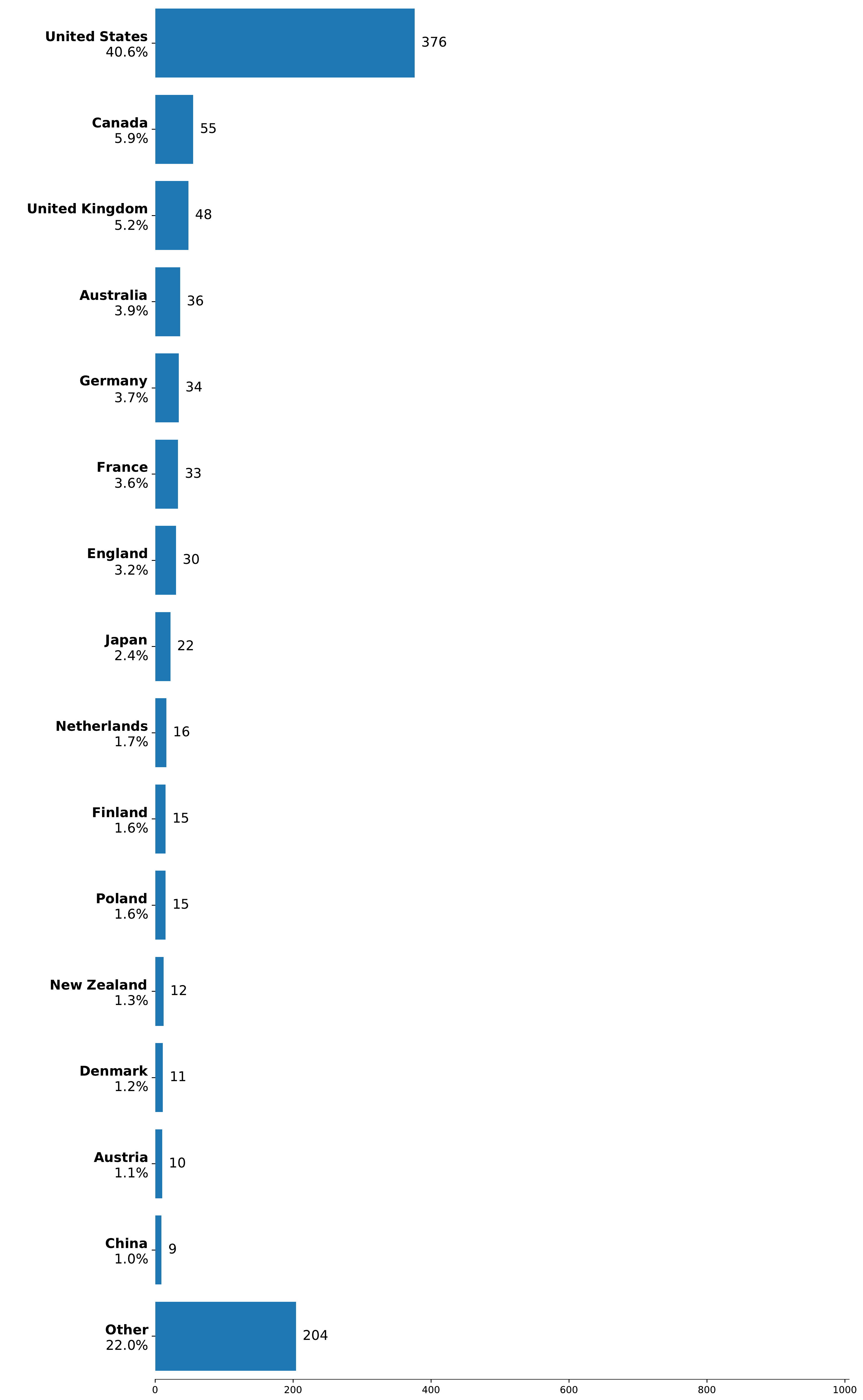}

\eject

\noindent \textbf{Location.} What is the name of state or territory in which you most often play Beat Saber?
\\ \noindent \includegraphics[width=\linewidth]{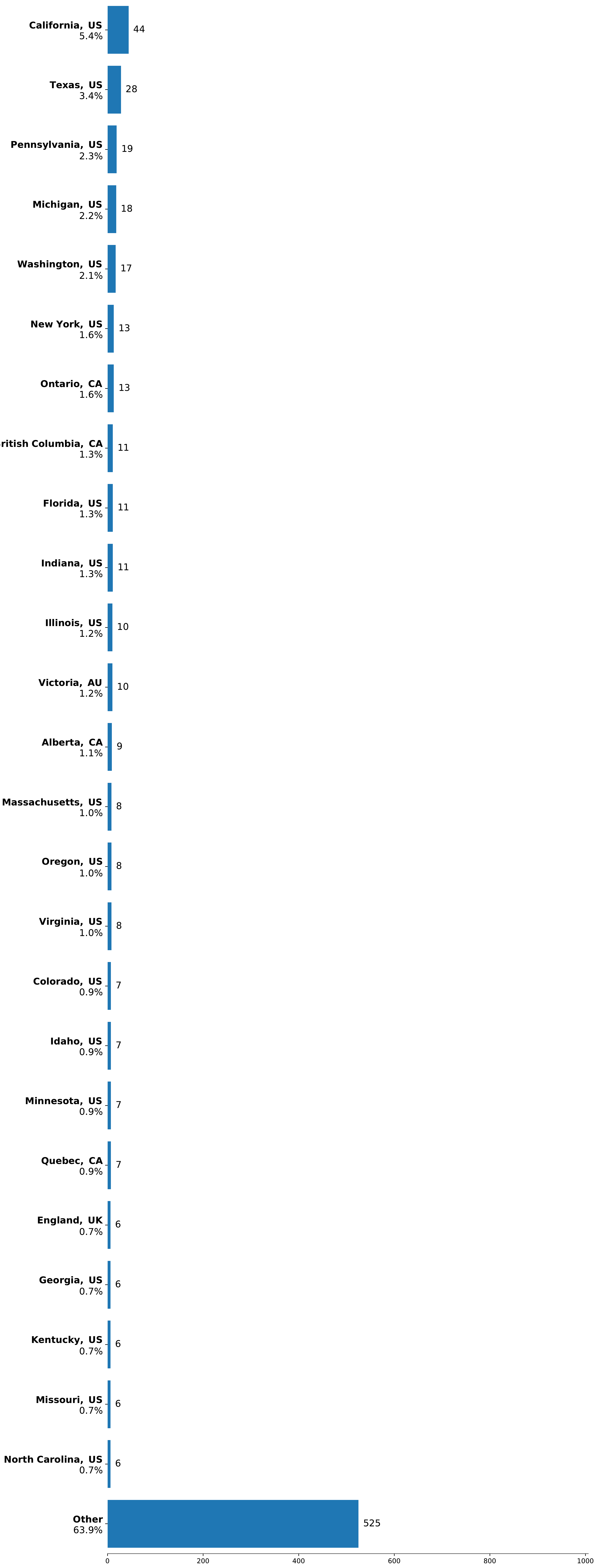}

\eject

\noindent \textbf{Location.} What is the name of the city in which you most often play Beat Saber?
\\ \noindent \includegraphics[width=\linewidth]{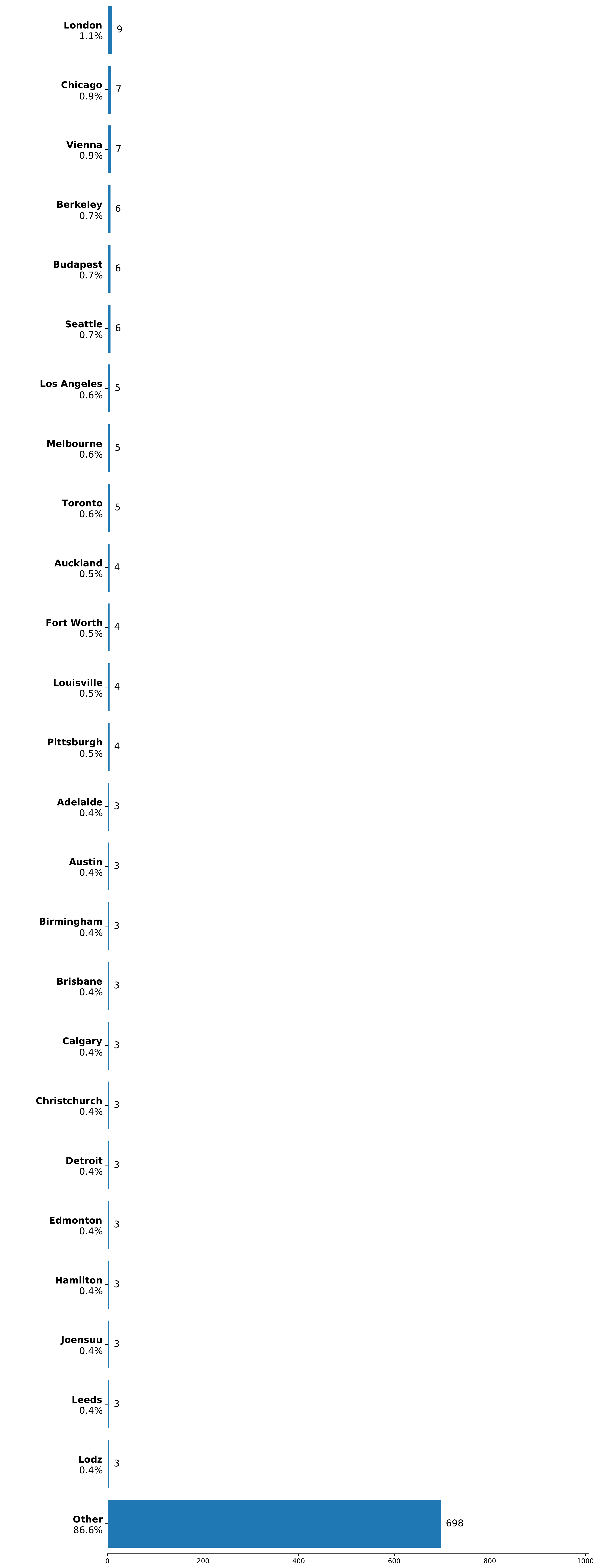}

\eject

\subsection{Anthropometrics}

Next, we asked participants to measure various dimensions of their body in order to asses the extent to which physical proportions affect their virtual reality experience.

\smallskip

\noindent \includegraphics[width=\linewidth]{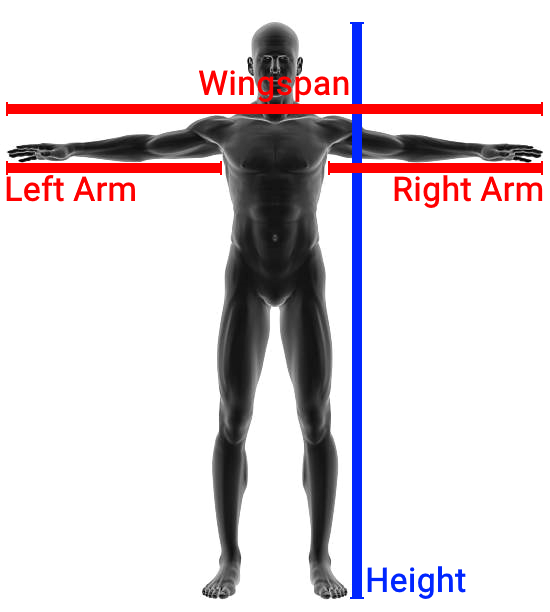}

\bigskip

\noindent \textbf{Height.} What is your exact height in centimeters?
\\ \noindent \includegraphics[width=\linewidth]{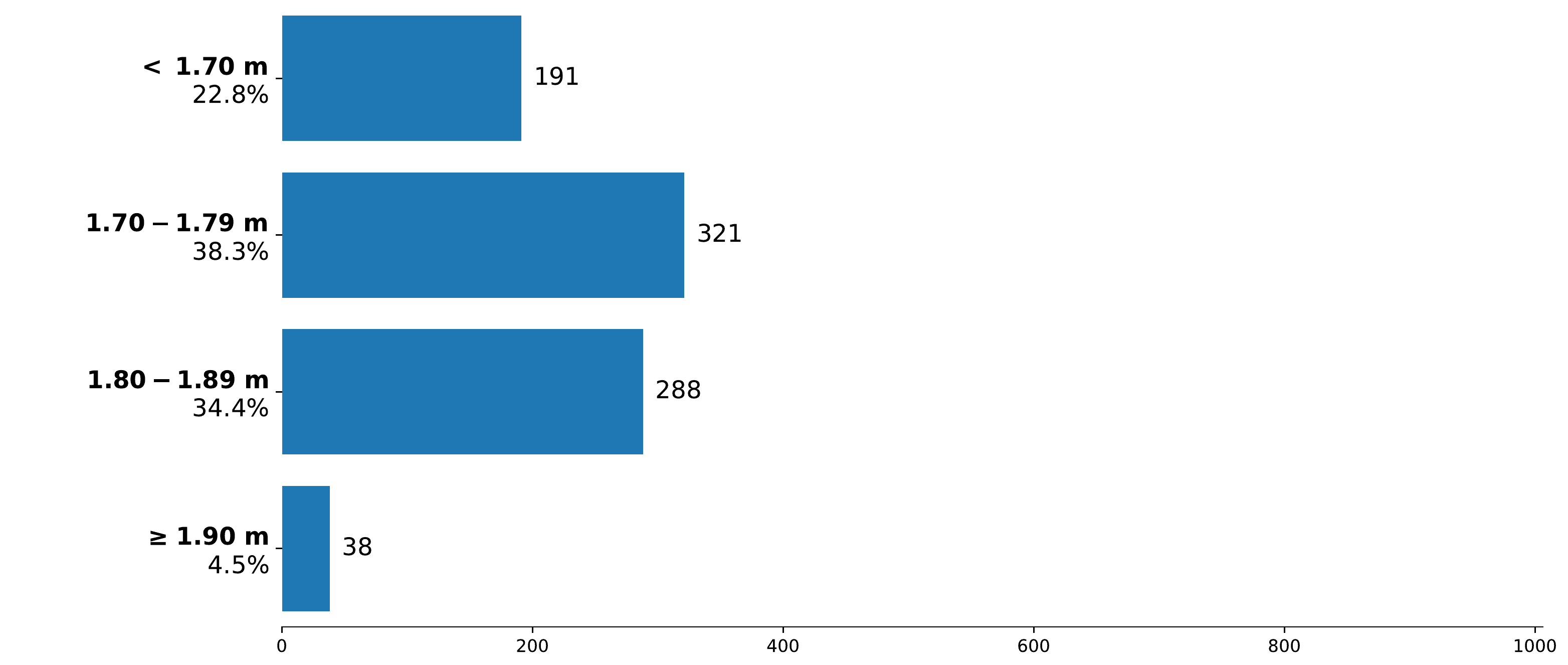}

\noindent \textbf{Left Arm.} What is the exact length of your left arm in centimeters?
\\ \noindent \includegraphics[width=\linewidth]{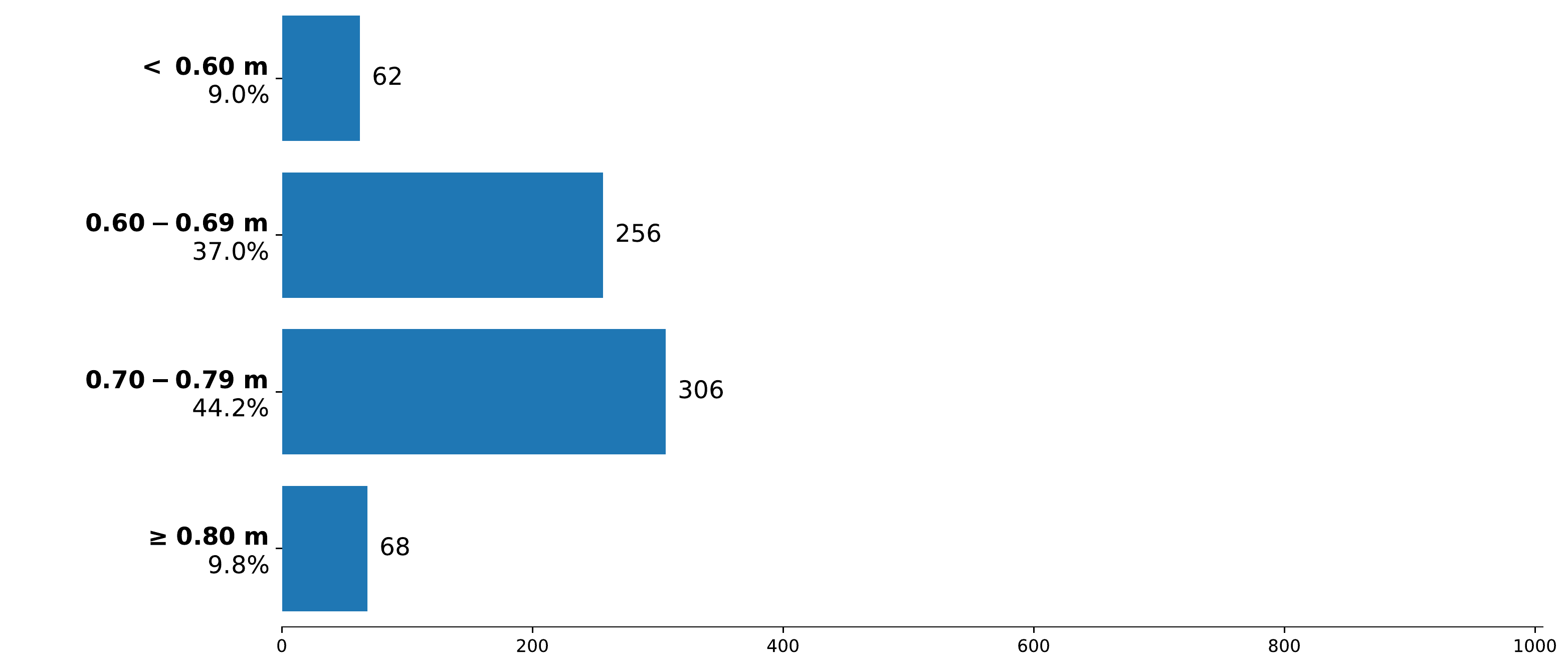}

\noindent \textbf{Right Arm.} What is the exact length of your right arm in centimeters?
\\ \noindent \includegraphics[width=\linewidth]{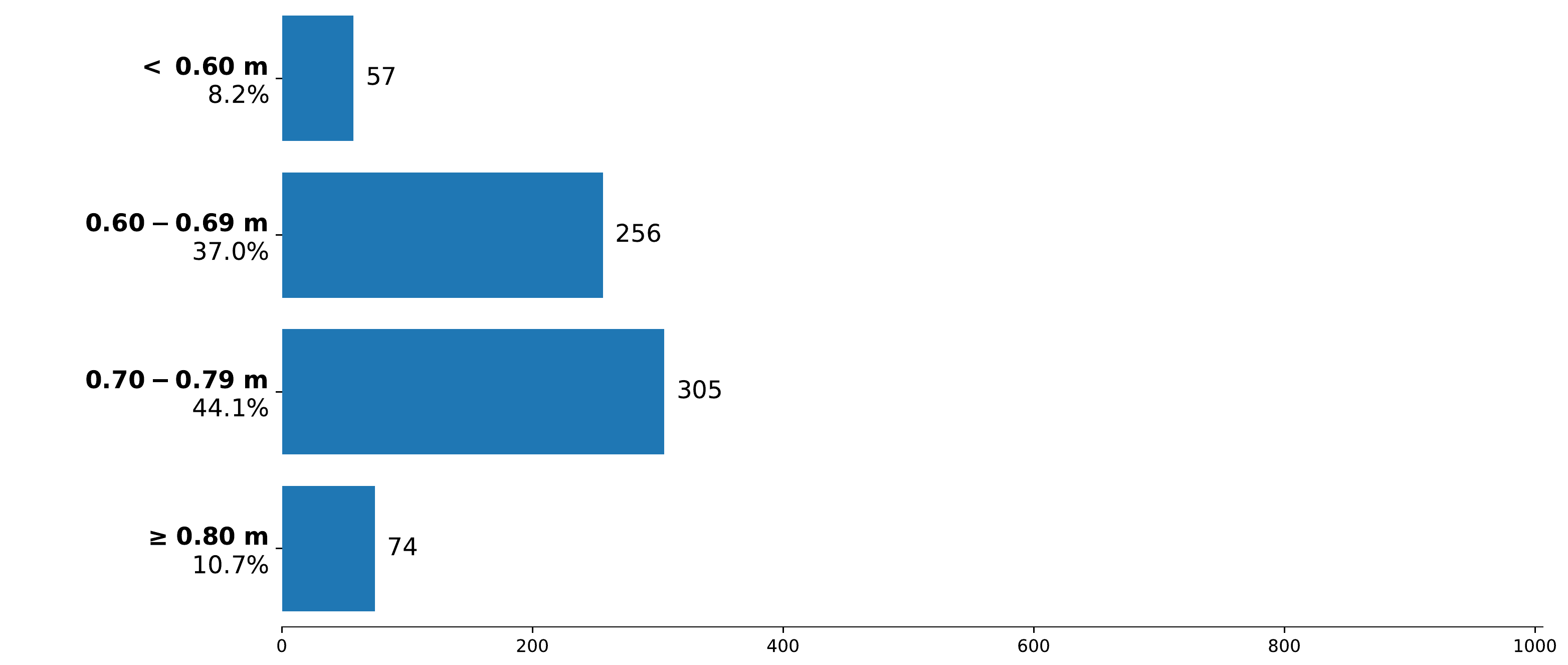}

\eject

\noindent \textbf{Wingspan.} What is your exact wingspan in centimeters?
\\ \noindent \includegraphics[width=\linewidth]{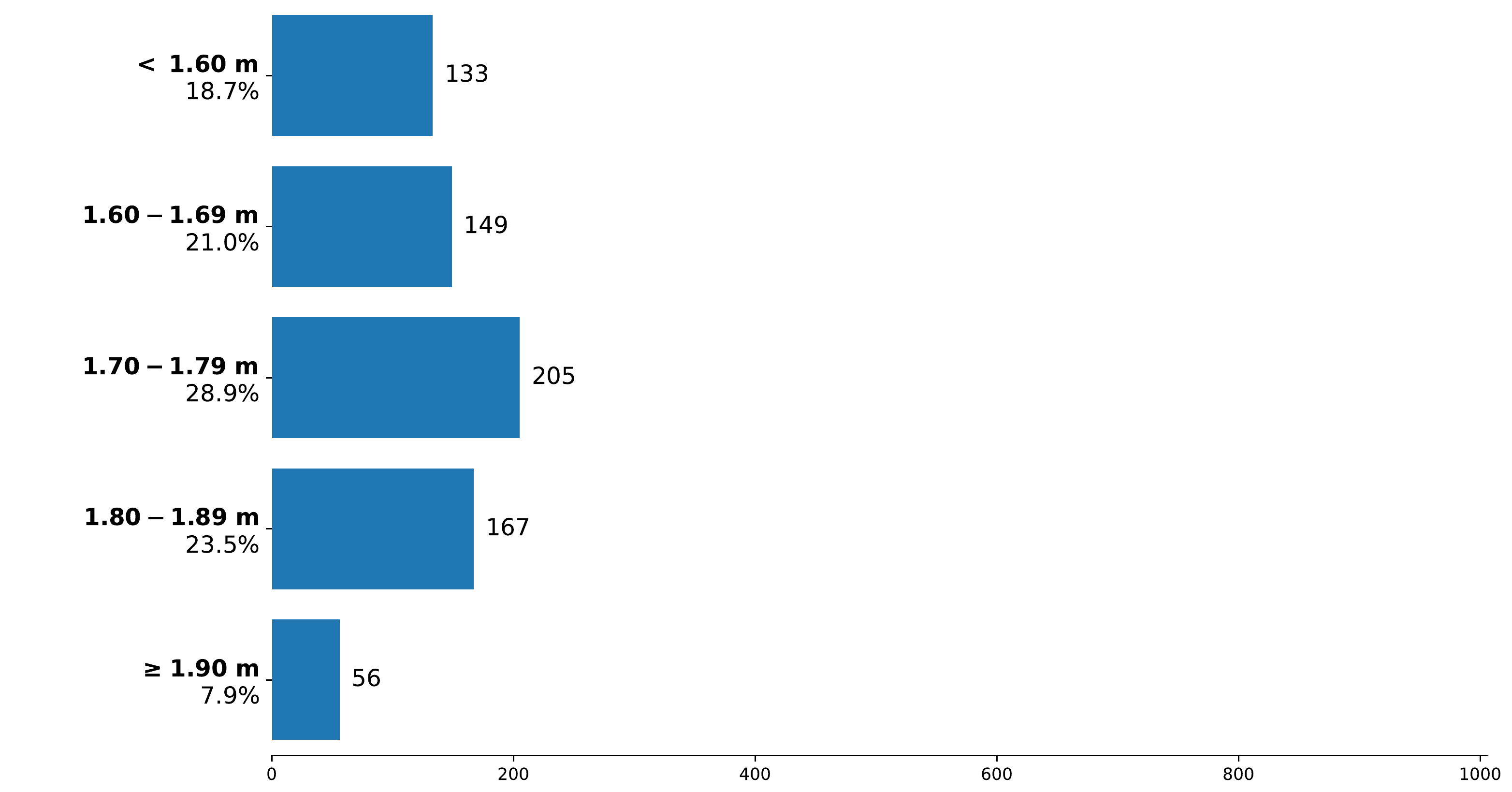}

\noindent \textbf{Handedness.} Are you left or right handed?
\\ \noindent \includegraphics[width=\linewidth]{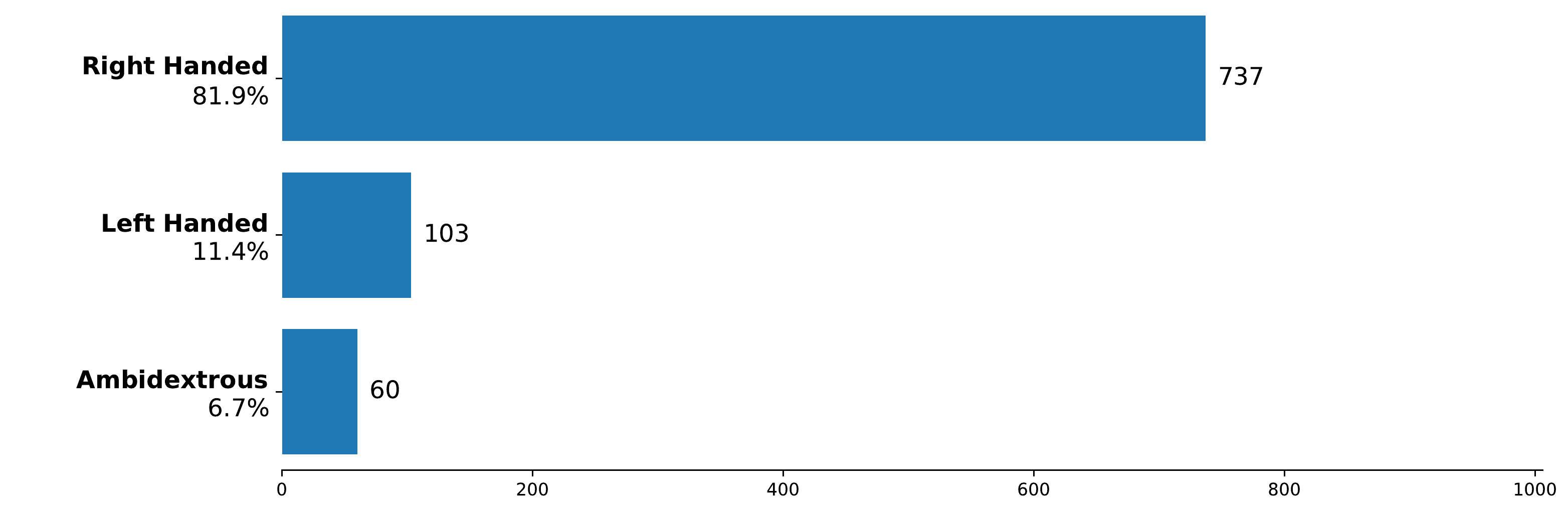}

\noindent \textbf{Weight.} What is your approximate weight in kilograms?
\\ \noindent \includegraphics[width=\linewidth]{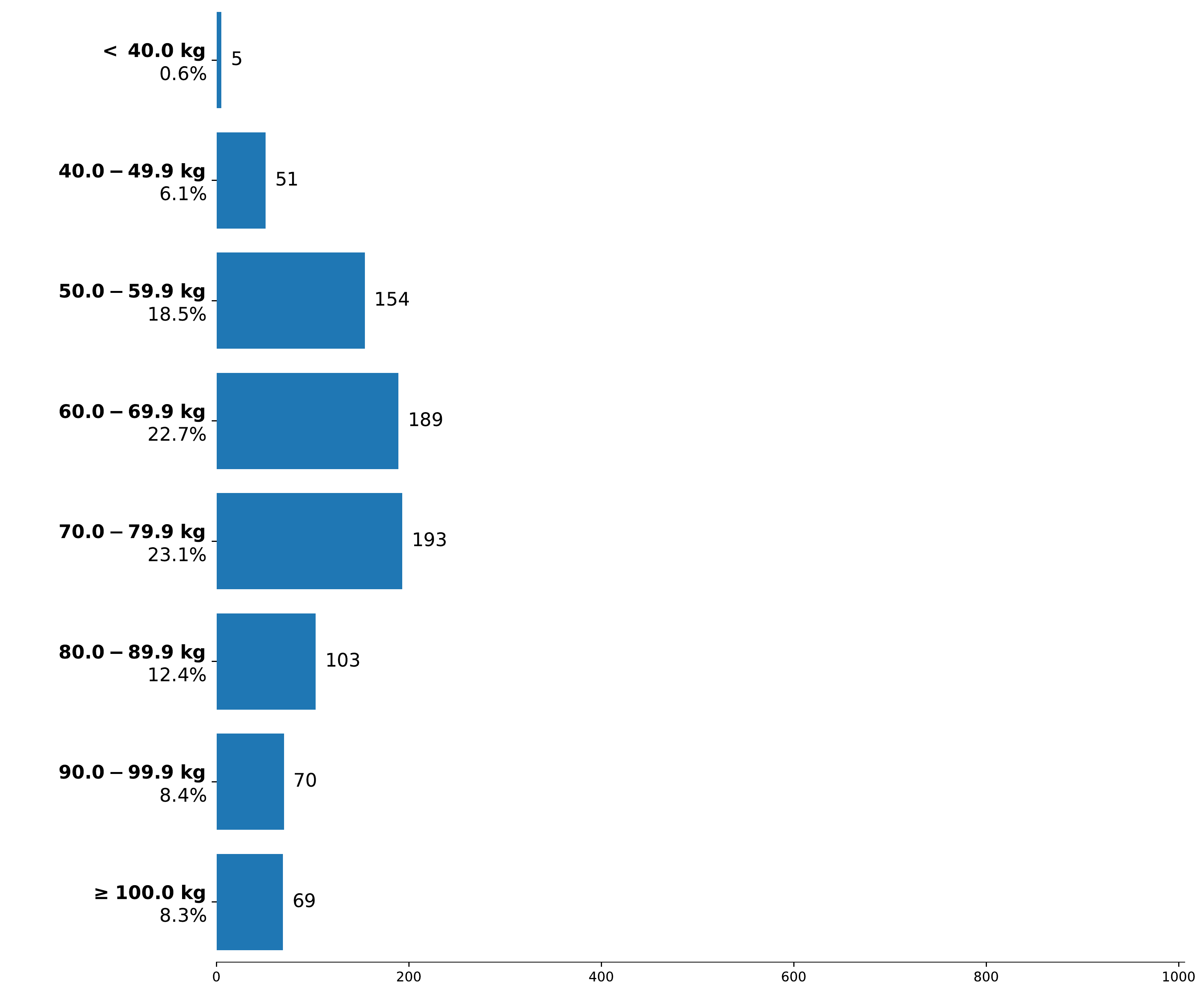}

\noindent \textbf{Interpupillary Distance.} What is your exact interpupillary distance (IPD) in millimeters?
\\ \noindent \includegraphics[width=\linewidth]{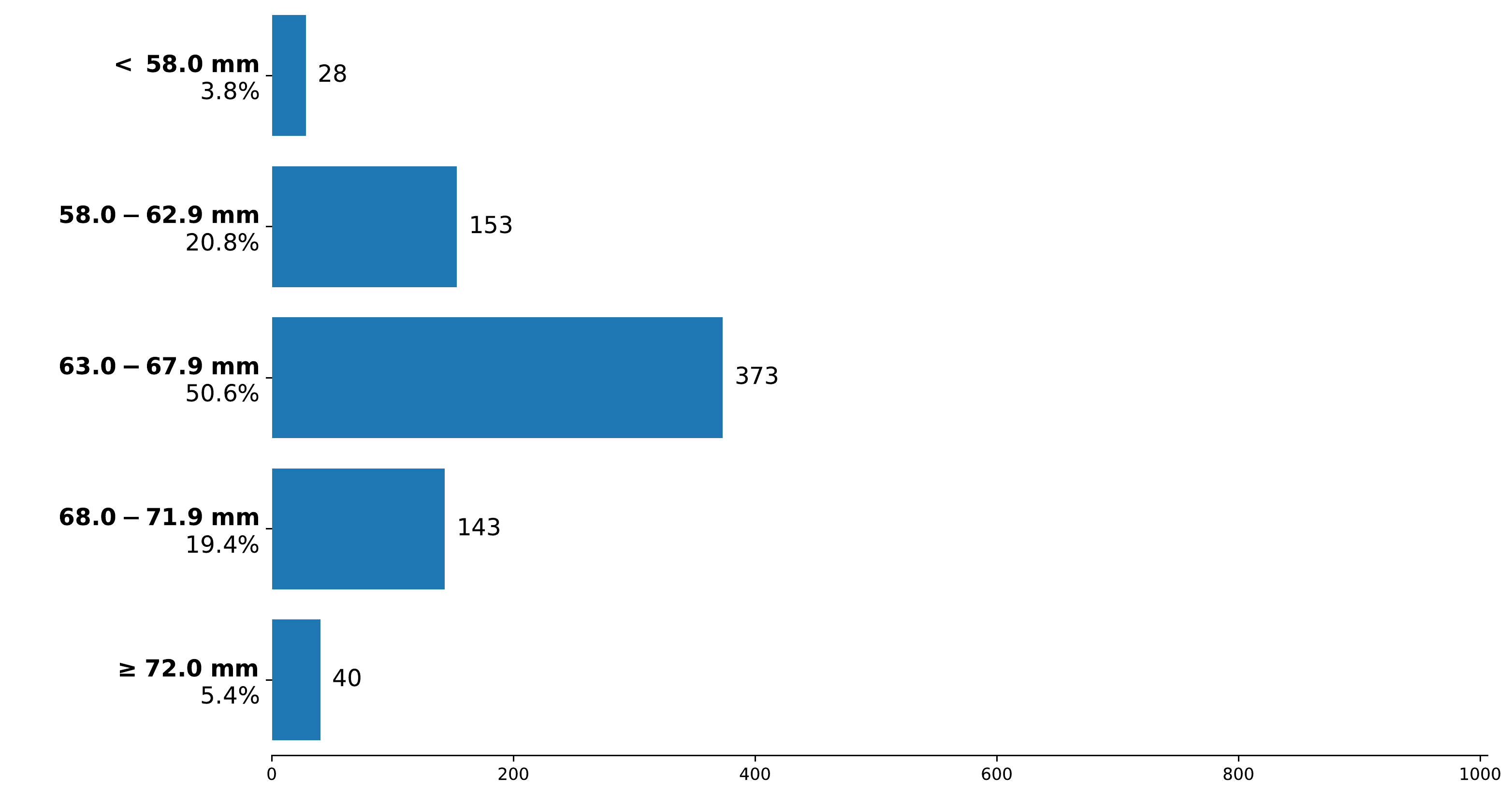}

\eject

\noindent \textbf{Foot Size.} What is the exact length of your foot in centimeters?
\\ \noindent \includegraphics[width=\linewidth]{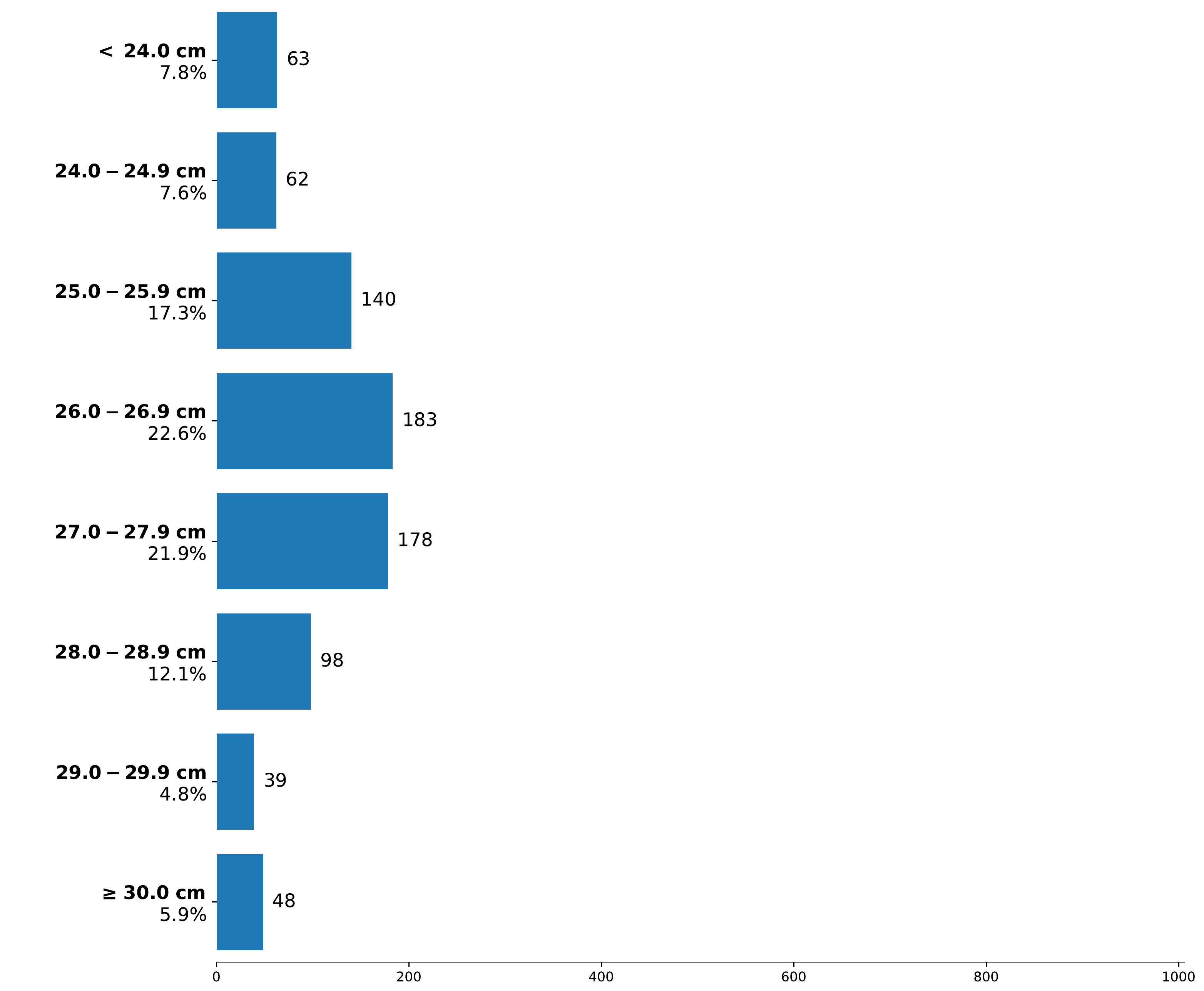}

\noindent \textbf{Hand Length.} What is the exact length of your hand in centimeters?
\\ \noindent \includegraphics[width=\linewidth]{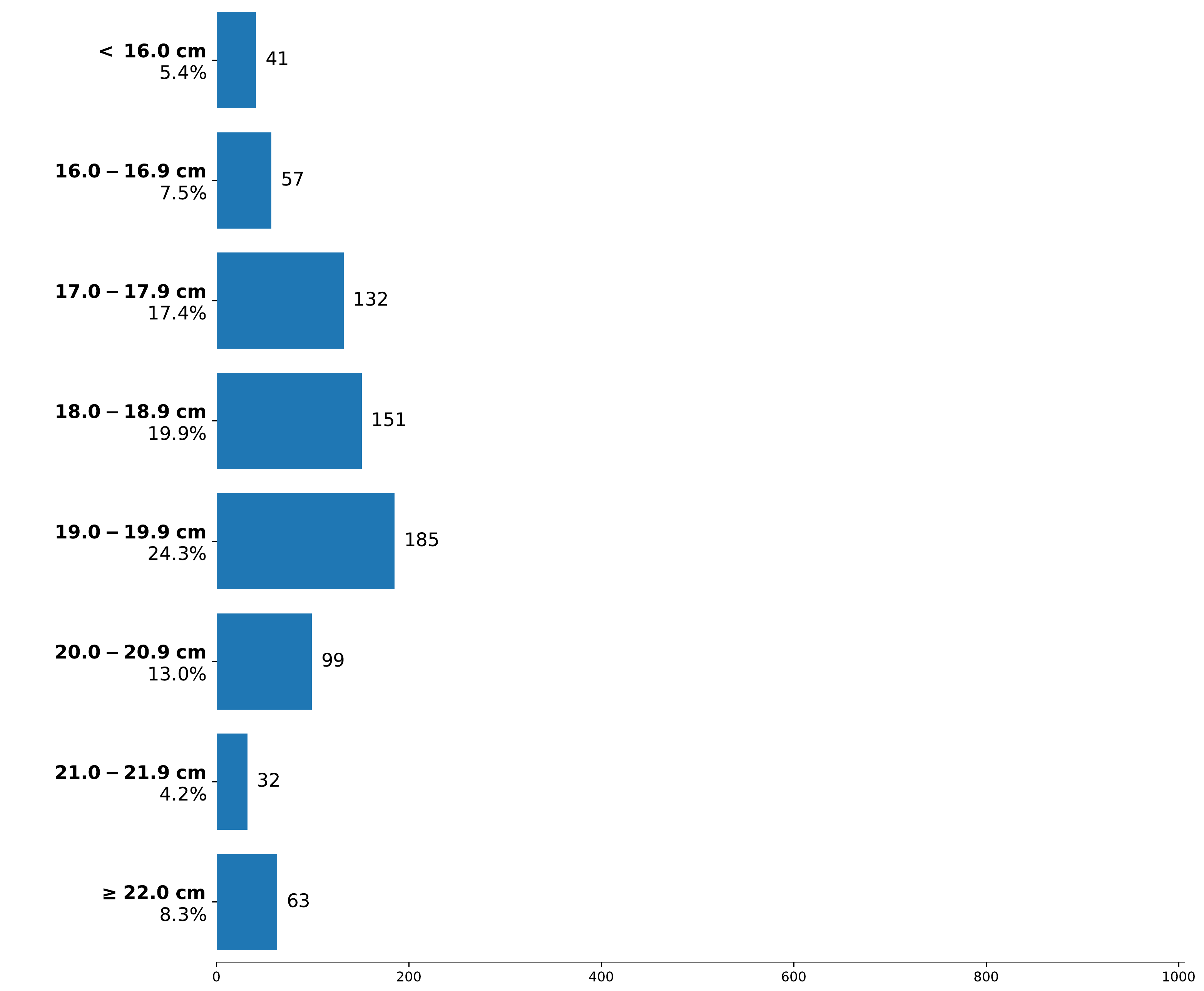}

\noindent \textbf{Reaction Time.} What is your average reaction time in milliseconds?
\\ \noindent \includegraphics[width=\linewidth]{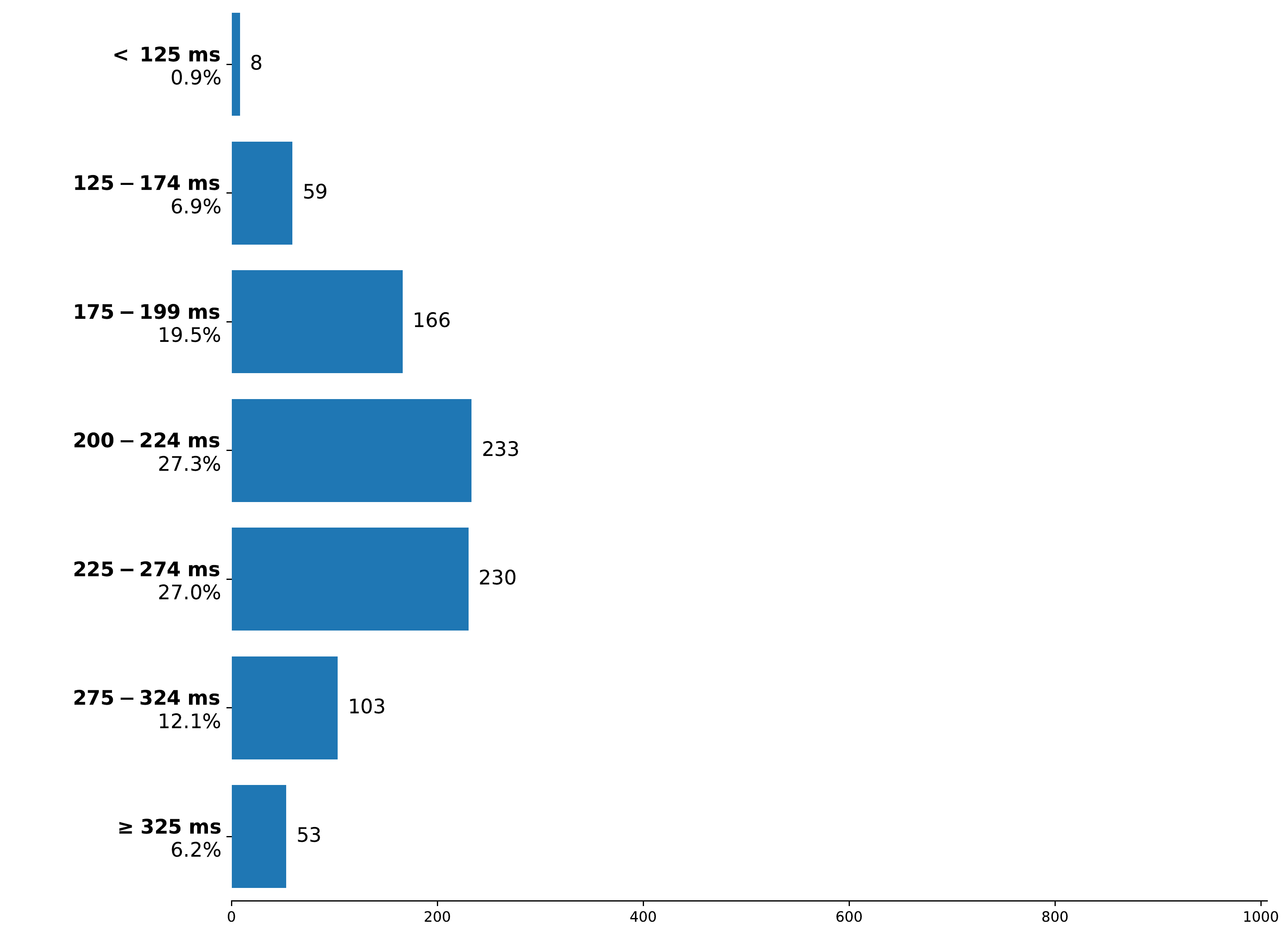}

\eject

\subsection{Clothing}

Finally, we asked participants a number of questions about the clothing they typically wear while playing Beat Saber, in order to asses whether certain types of clothing may encumber free movement while playing the game.

\bigskip

\noindent \textbf{Lower Body.} What clothing, if any, do you typically wear on your lower body when playing Beat Saber?
\\ \noindent \includegraphics[width=\linewidth]{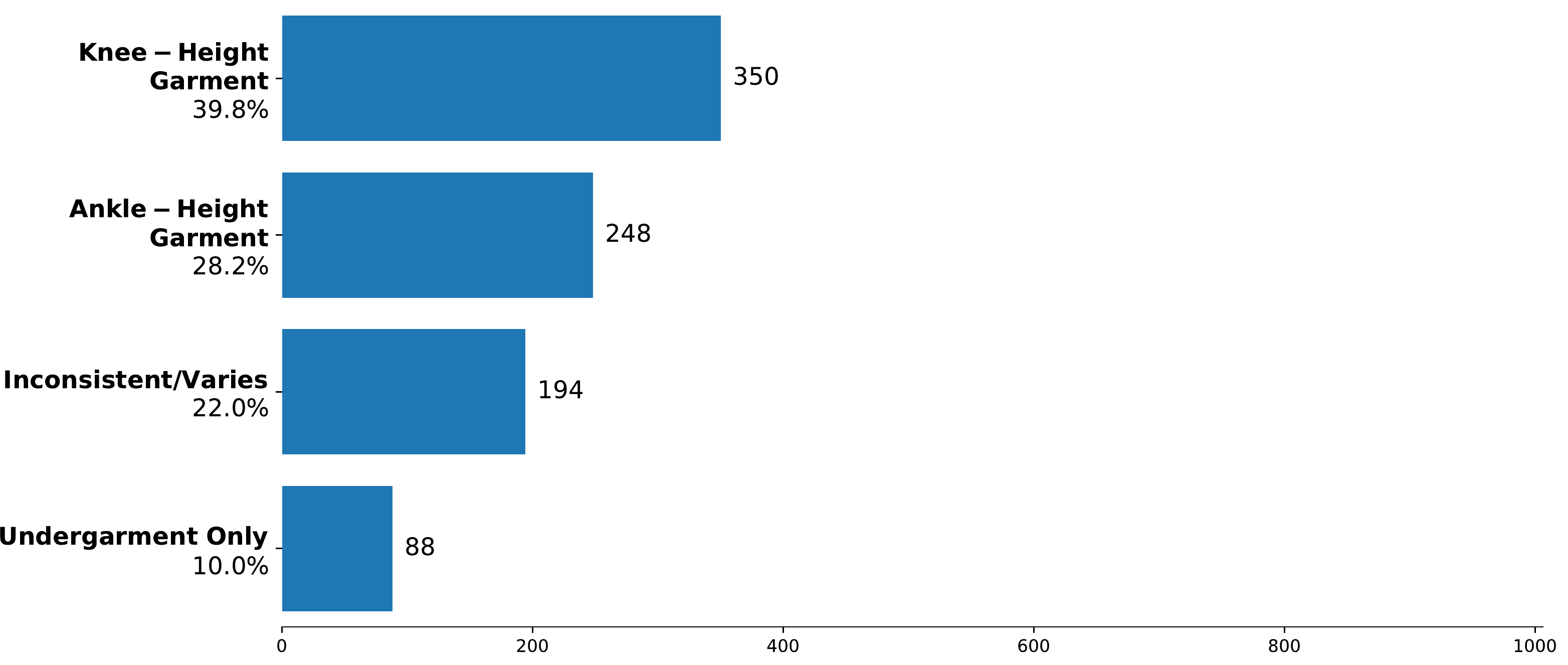}

\bigskip

\noindent \textbf{Upper Body.} What clothing, if any, do you typically wear on your upper body when playing Beat Saber?
\\ \noindent \includegraphics[width=\linewidth]{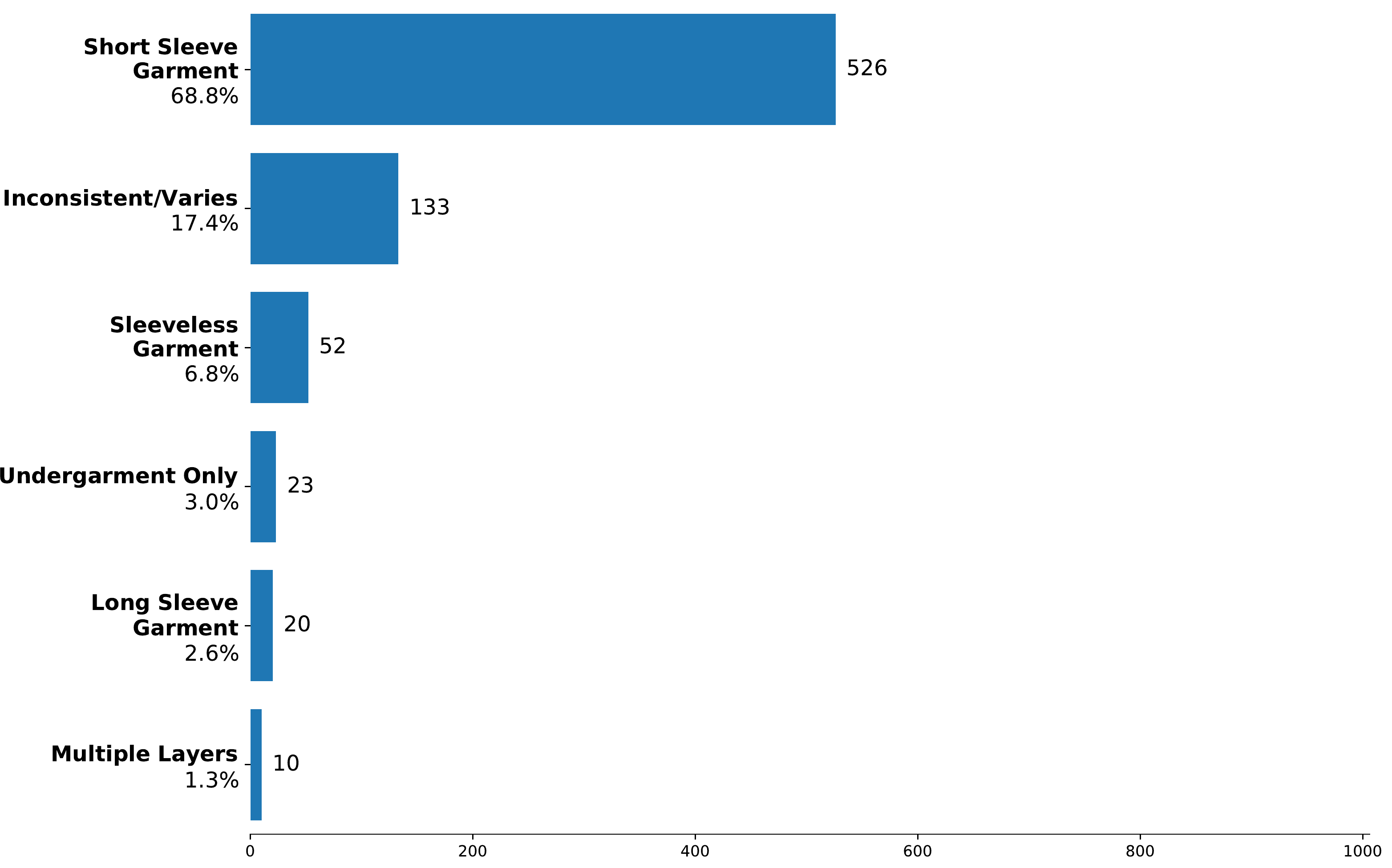}

\bigskip

\noindent \textbf{Footwear.} What footwear, if any, do you typically wear when playing Beat Saber?
\\ \noindent \includegraphics[width=\linewidth]{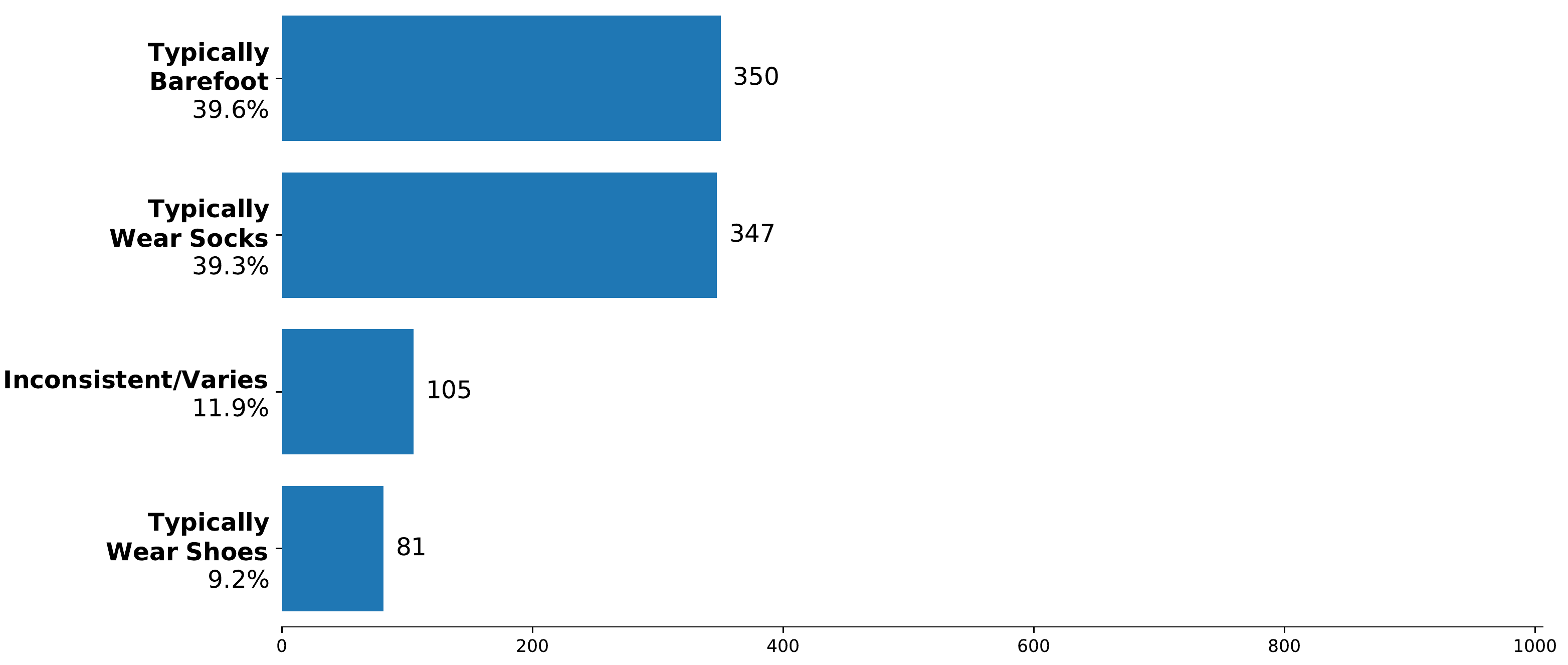}

\eject

\section{Selected Crosstabs}

In this section, we identify and discuss correlations between several of the survey responses and Beat Saber performance. For each of the following variables, we display the effect of each selection on performance points (PP), a standardized measure of overall Beat Saber skill that is used to produce global player rankings.

Cross-tabs for all surveyed variables are not included; rather, we have chosen to highlight correlations that are interesting, noteworthy, or indicative of counter-intuitive trends, particularly those with high statistical significance.

\subsection{Surveyed Population}

First, we include the unsurprising positive correlation between play time and performance. This result provides a rough benchmark for the remaining correlations, as it allows one to understand the size of each effect in terms of the equivalent number of hours of experience necessary to achieve an equivalent performance delta.

\medskip

\noindent \textbf{Play Time.} To the nearest hour, how many total hours have you spent playing Beat Saber?
\\ \noindent \includegraphics[width=\linewidth]{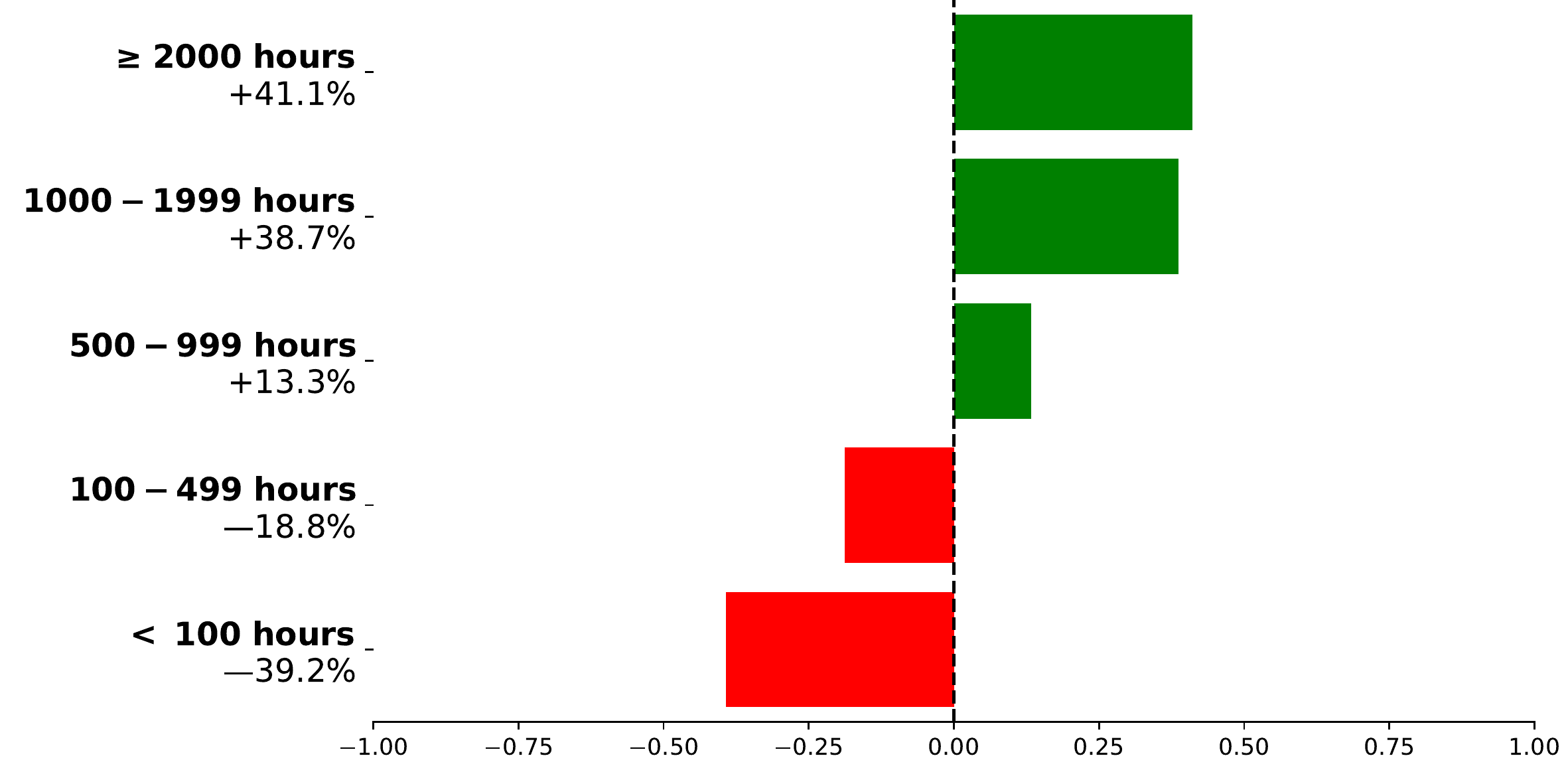}

\subsection{Demographics}

There are several demographics which appear to strongly affect Beat Saber performance, with many likely being a proxy for the primary factor of age.

\medskip

\noindent \textbf{Age.} What is your age in years?
\\ \noindent \includegraphics[width=\linewidth]{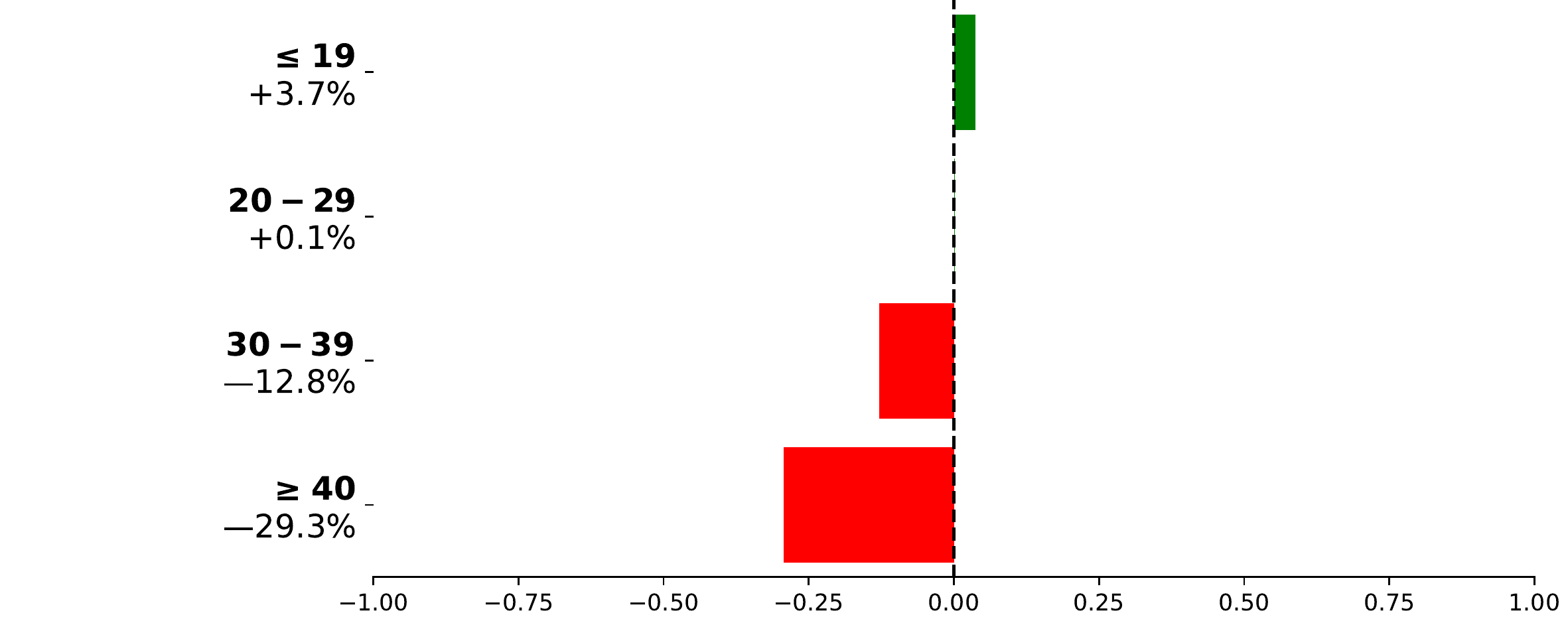}

\noindent \textbf{Employment Status.} Which of the following options best represents your current employment status?
\\ \noindent \includegraphics[width=\linewidth]{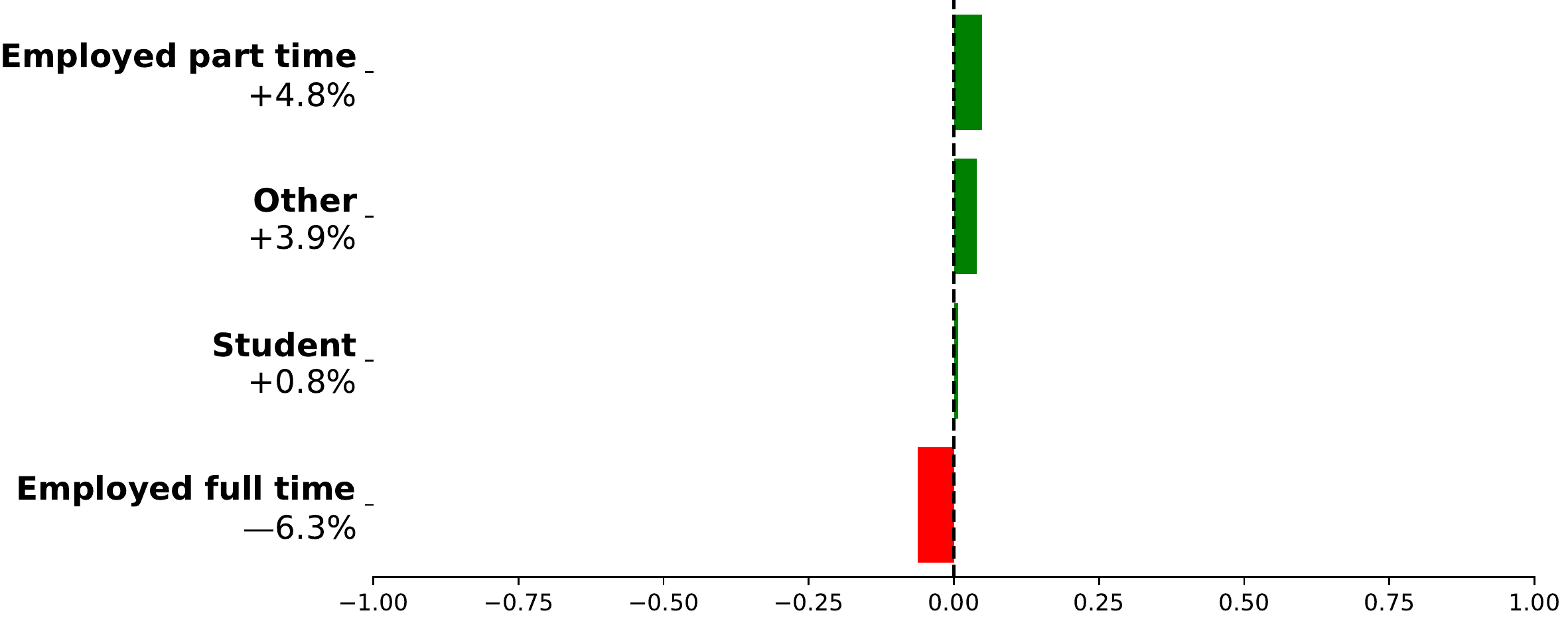}

\eject

\noindent \textbf{Educational Status.} What is the highest degree or level of school you have completed?
\\ \noindent \includegraphics[width=\linewidth]{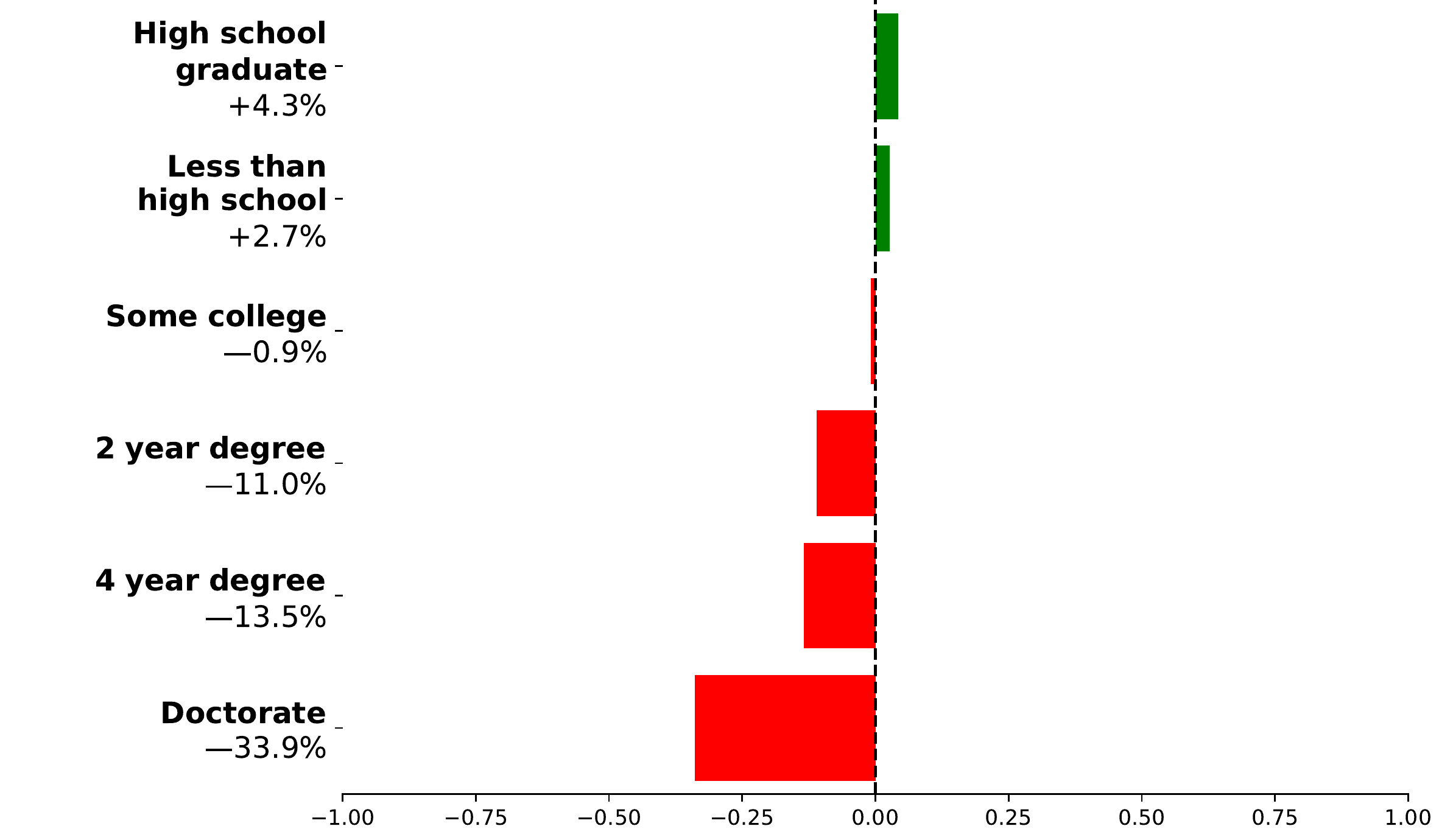}

\noindent \textbf{Marital Status.} Which of the following options best represents your current marital status?
\\ \noindent \includegraphics[width=\linewidth]{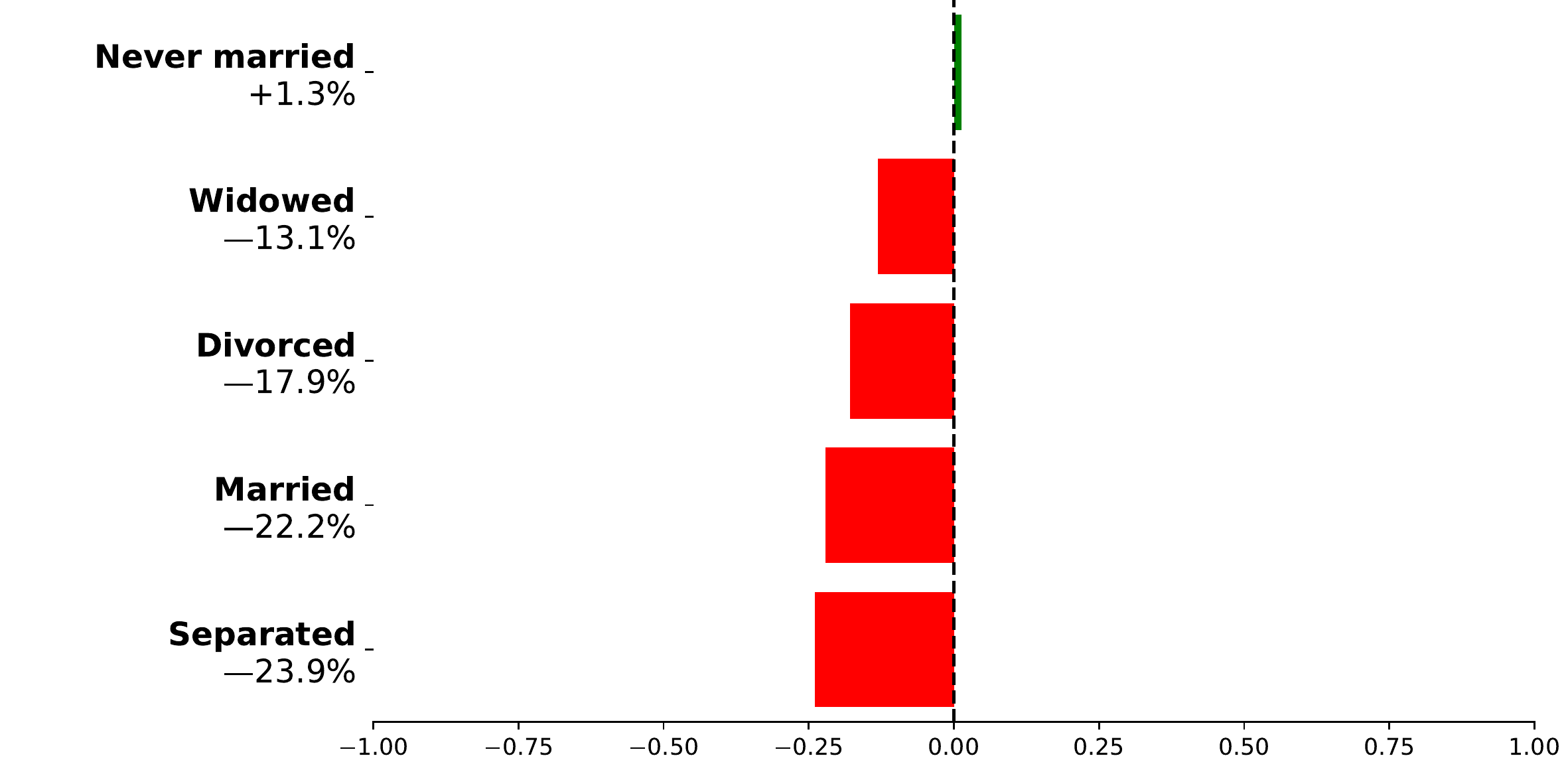}

\noindent \textbf{Political Orientation.} Which of the following generally best represents your political views?
\\ \noindent \includegraphics[width=\linewidth]{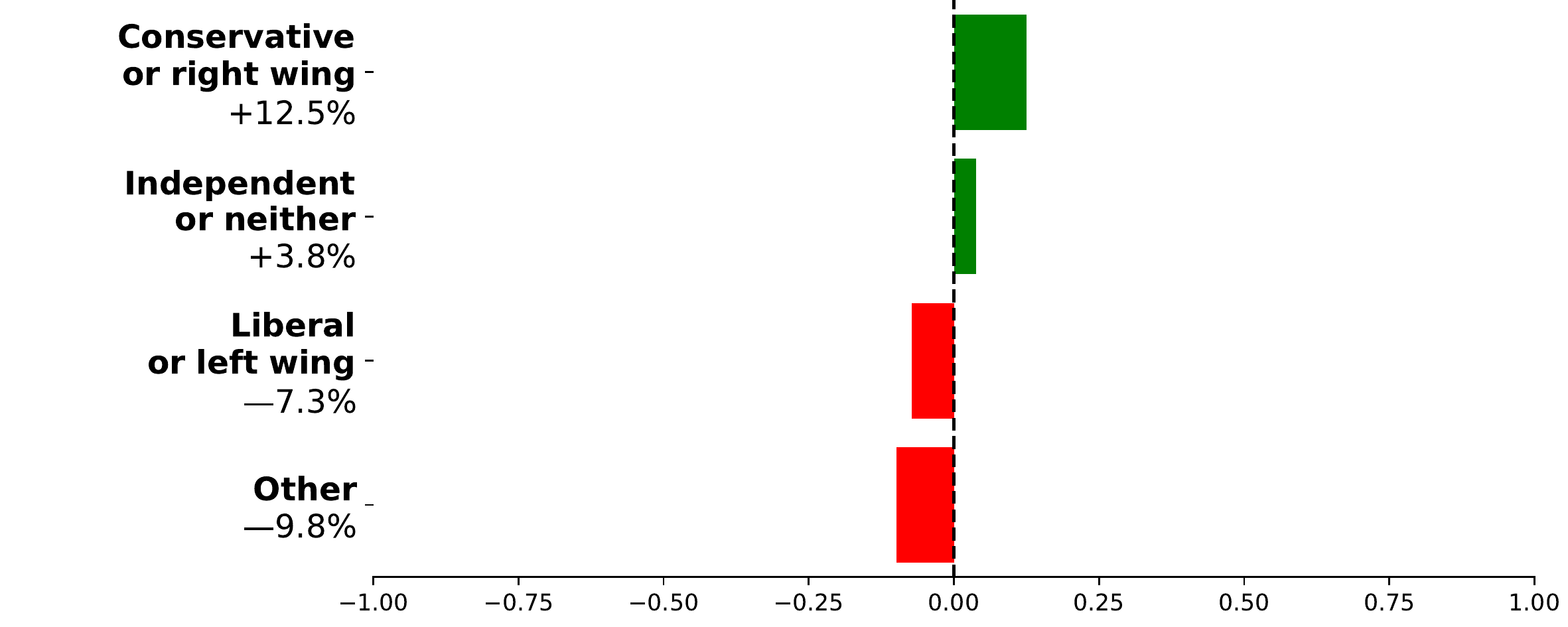}

\noindent \textbf{Languages.} Which languages do you speak fluently? If multiple, list all languages spoken in order of proficiency.
\\ \noindent \includegraphics[width=\linewidth]{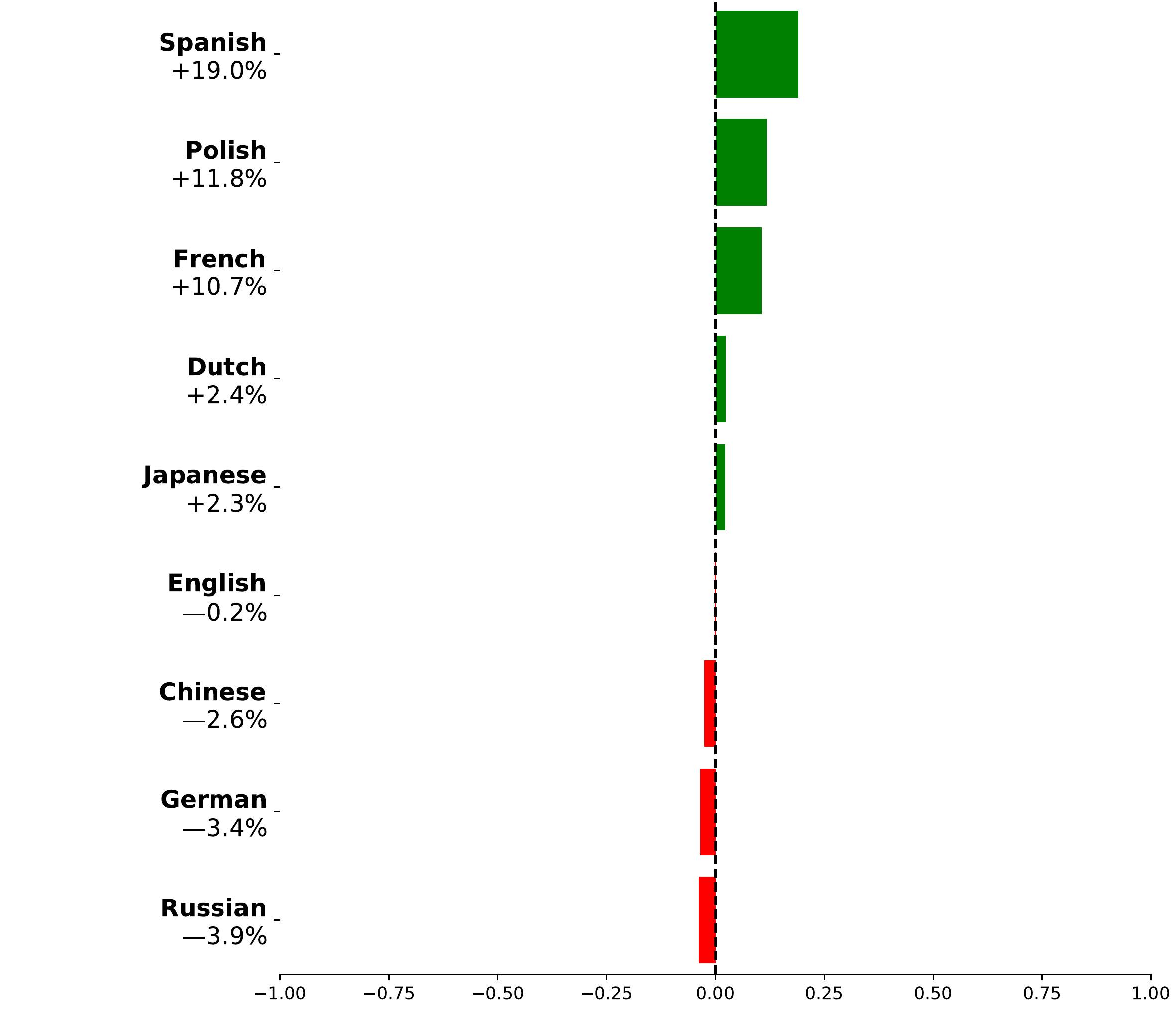}

\eject

\subsection{Technical Specifications}

For tethered VR users, technical specifications are another important predictor of Beat Saber performance. In particular, low-fidelity setups appear to be a significant impediment to performance across all examined specifications, with diminishing returns for added resources in areas like system memory and CPU speed.

\bigskip

\noindent \textbf{Physical Cores.} According to the Steam system report, how many physical CPU cores are in the user's PC?
\\ \noindent \includegraphics[width=\linewidth]{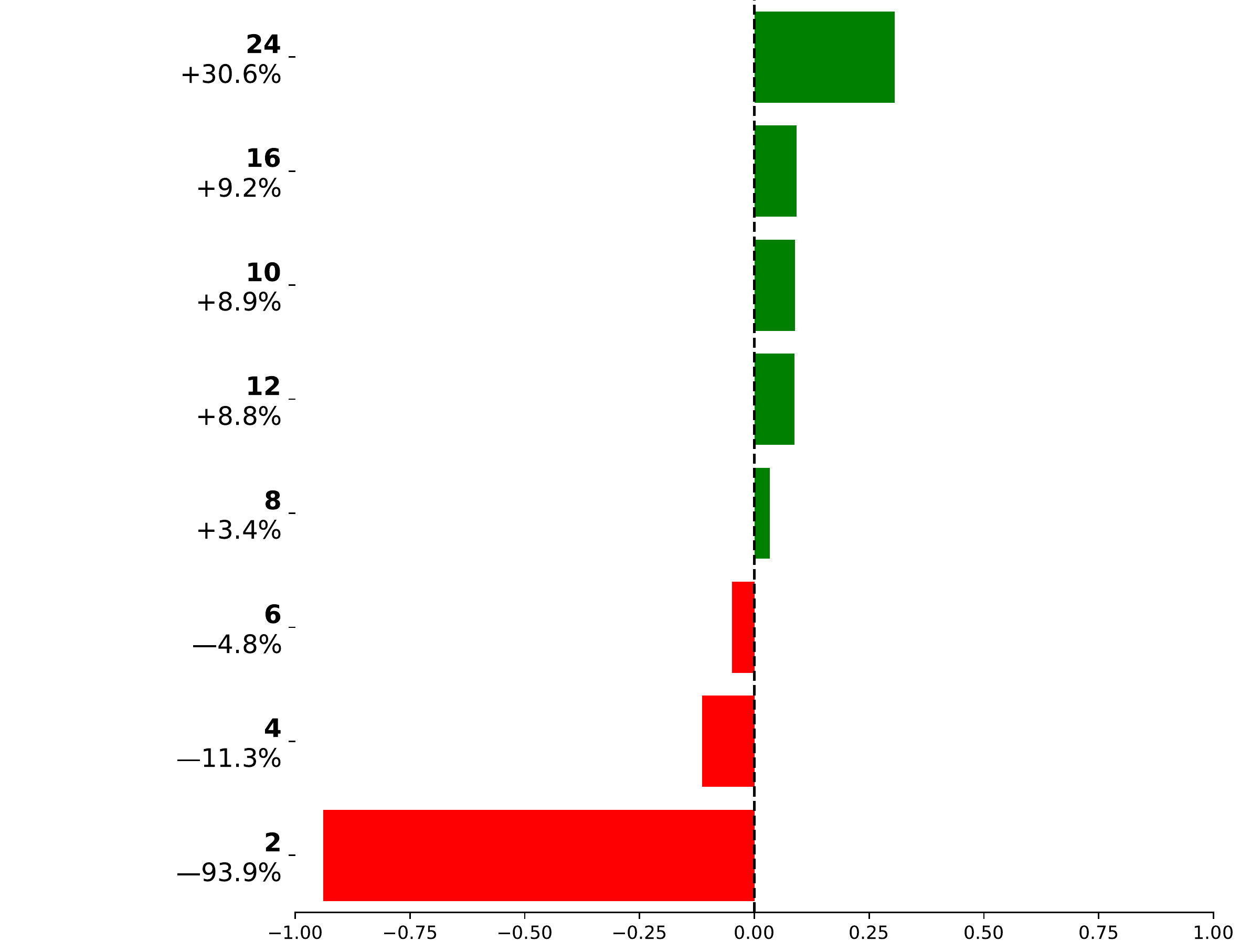}

\medskip

\noindent \textbf{CPU Speed.} According to the Steam system report, what is the base CPU clock speed in the user's PC?
\\ \noindent \includegraphics[width=\linewidth]{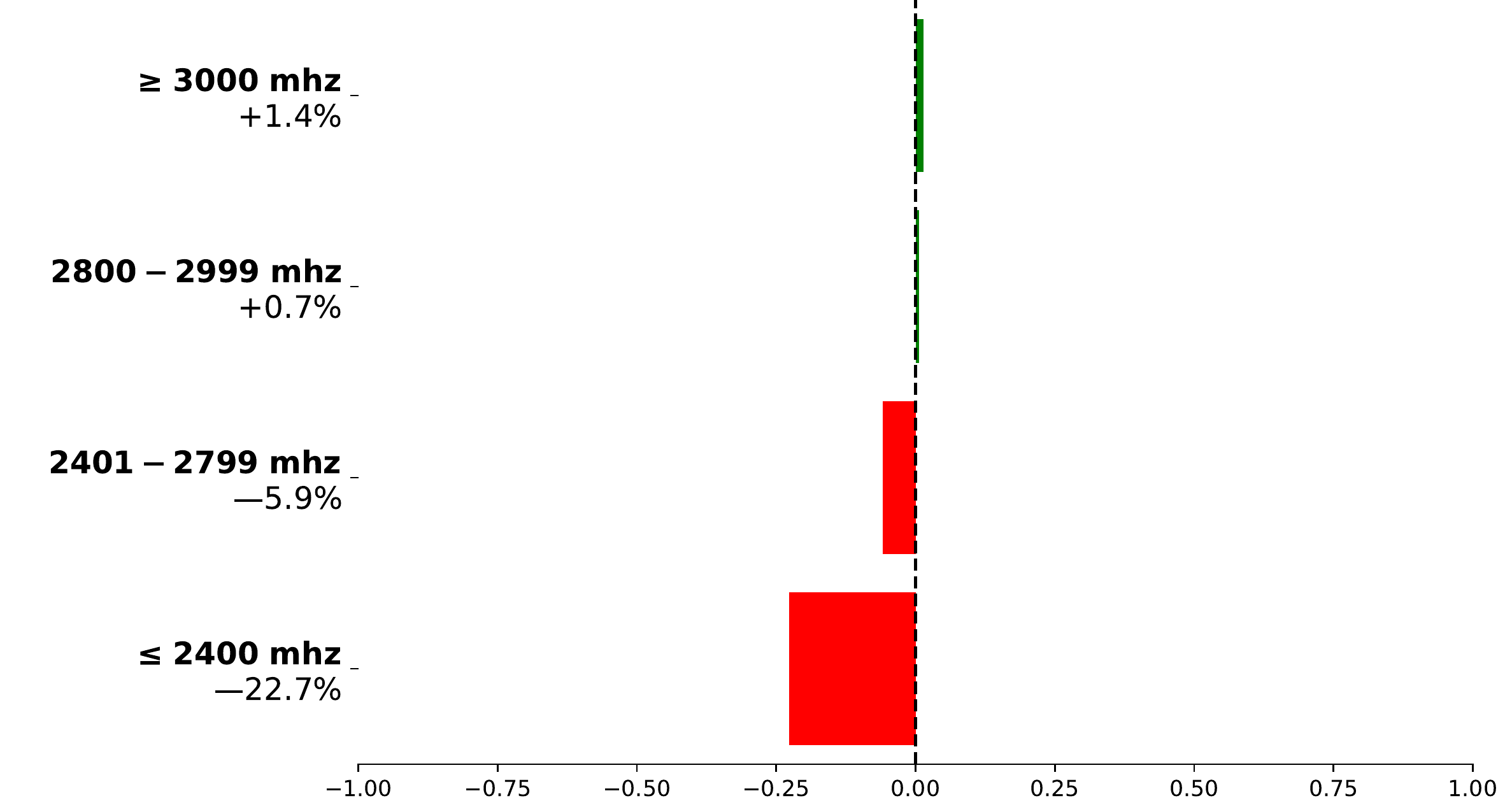}

\medskip

\noindent \textbf{CPU Brand.} According to the Steam system report, what is the vendor of the CPU in the user's PC?
\\ \noindent \includegraphics[width=\linewidth]{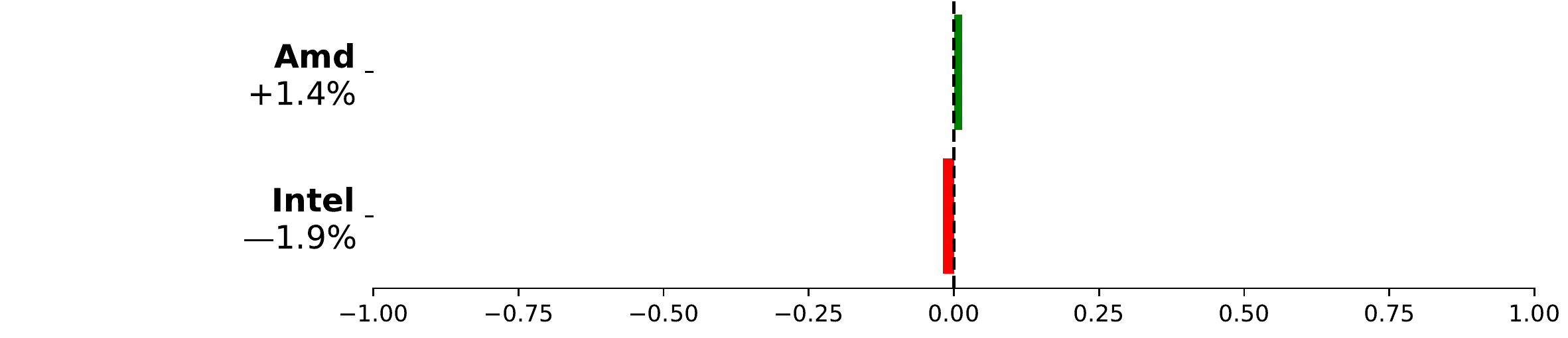}

\eject

\noindent \textbf{Form Factor.} According to the Steam system report, is the user's PC a laptop or desktop?
\\ \noindent \includegraphics[width=\linewidth]{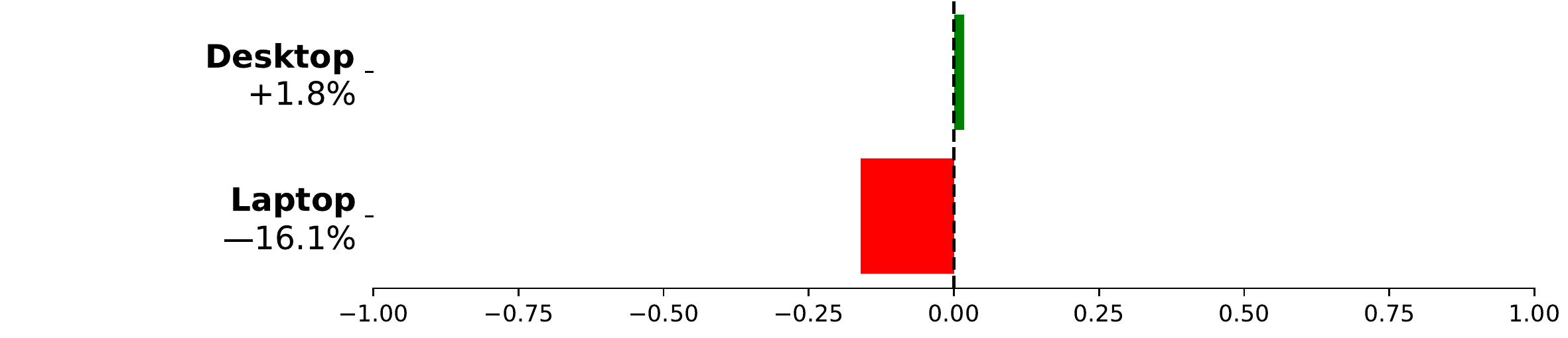}

\medskip

\noindent \textbf{Operating System.} According to the Steam system report, what is the operating system of the user's PC?
\\ \noindent \includegraphics[width=\linewidth]{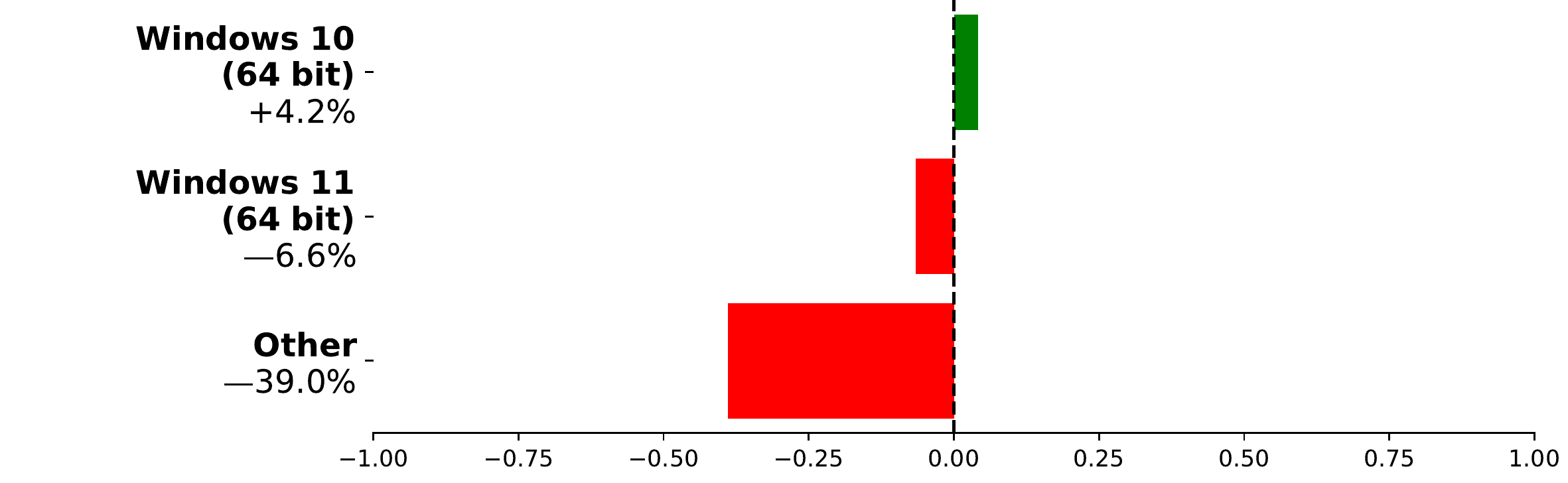}

\medskip

\noindent \textbf{System Memory.} According to the Steam system report, how much RAM is in the user's PC?
\\ \noindent \includegraphics[width=\linewidth]{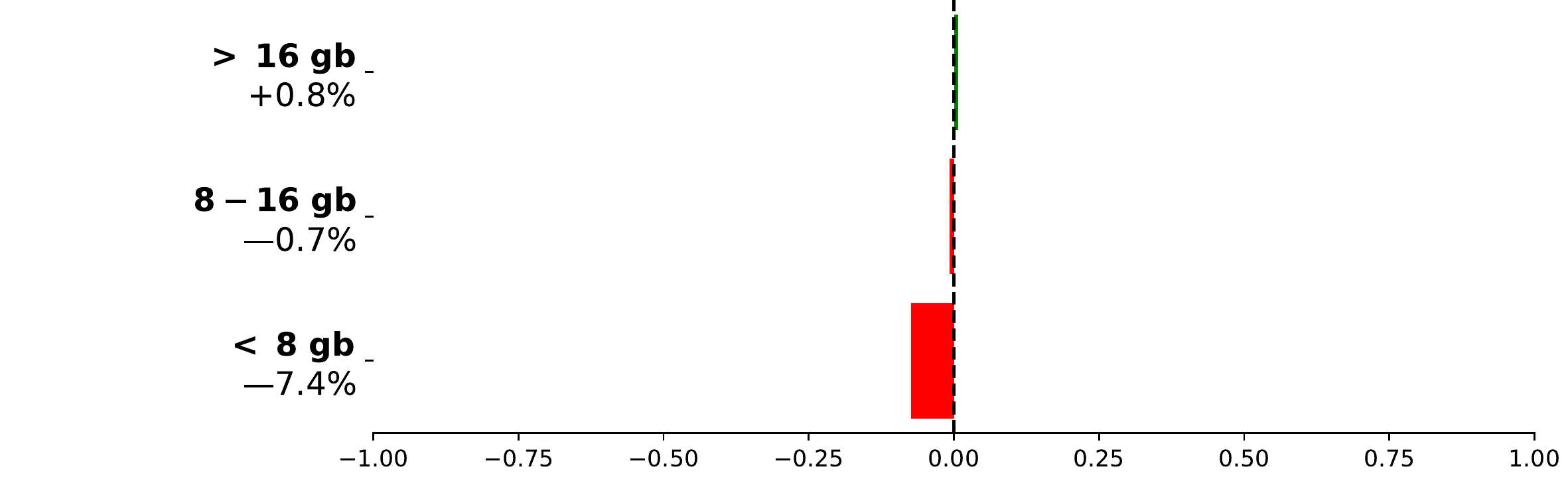}

\medskip

\noindent \textbf{Base Stations.} According to the Steam system report, how many lighthouses or base stations does the user have?
\\ \noindent \includegraphics[width=\linewidth]{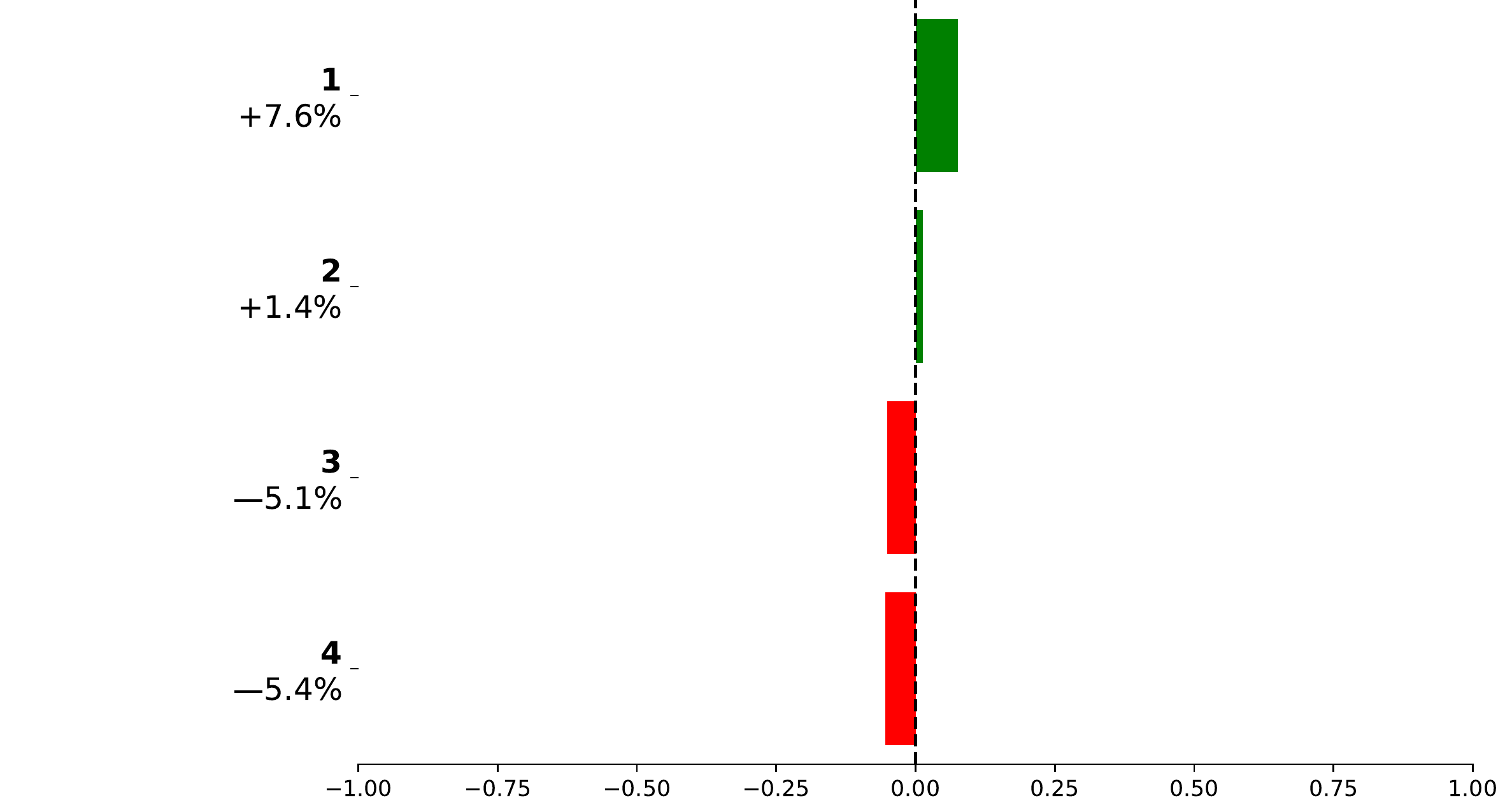}

\medskip

The inverse correlation between the number of base stations of SteamVR users and their Beat Saber performance is potentially counter-intuitive. Conventional wisdom suggests that adding more base stations reduces tracking blindspots and decreases the likelihood of tracking errors. However, this may be counteracted by the reduced polling rate observed when a higher number of base stations are utilized.

\eject

\subsection{Background}

The background questions demonstrated an expected effect with respect to an athletic background improving Beat Saber performance, but also an unexpected negative effect of most forms of musical background.

\bigskip

\noindent \textbf{Music.} Have you ever skillfully played a musical instrument?
\\ \noindent \includegraphics[width=\linewidth]{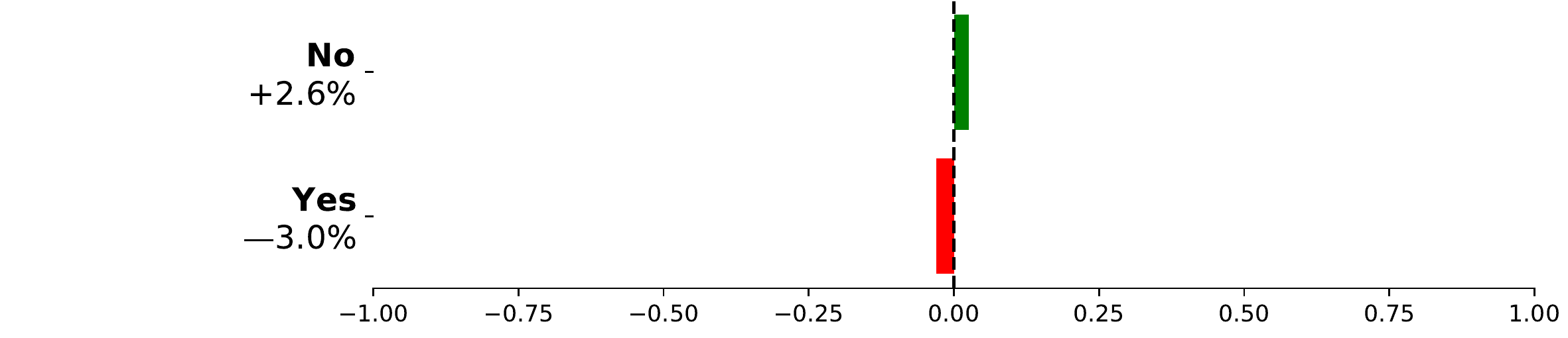}

\medskip

\noindent \textbf{Music.} If you have ever skillfully played a musical instrument, list the instrument(s).
\\ \noindent \includegraphics[width=\linewidth]{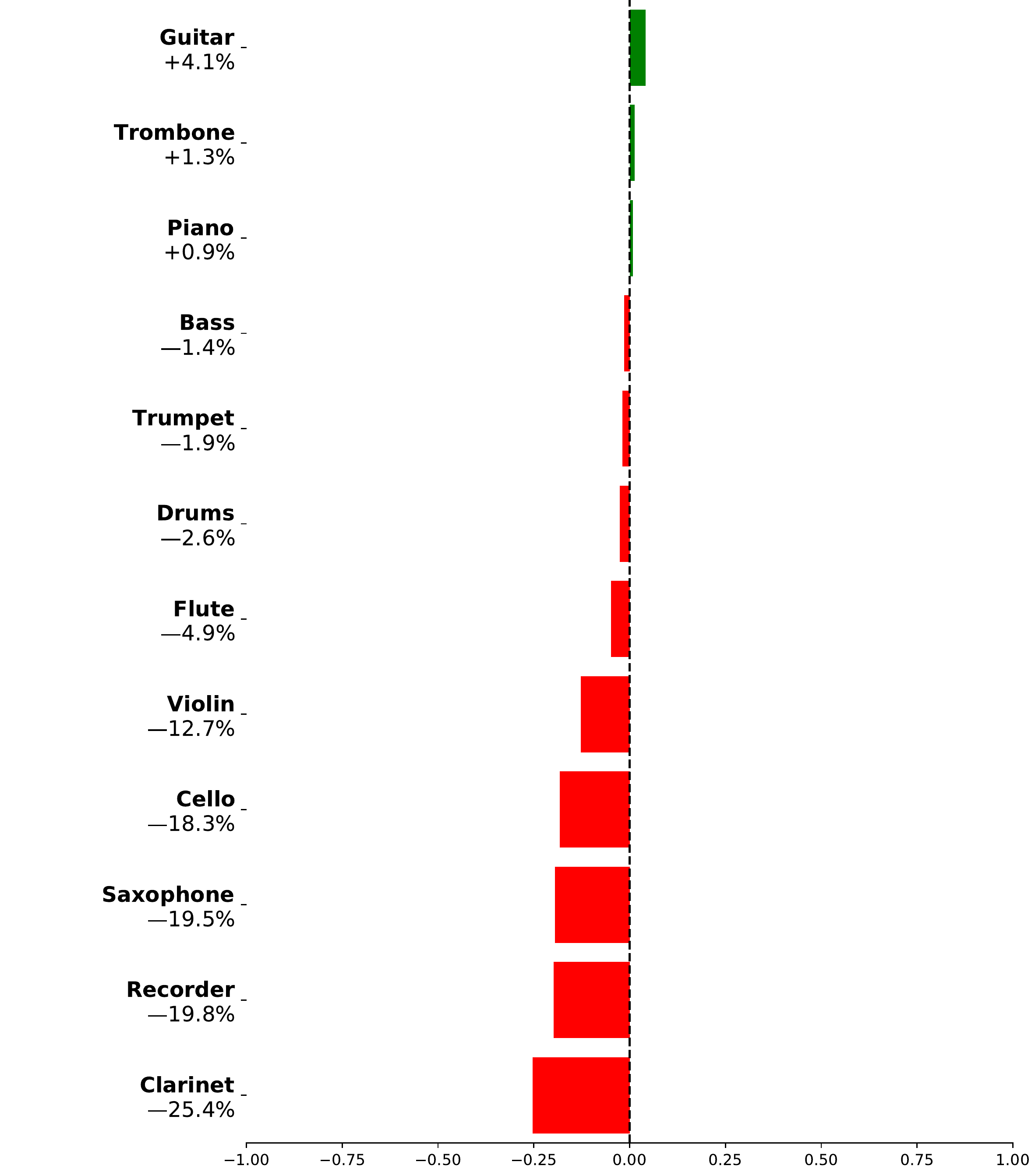}

\medskip

\noindent \textbf{Dance.} Have you ever skillfully practiced or exhibited a recognized form of dance?
\\ \noindent \includegraphics[width=\linewidth]{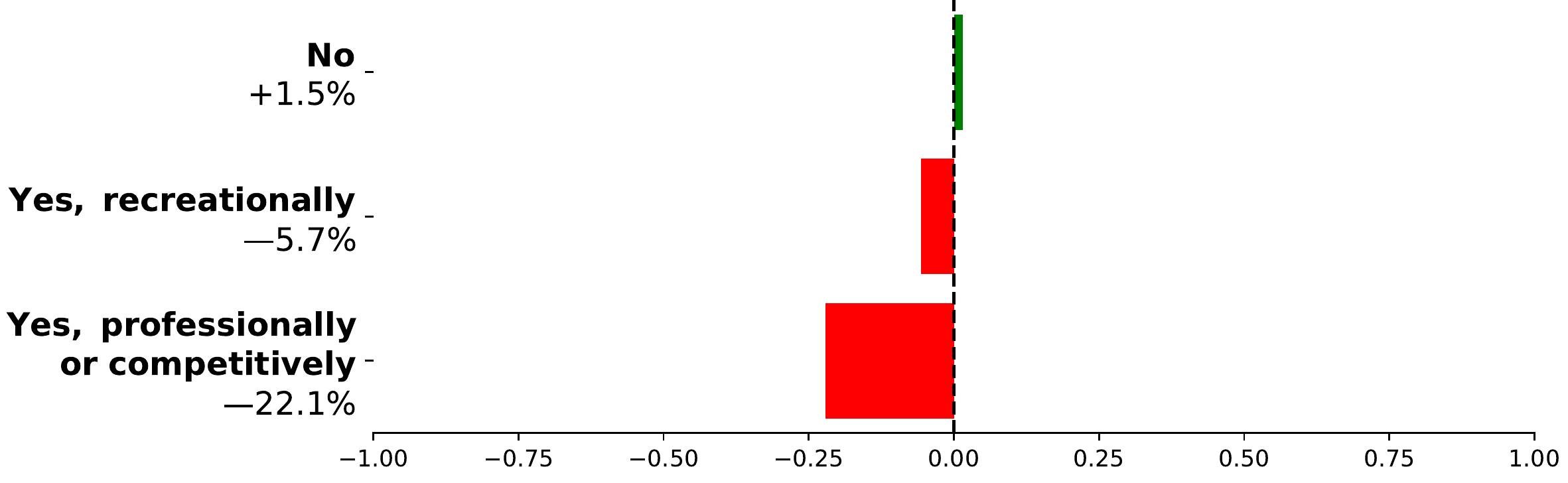}

\medskip

\noindent \textbf{Rhythm Games.} Have you ever played a rhythm game other than Beat Saber?
\\ \noindent \includegraphics[width=\linewidth]{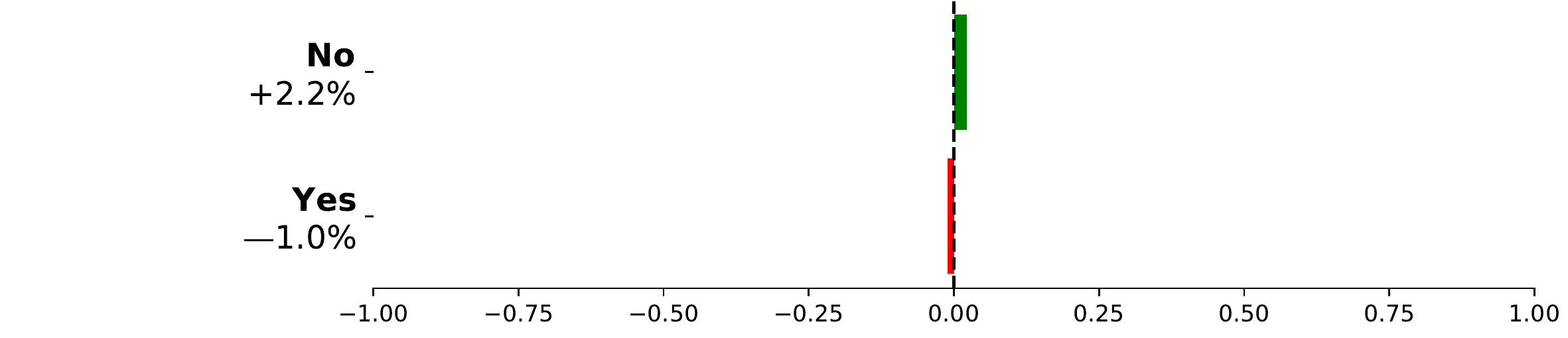}

\eject

\noindent \textbf{Rhythm Games.} If you have ever played a rhythm game other than Beat Saber, list the game(s).
\\ \noindent \includegraphics[width=\linewidth]{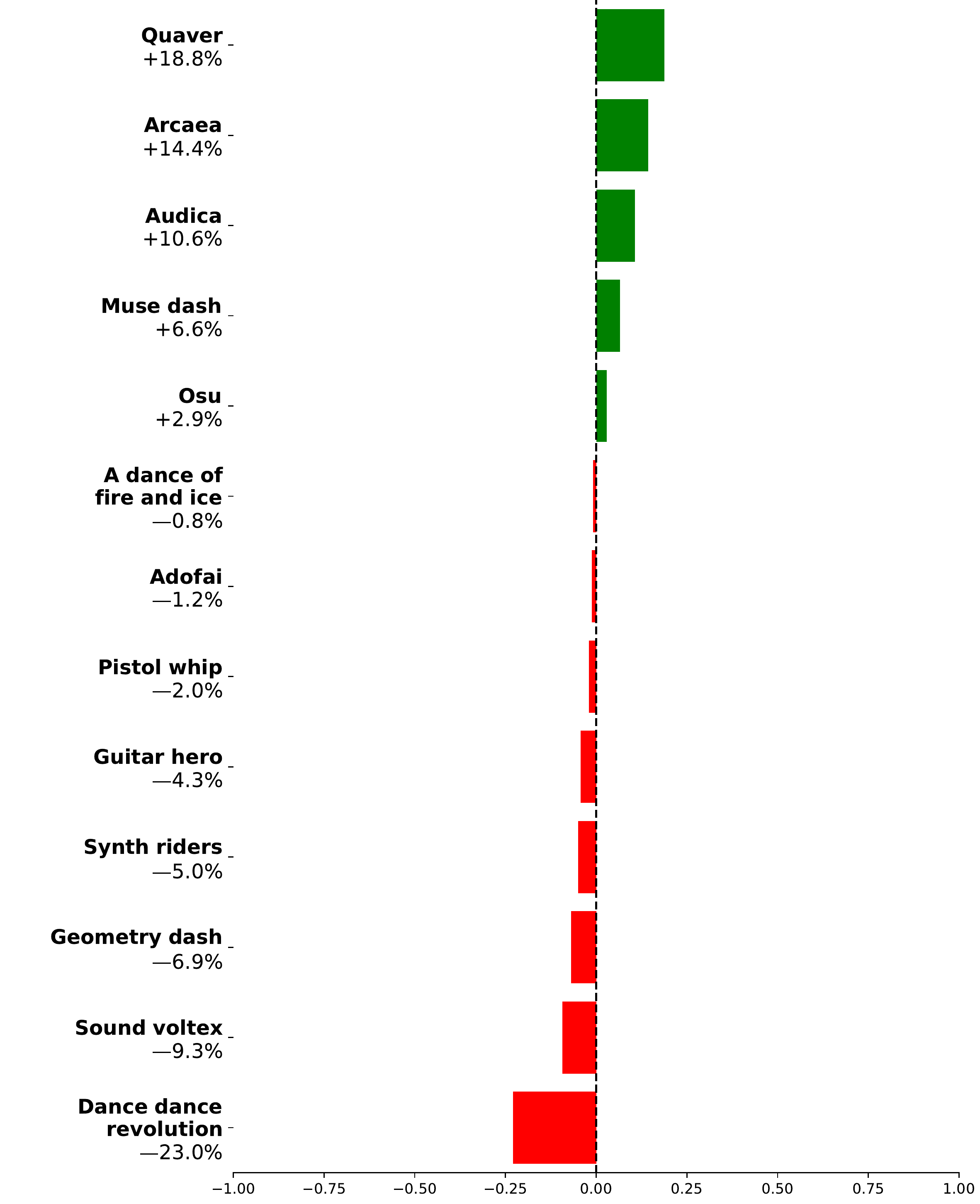}

\noindent \textbf{Athletics.} Have you ever competitively participated in an individual or team-based athletic sport?
\\ \noindent \includegraphics[width=\linewidth]{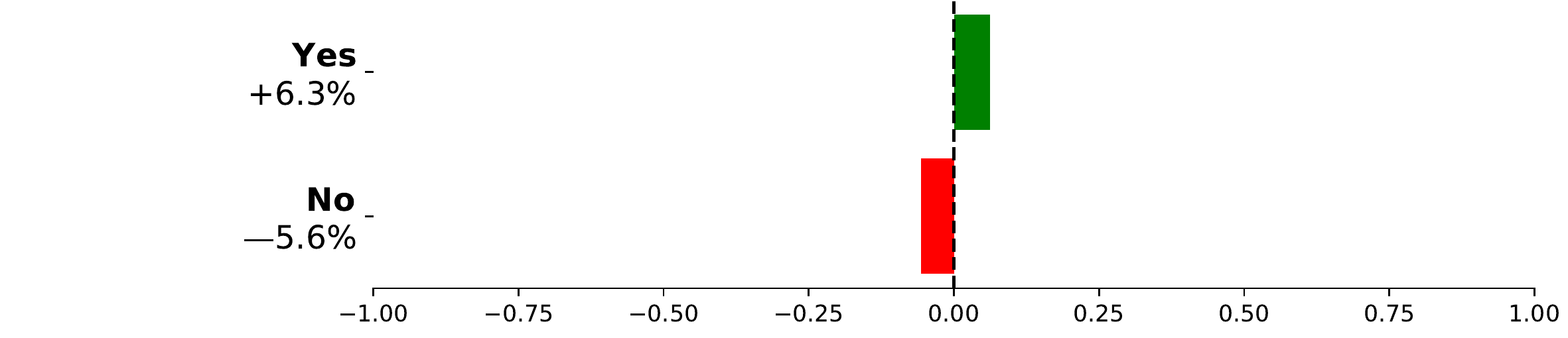}

\noindent \textbf{Athletics.} If you have ever competitively participated in an individual or team-based athletic sport, list the sport(s).
\\ \noindent \includegraphics[width=\linewidth]{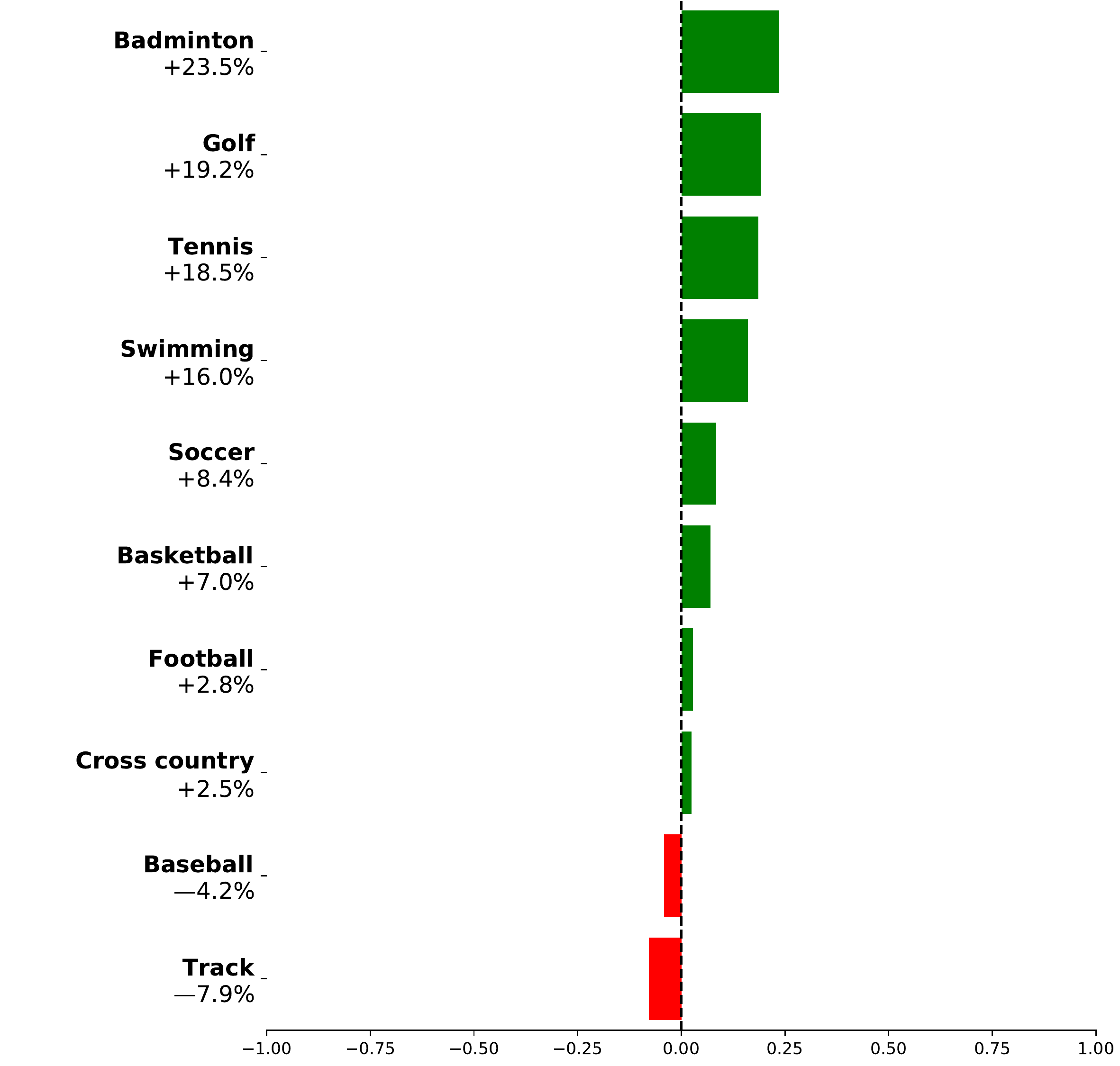}

\eject

\subsection{Health}

Unsurprisingly, visual acuity is an important factor impacting Beat Saber gameplay, with prescription lens users and color blind individuals experiencing worse performance on average. Participants who reported inconsistent use of lenses demonstrated the worst performance.

\bigskip

\noindent \textbf{Eyesight.} Do you regularly wear prescription glasses or contact lenses?
\\ \noindent \includegraphics[width=\linewidth]{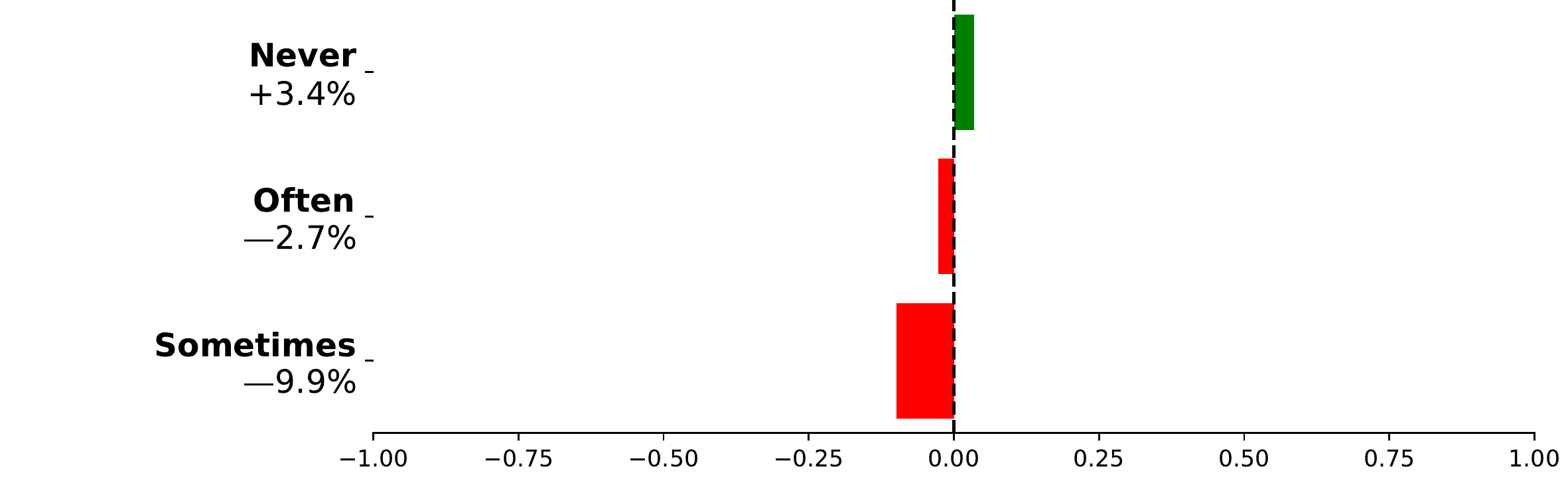}

\medskip

\noindent \textbf{Lenses.} Do you usually wear prescription  glasses or contact lenses while playing Beat Saber?
\\ \noindent \includegraphics[width=\linewidth]{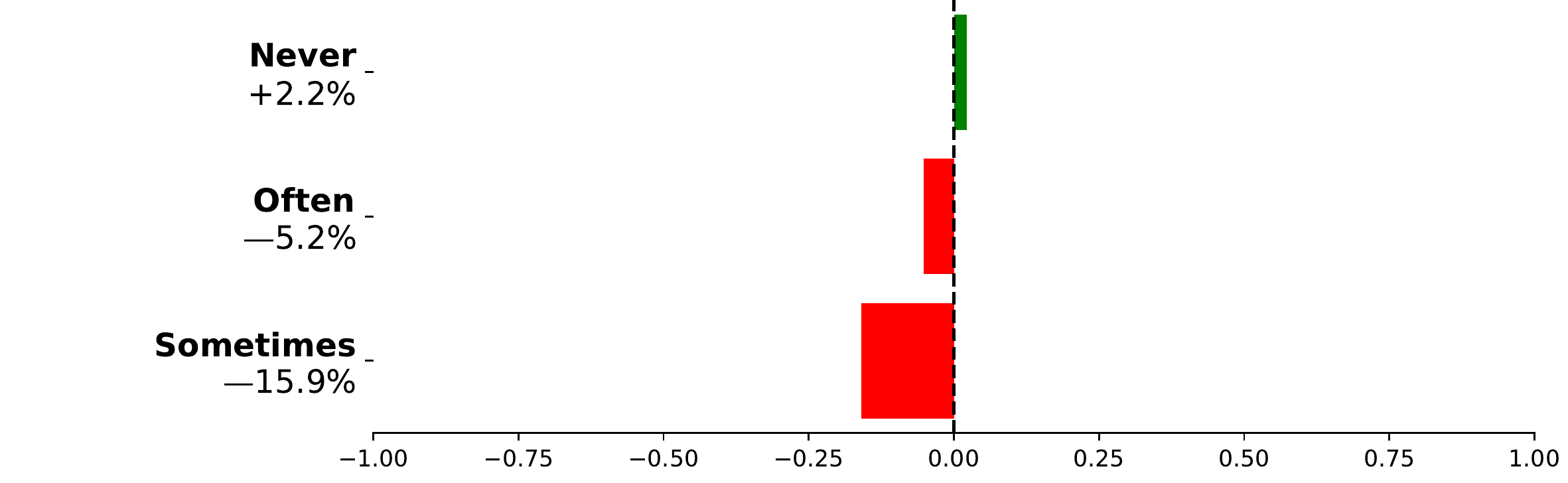}

\medskip

\noindent \textbf{Color Blindness.} Do you have any form of color blindness or color vision deficiency?
\\ \noindent \includegraphics[width=\linewidth]{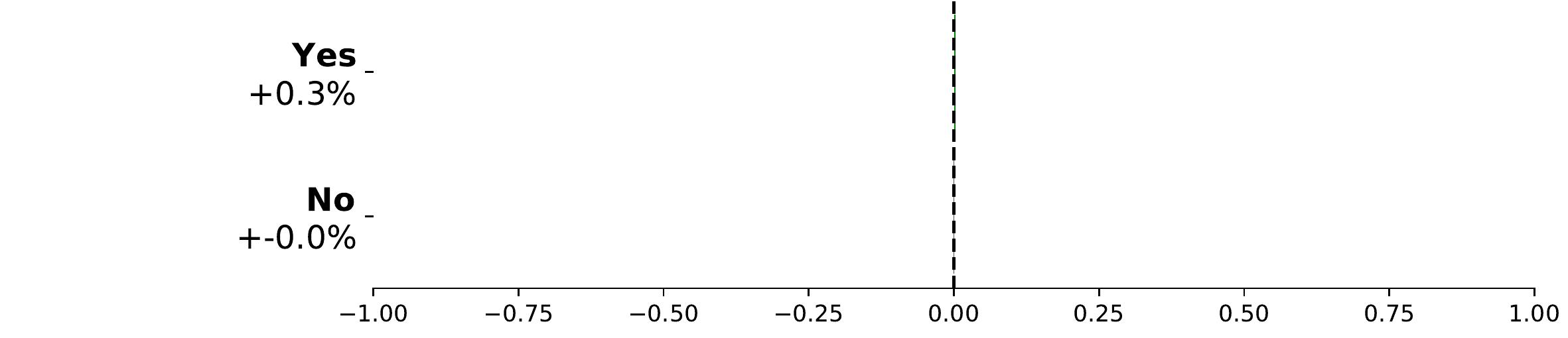}

\subsection{Behavior Patterns}

The effects illustrated in this section are amongst the strongest in magnitude presented in this study. In particular, the choice of grip and preparation method have a strong impact on Beat Saber performance.

\bigskip

\noindent \textbf{Grip.} Which of the following grips do you prefer to use on standalone VR devices (e.g., Oculus Quest 2, PICO Neo3, etc.)?
\\ \noindent \includegraphics[width=\linewidth]{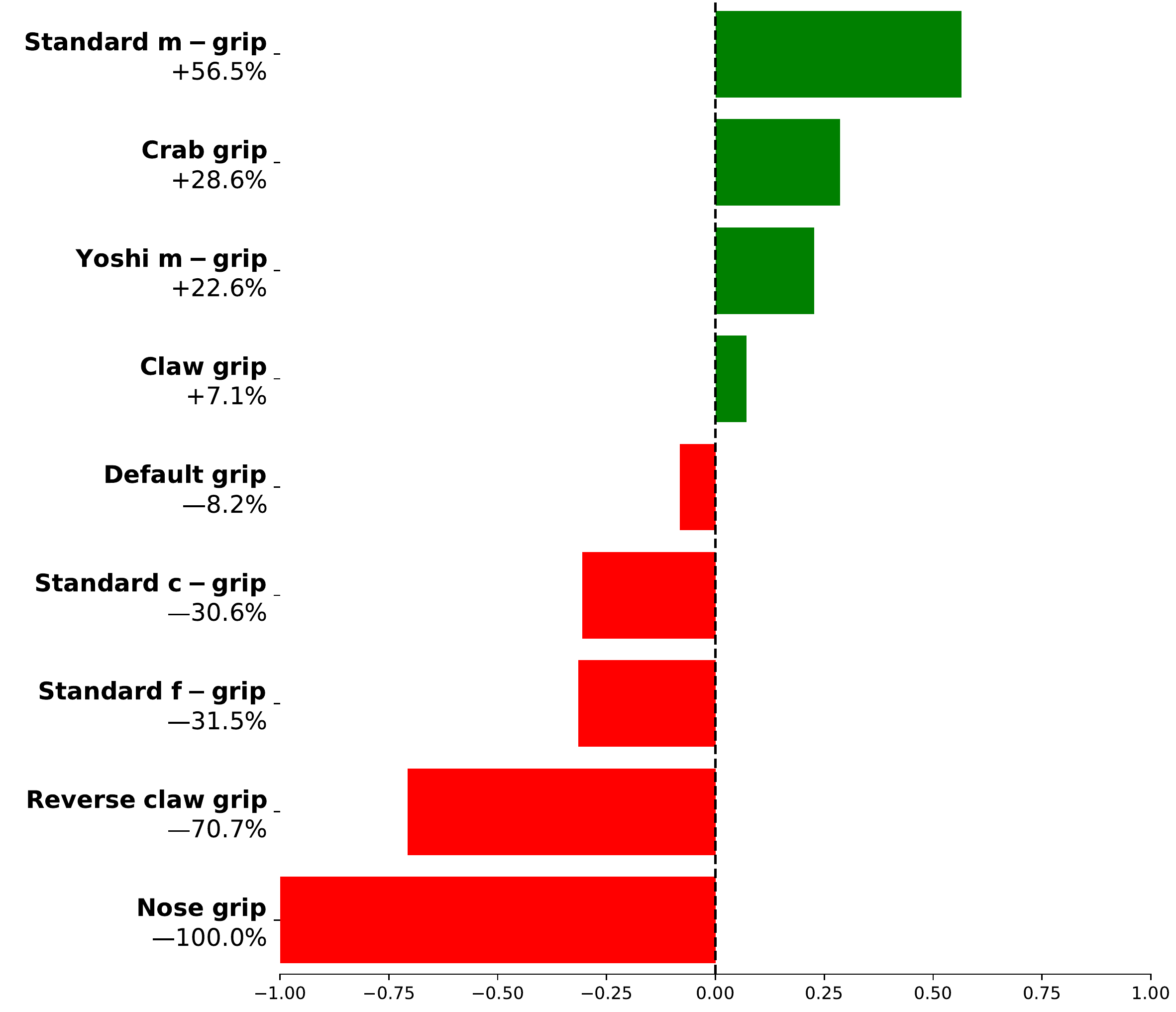}

\eject

\noindent \textbf{Caffinated Items.} Approximately how many caffeinated foods or beverages (e.g., Coffee, Black Tea, Energy Drinks, etc.) do you consume on a regular basis?
\\ \noindent \includegraphics[width=\linewidth]{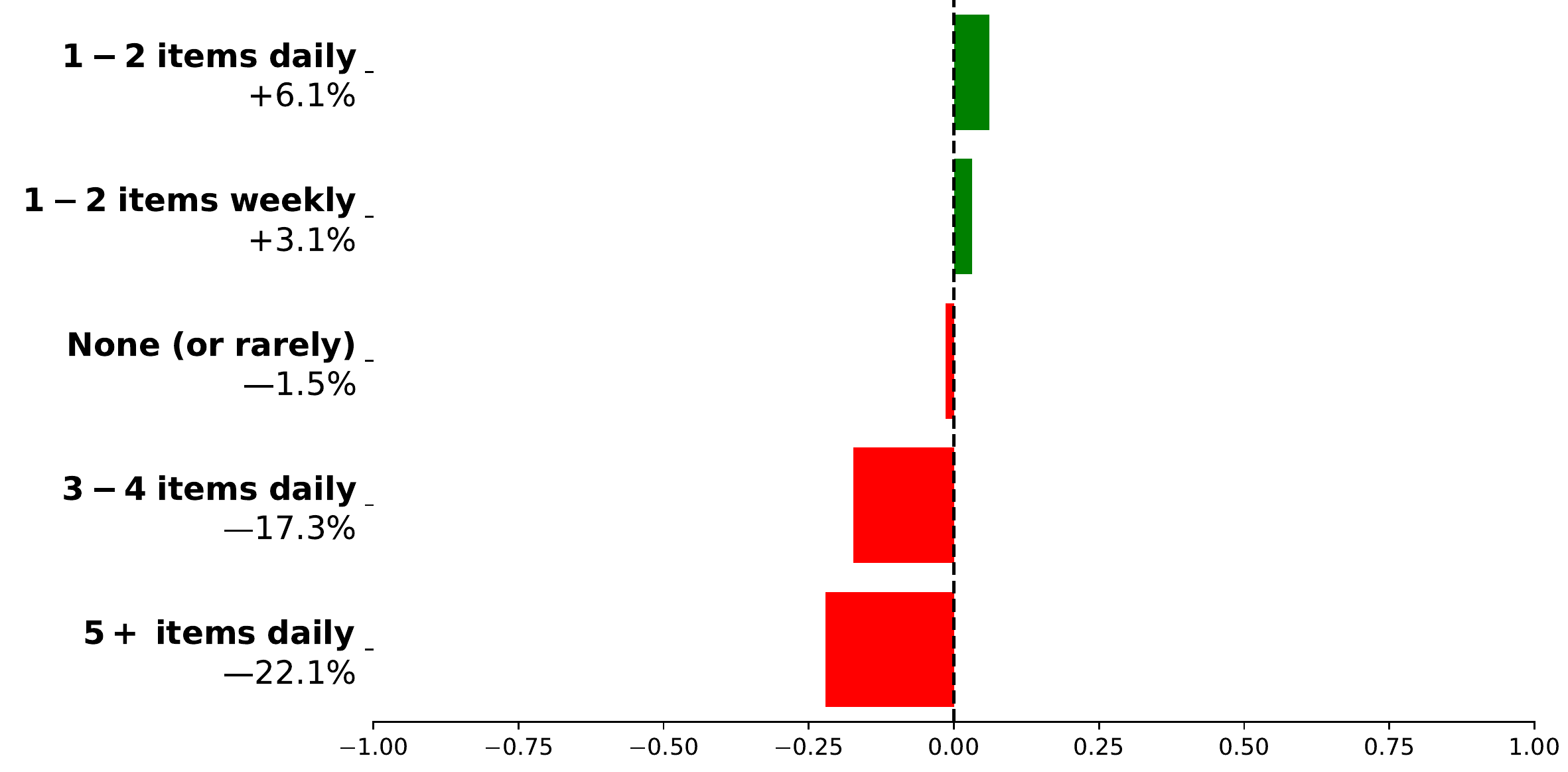}

\noindent \textbf{Caffeine Consumption.} Do you usually consume caffeine in the 3 hours before starting to play Beat Saber?
\\ \noindent \includegraphics[width=\linewidth]{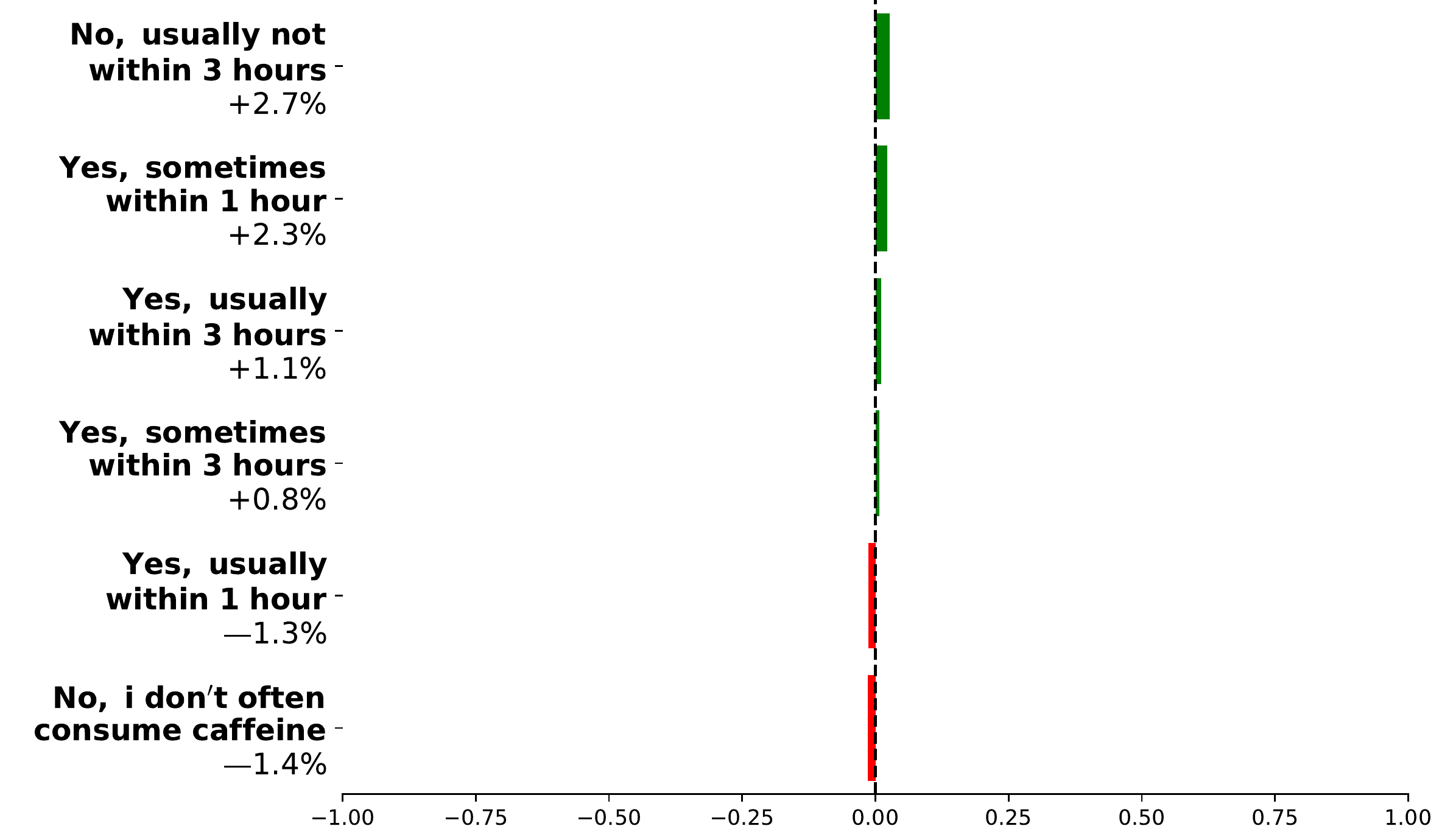}

\noindent \textbf{Substance Use.} How often do you play Beat Saber while under the influence of an intoxicating substance?
\\ \noindent \includegraphics[width=\linewidth]{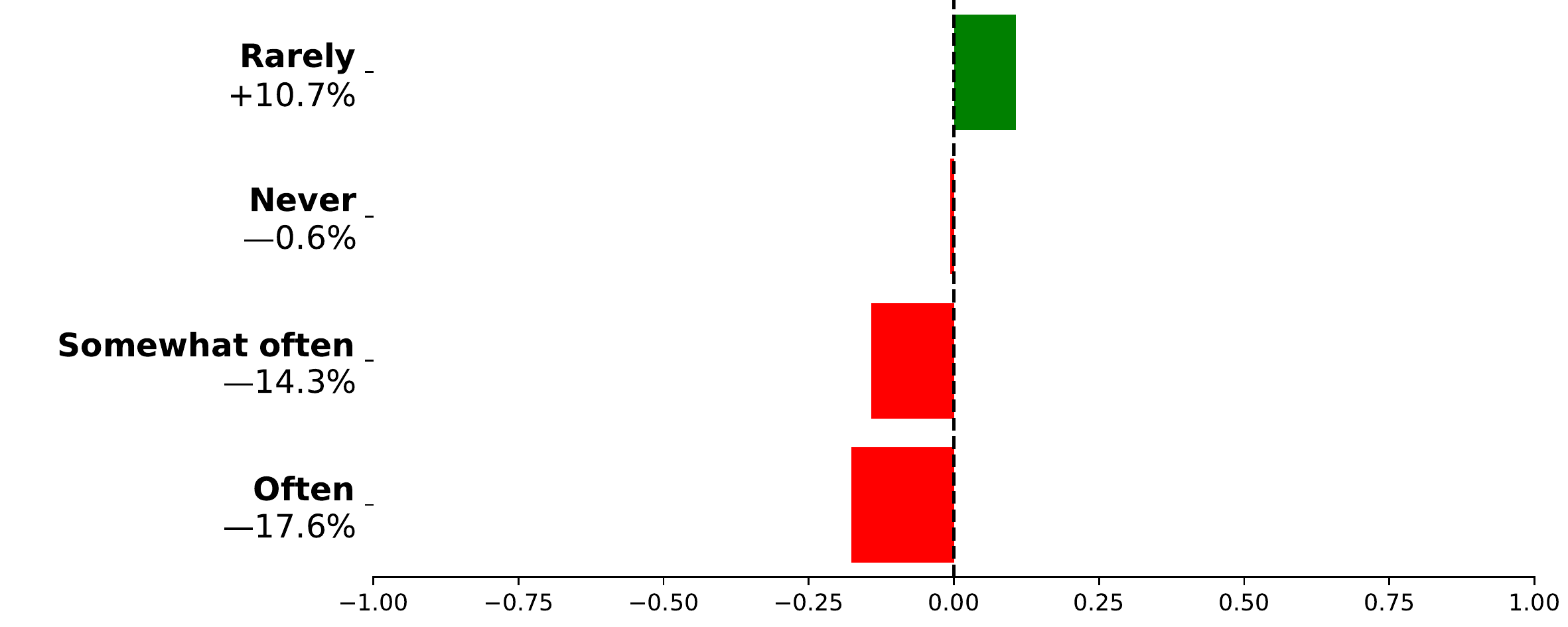}

\noindent \textbf{Physical Fitness.} How would you rate your current level of overall physical fitness?
\\ \noindent \includegraphics[width=\linewidth]{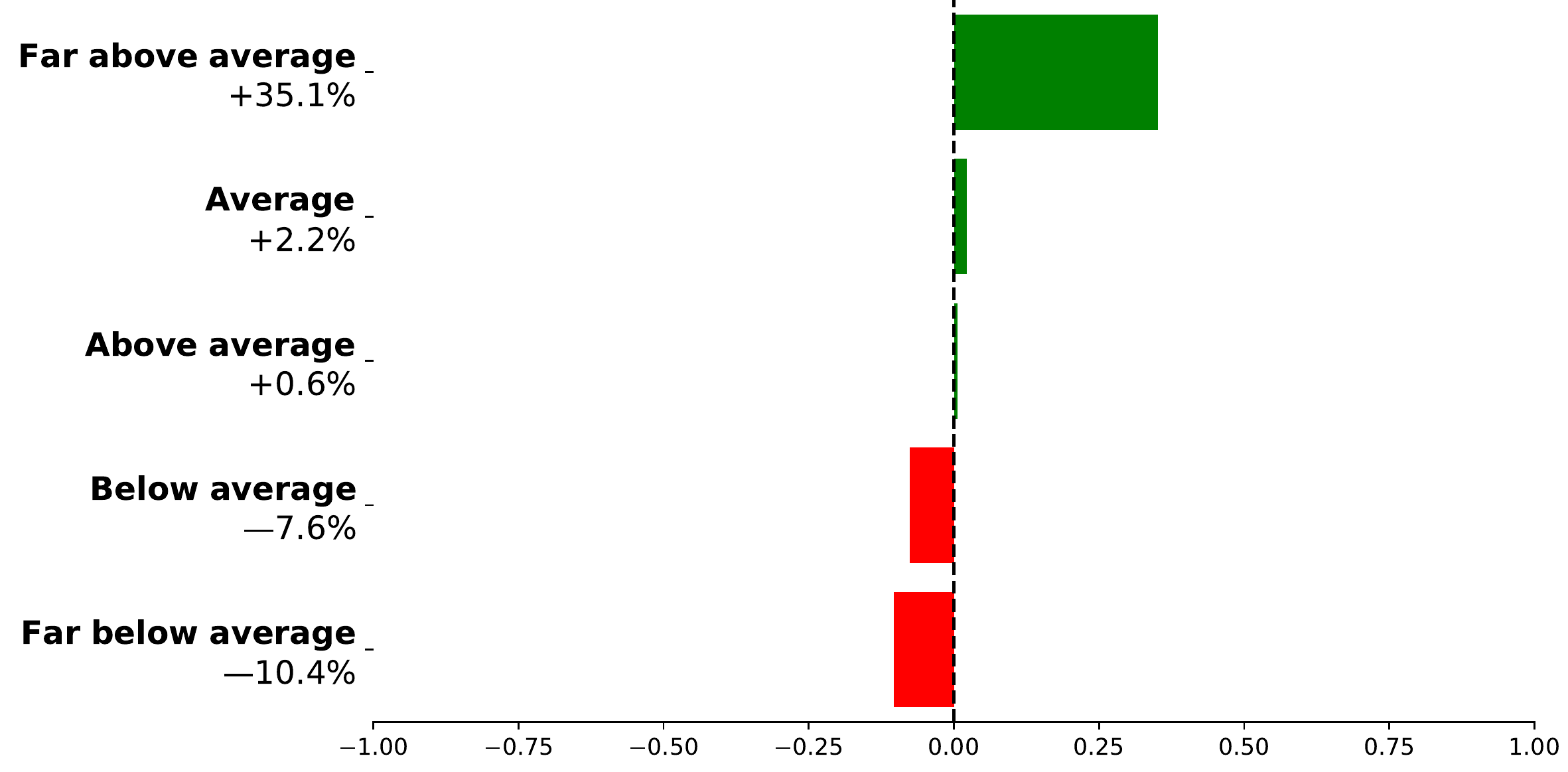}

\eject

\noindent \textbf{Preparation.} Which of the following activities, if any, do you perform immediately before playing Beat Saber?
\\ \noindent \includegraphics[width=\linewidth]{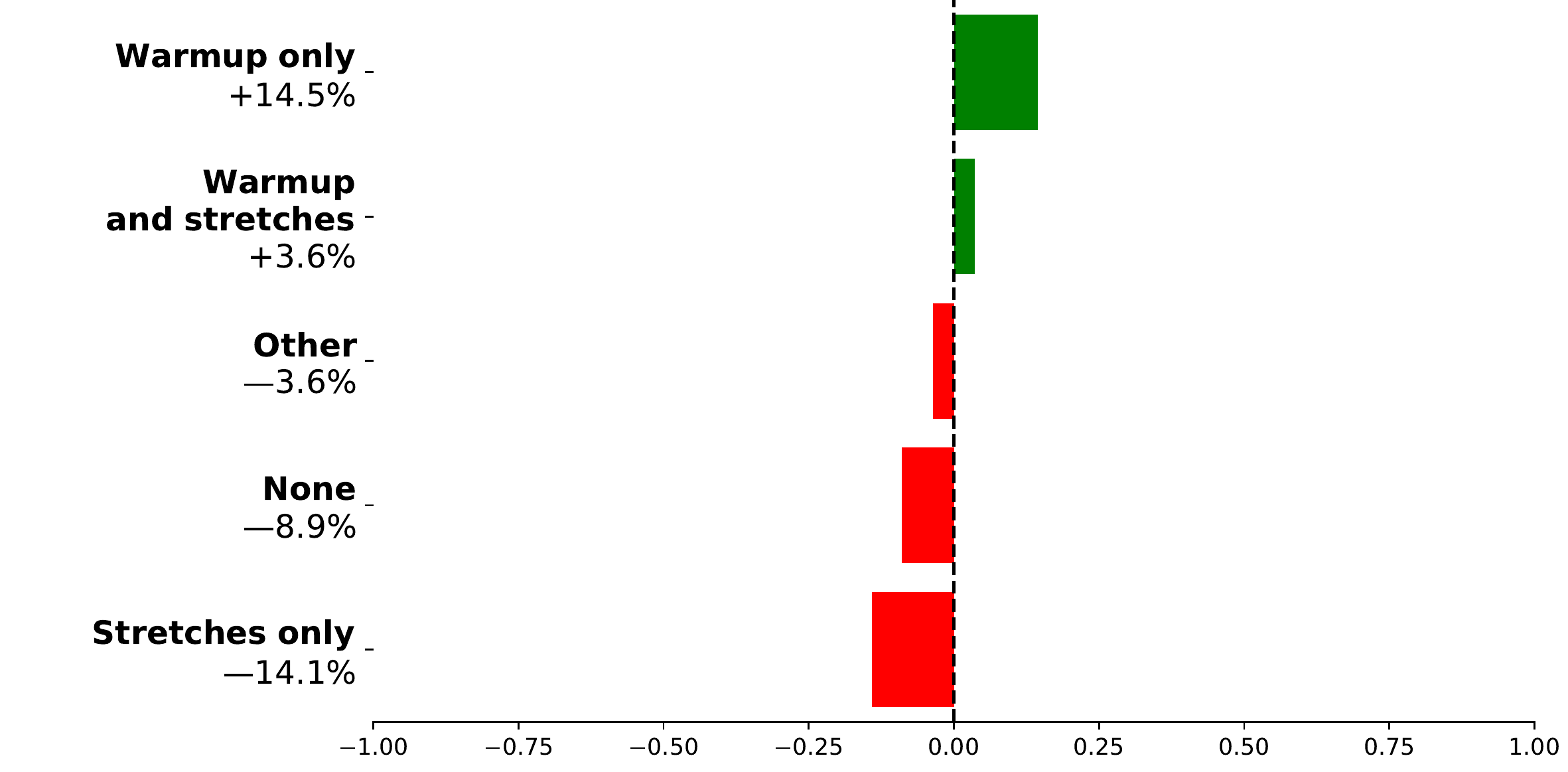}

\medskip

Interestingly, while warm-up activities have a positive effect on Beat Saber performance, stretching appears to have a negative effect, perhaps due to a secondary correlation with age; individuals who reported only stretching before playing demonstrated worse performance than those performing no preperatory activities at all.

\subsection{Environment}

Next, we evaluate the effect of environment on Beat Saber performance. Here, the trend points to better performance associated with individuals playing comfortably in their own homes, with limited obstructive clothing.

\bigskip

\noindent \textbf{Venue.} In which location do you most often play Beat Saber?
\\ \noindent \includegraphics[width=\linewidth]{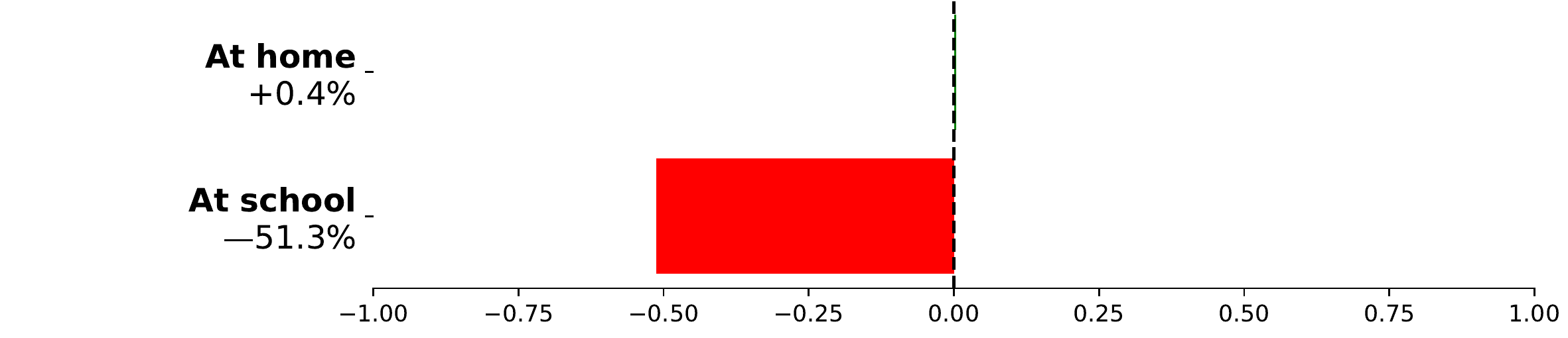}

\medskip

\noindent \textbf{Lower Body.} What clothing, if any, do you typically wear on your lower body when playing Beat Saber?
\\ \noindent \includegraphics[width=\linewidth]{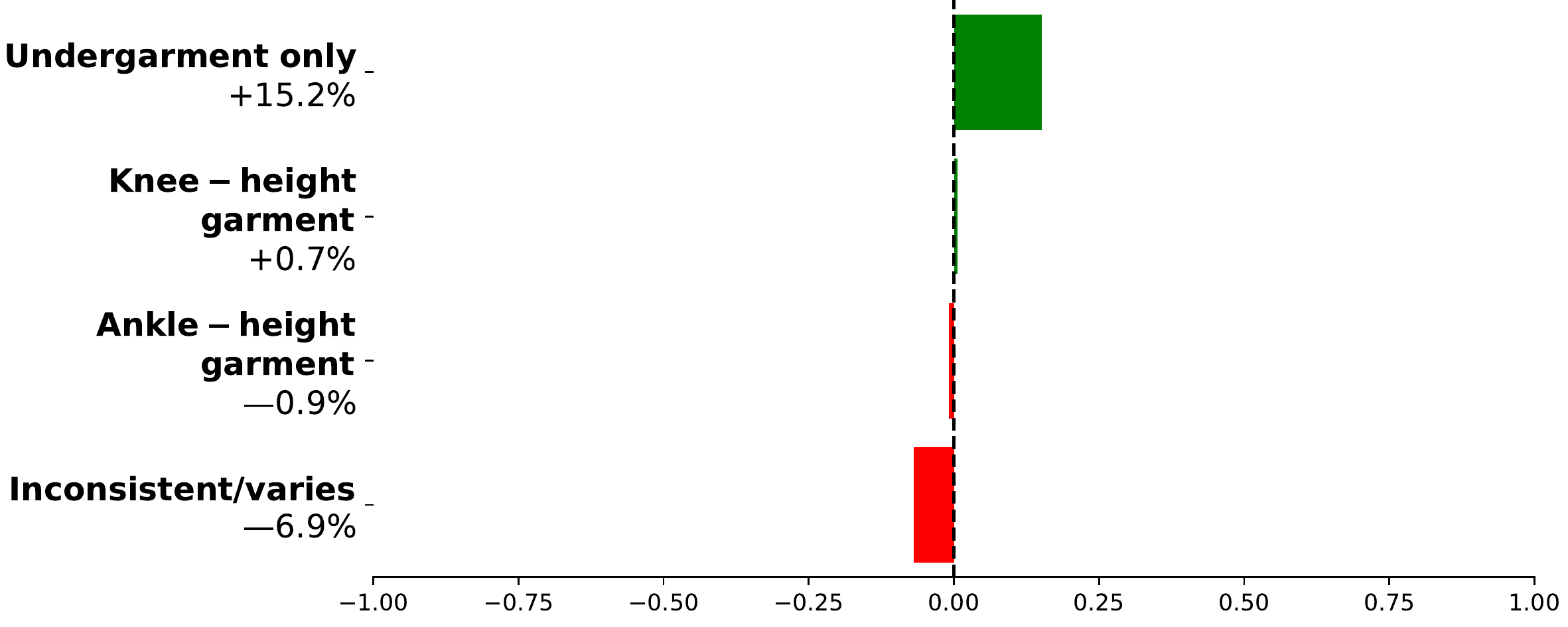}

\medskip

\noindent \textbf{Upper Body.} What clothing, if any, do you typically wear on your upper body when playing Beat Saber?
\\ \noindent \includegraphics[width=\linewidth]{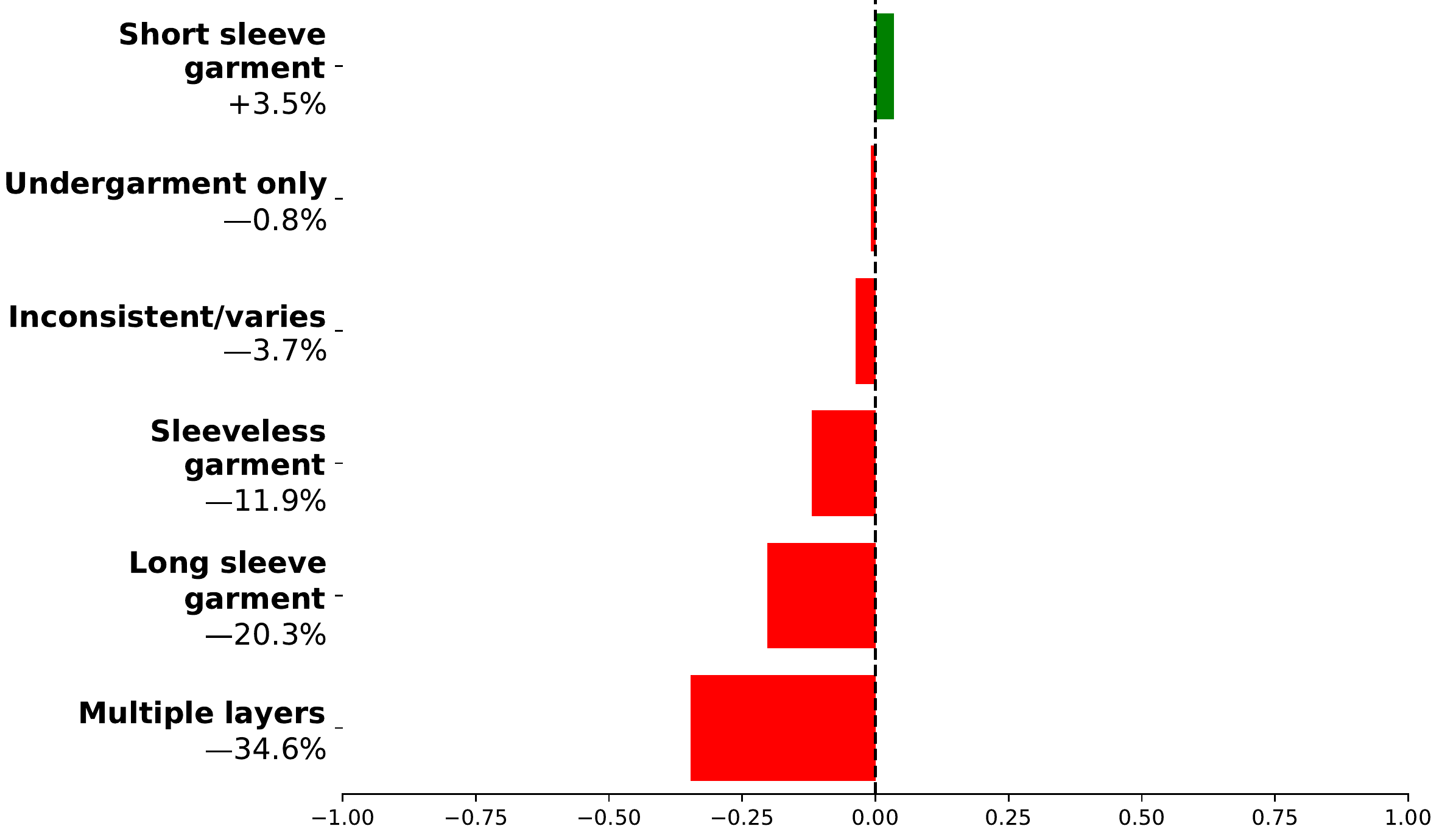}

\eject

\noindent \textbf{Footwear.} What footwear, if any, do you typically wear when playing Beat Saber?
\\ \noindent \includegraphics[width=\linewidth]{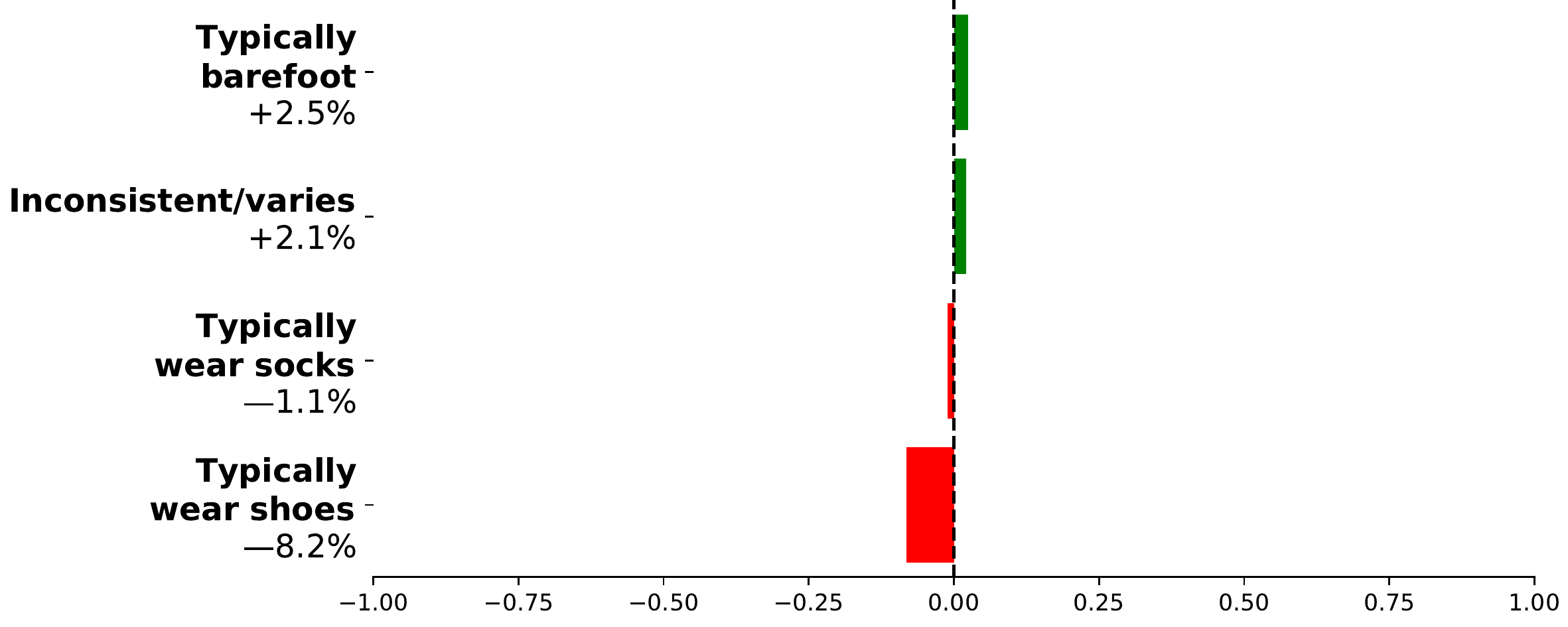}

\medskip

\subsection{Anthropometrics}

Finally, we evaluate the effect of anthropometric measurements on in-game performance. In general, larger individuals seem to have a performance advantage, with an additional clear advantage for ambidextrous players.

\bigskip

\noindent \textbf{Wingspan.} What is your exact wingspan in centimeters?
\\ \noindent \includegraphics[width=\linewidth]{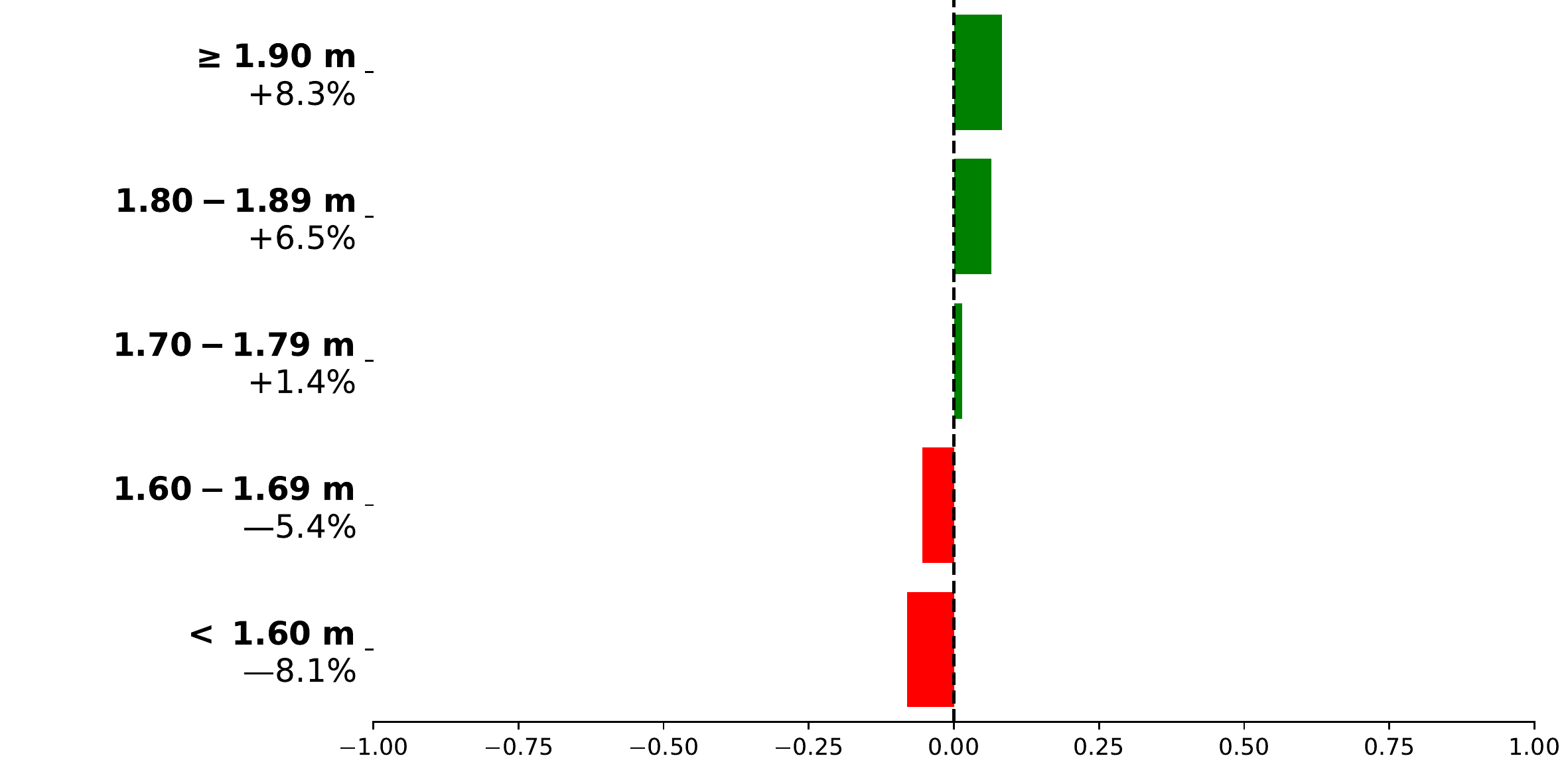}

\medskip

\noindent \textbf{Hand Length.} What is your hand length in centimeters?
\\ \noindent \includegraphics[width=\linewidth]{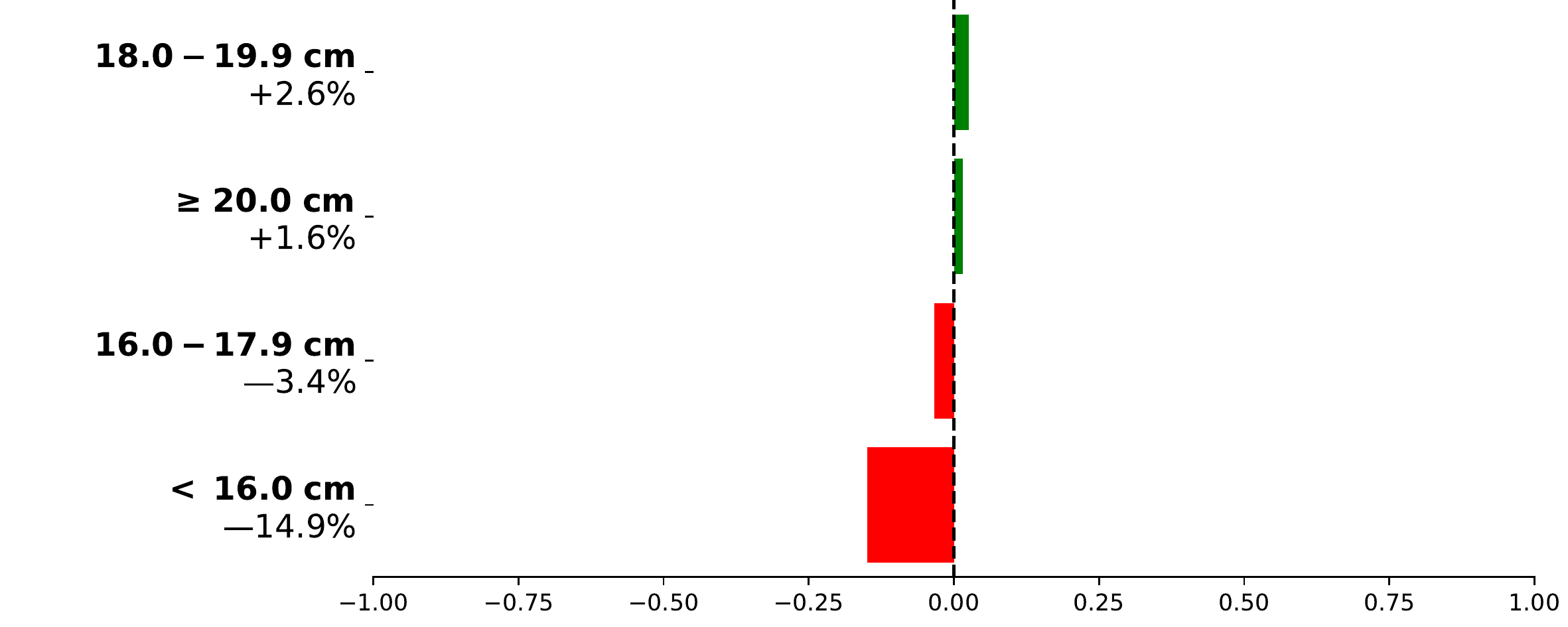}

\medskip

\noindent \textbf{Interpupillary Distance.} What is your exact interpupillary distance (IPD) in millimeters?
\\ \noindent \includegraphics[width=\linewidth]{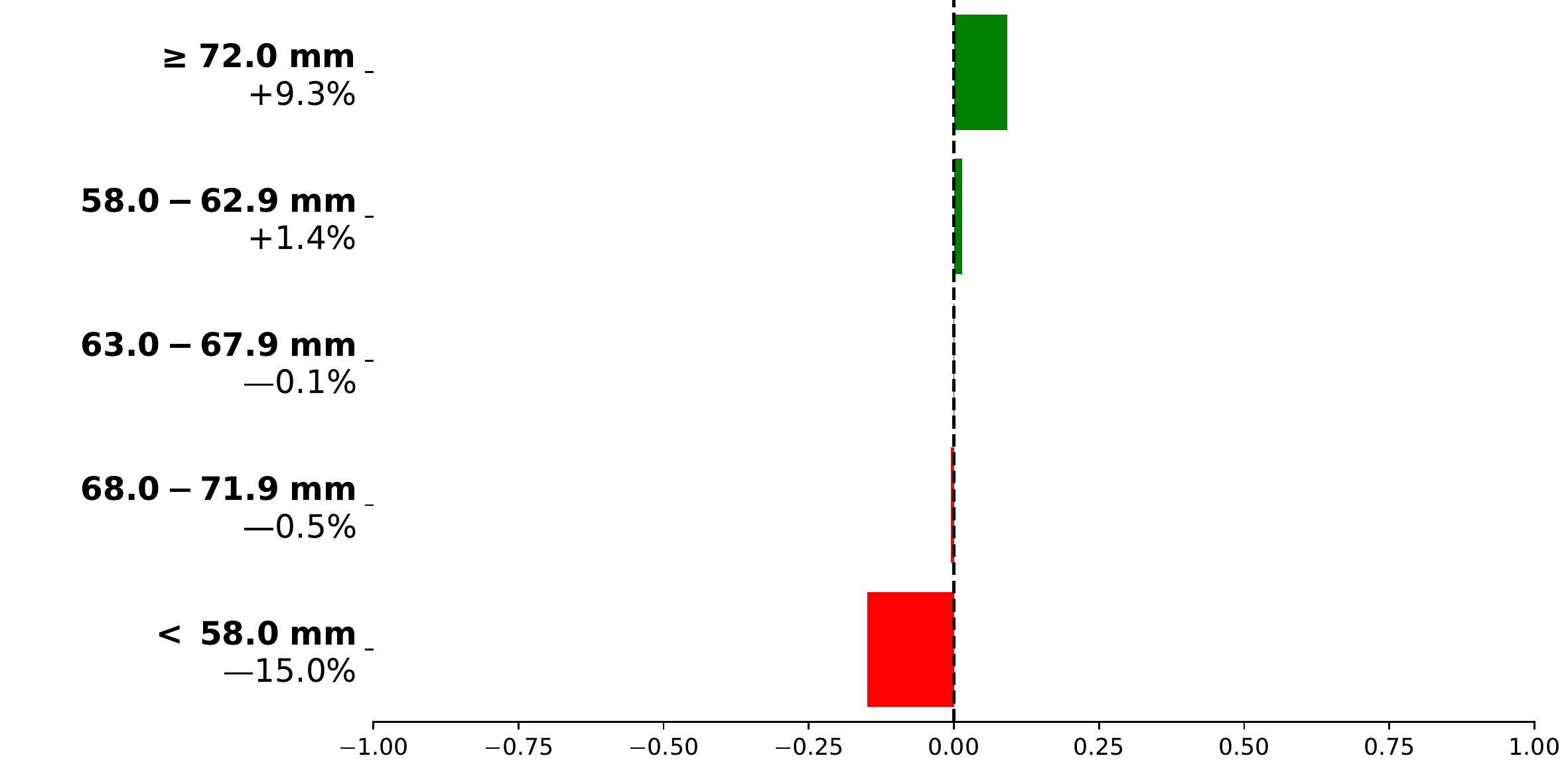}

\medskip

\noindent \textbf{Handedness.} Are you left or right handed?
\\ \noindent \includegraphics[width=\linewidth]{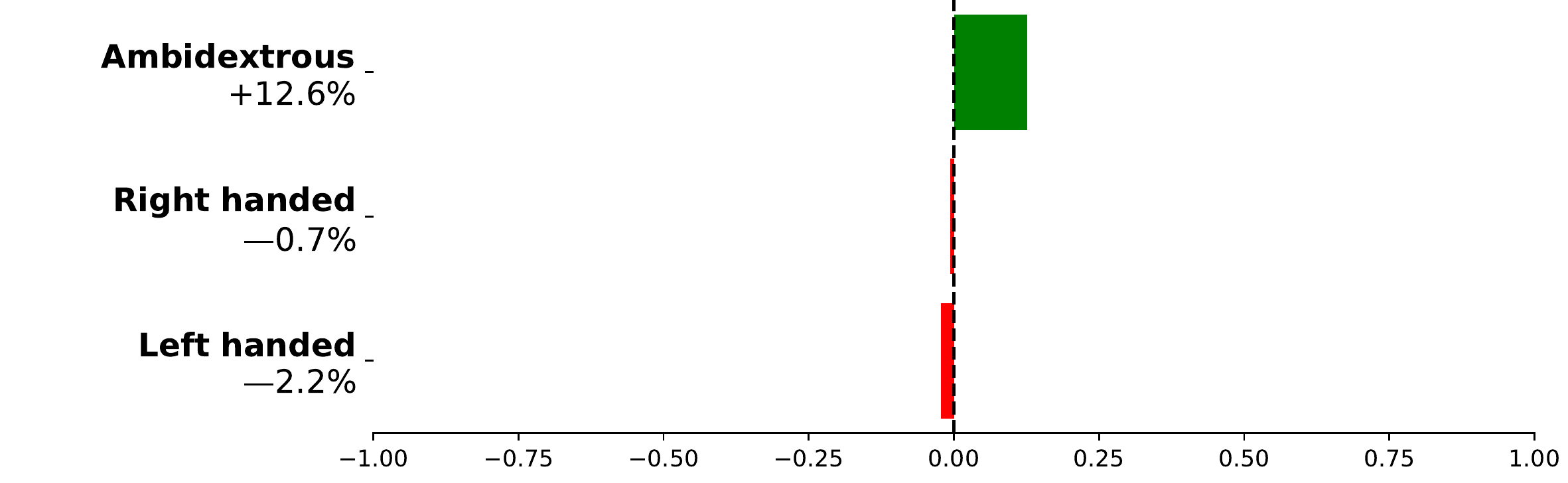}

\eject

\section{Key Findings}

In this section, we summarize the most notable results of the preceding analysis as ten key findings. For each finding, we performed an appropriate test of statistical significance to accompany the result and commentary.

\medskip

\noindent \textbf{(KF1)} Age is inversely correlated with performance.

\begin{itemize}[leftmargin=1.5em]
	\renewcommand{\labelitemi}{$\Rightarrow$}
	\item Two-Tailed Wald Test: N=833, r=-0.120, p=0.0005
\end{itemize}

There is a clear trend of younger players outperforming older players on average, with 18-19 being the best performing age group. Possible explanations include a reduction with age of eyesight, reaction time, physical fitness, and neuroplasticity. A number of other correlations, including employment status, marital status, and educational status, are likely being observed, at least in part, as a proxy for age. \\

\noindent \textbf{(KF2)} Political orientation is correlated with performance.

\begin{itemize}[leftmargin=1.5em]
	\renewcommand{\labelitemi}{$\Rightarrow$}
	\item Two-Tailed Paired T-Test: N=326, t=-2.585, p=0.0102
\end{itemize}

Self-identified conservatives out-perform self-identified liberals by an average of 20.8\%. This trend is particularly significant in that it runs contrary to KF1. \\

\noindent \textbf{(KF3)} System type is correlated with performance.

\begin{itemize}[leftmargin=1.5em]
	\renewcommand{\labelitemi}{$\Rightarrow$}
	\item Two-Tailed Paired T-Test: N=526, t=-2.429, p=0.0155
\end{itemize}

Laptop users underperform desktop users by an average of 17.0\%. This aligns with the overall trend of PC specifications, such as CPU speed, core count, and system memory, correlating with performance. \\

\noindent \textbf{(KF4)} Operating system is correlated with performance.

\begin{itemize}[leftmargin=1.5em]
	\renewcommand{\labelitemi}{$\Rightarrow$}
	\item One-Way ANOVA: N=531, f=5.681, p=0.0036
\end{itemize}

Windows 10 users outperform Windows 11 users by an average of 10.2\%. A possible explanation for this may be that productivity-oriented users are more likely to use Windows 11, while those who primarily use their PC for gaming may prefer Windows 10. \\

\noindent \textbf{(KF5)} Athletic experience is correlated with performance.

\begin{itemize}[leftmargin=1.5em]
	\renewcommand{\labelitemi}{$\Rightarrow$}
	\item Two-Tailed Paired T-Test: N=856, t=3.208, p=0.0014
\end{itemize}

Users with any experience playing an individual or team sport perform 11.9\% better than those without such experience on average. Sports requiring a high degree of hand-eye coordination, such as Badminton, Tennis, and Golf, appear to have the highest positive effect on Beat Saber performance. In general, sports involving arm exercise, such as Swimming, also have a positive effect. In this respect, Beat Saber is more like an athletic sport than a typical rhythm game. In fact, experience with musical instruments, dancing, or other rythm games has a slight negative effect on Beat Saber performance.

\eject

\noindent \textbf{(KF6)} Controller grip is strongly correlated with performance.

\begin{itemize}[leftmargin=1.5em]
	\renewcommand{\labelitemi}{$\Rightarrow$}
	\item One-Way ANOVA: N=363, f=3.599, p=0.0005
\end{itemize}

Players utilizing the two-finger ``Claw Grip'' and three-finger ``M-Grip'' (and similar ``Yoshi M-Grip'') demonstrate the best performance, with up to 64.5\% better performance than the default grip on Oculus Quest controllers. \\

\noindent \textbf{(KF7)} Wearing undergarments only is strongly correlated with better performance.

\begin{itemize}[leftmargin=1.5em]
	\renewcommand{\labelitemi}{$\Rightarrow$}
	\item Two-Tailed Paired T-Test: N=247, t=3.167, p=0.0017
\end{itemize}

In general, less clothing appears to correlate with better performance, with fewer garments on the upper body, lower body, and feet all being associated with better average scores. This could be explained by the fact that clothing may serve to limit users' range of motion, or to reduce heat dissipation. \\

\noindent \textbf{(KF8)} Wingspan is positively correlated with performance.

\begin{itemize}[leftmargin=1.5em]
	\renewcommand{\labelitemi}{$\Rightarrow$}
	\item Two-Tailed Wald Test: N=635, r=-0.113, p=0.0045
\end{itemize}

Players with a wider wingspan consistently achieve better peak performance than those with a smaller wingspan. Having larger hands also appears to be beneficial. Height, by contrast, does not correlate as strongly with Beat Saber performance, perhaps because an in-game height adjustment setting exists. \\
	
\noindent \textbf{(KF9)} Ambidextrous players have a significant advantage.

\begin{itemize}[leftmargin=1.5em]
	\renewcommand{\labelitemi}{$\Rightarrow$}
	\item Two-Tailed Paired T-Test: N=792, t=1.736, p=0.0830
\end{itemize}

Ambidextrous players appear to score significantly better than both right-handed and left-handed users by an average of 12.6\%. Left-handed users tend to slightly underperform right-handed users, by a margin of about 1.2\%. This may be because most maps are designed for right-handed users, and the game's ``left-handed mode,'' which mirrors levels, is not always utilized. \\

\noindent \textbf{(KF10)} Preparation is highly correlated with performance.

\begin{itemize}[leftmargin=1.5em]
	\renewcommand{\labelitemi}{$\Rightarrow$}
	\item One-Way ANOVA: N=723, f=7.955, p=2.8161e-06
\end{itemize}

The chosen method of preparation appears to have a significant impact on performance. Warm-up activities, such as playing easier than normal maps at the start of a session, improve scores by as much as 14.1\%, while stretches appear to have a negative effect, decreasing scores by 14.5\%.

\medskip

We note that the distribution of performance points appears to be non-uniform (see below); a linear increase in PP may indicate a non-linear increase in skill, and the PP effects noted above should be interpreted accordingly.

\eject

\twocolumn[\section{Model Beat Saber Players}]

\subsection{Hypothetical Median Beat Saber Player}

\noindent The following illustrates a fictional Beat Saber user with median values across all sampled attributes:

\bigskip

\noindent \includegraphics[width=\linewidth]{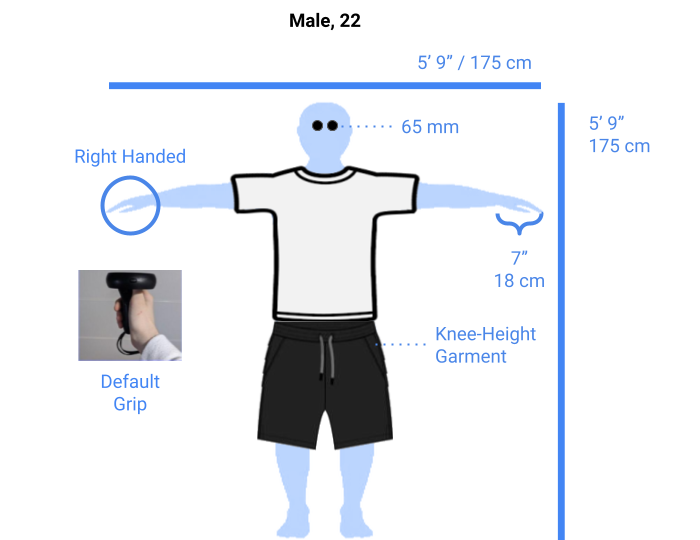}

\bigskip

\begin{itemize}
    \item \textbf{Employment Status}: Student
    \item \textbf{Educational Status}: Less than high school
    \item \textbf{Political Orientation}: Independent or neither
    \item \textbf{Primary Language}: English
    \item \textbf{Base Stations}: 2
    \item \textbf{Rhythm Games}: Yes
    \item \textbf{Athletics}: No
    \item \textbf{Caffinated Items}: None (or rarely)
    \item \textbf{Physical Fitness}: Average
    \item \textbf{Preparation}: None
\end{itemize}

\bigskip

The following histogram shows where this hypothetical user would likely fall on the overall distribution of performance points observed across the surveyed users:

\bigskip

\noindent \includegraphics[width=\linewidth]{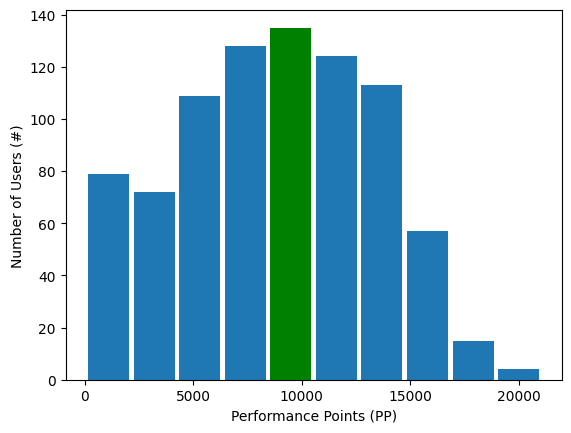}

\newpage

\subsection{Hypothetical Best-Performing Player}

\noindent The following illustrates a fictional Beat Saber user with the theoretically ``optimal'' value of each attribute:

\bigskip

\noindent \includegraphics[width=\linewidth]{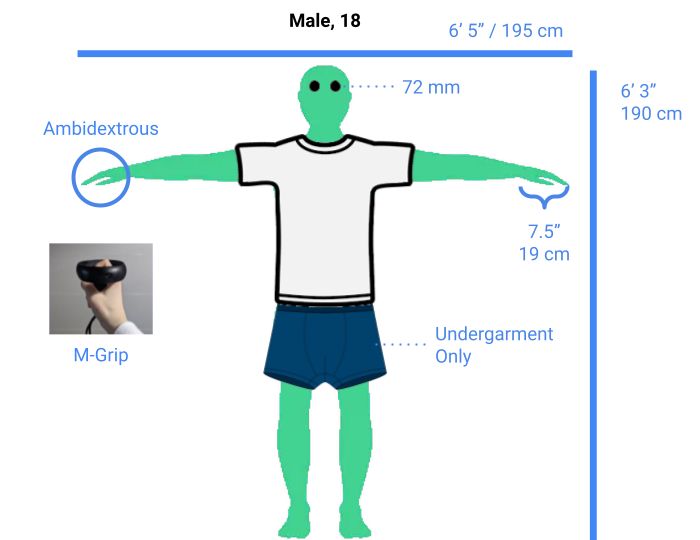}

\bigskip

\begin{itemize}
    \item \textbf{Employment Status}: Employed part time
    \item \textbf{Educational Status}: High school graduate
    \item \textbf{Political Orientation}: Conservative or right wing
    \item \textbf{Primary Language}: Spanish
    \item \textbf{Base Stations}: 1
    \item \textbf{Rhythm Games}: No
    \item \textbf{Athletics}: Yes
    \item \textbf{Caffinated Items}: 1-2 items daily
    \item \textbf{Physical Fitness}: Far above average
    \item \textbf{Preparation}: Warmup only
\end{itemize}

\bigskip

The following histogram shows where this hypothetical user would likely fall on the overall distribution of performance points observed across the surveyed users:

\bigskip

\noindent \includegraphics[width=\linewidth]{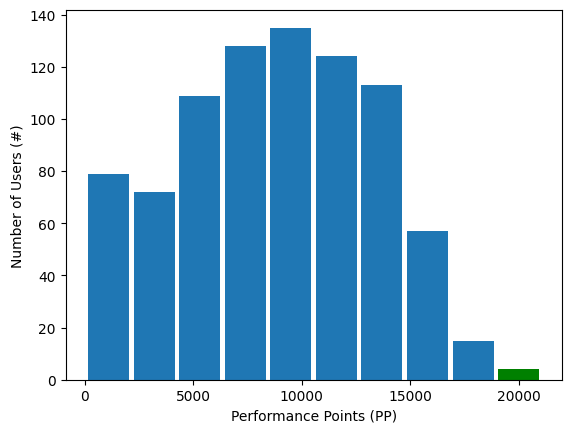}

\onecolumn

\section*{Interactive Report}

\noindent Readers are invited to explore an interactive version of this report at \url{https://www.beatleader.xyz/census2023}.

\bibliographystyle{plainurl}
\bibliography{references}


\end{document}